\documentclass{ptephy}

\preprintnumber{XXXX-XXXX}

\usepackage{lineno}
\usepackage{delarray}
\usepackage{multirow}
\usepackage{amsmath,amssymb}
\usepackage{wrapfig}
\usepackage{color}
\usepackage{subfig}
\usepackage{dcolumn}  
\usepackage{lscape}
\usepackage{setspace}
\usepackage{textcomp}
\usepackage{feynmp}
\usepackage{here}

\allowdisplaybreaks[3]

\newcommand{\mydif}{\mathrm{d}}
\newcommand{\bvec}[1]{\mbox{\boldmath $#1$}}

\begin{document}

\title{Measurement of the tau Michel parameters $\bar{\eta}$ and $\xi\kappa$
 in the radiative leptonic decay $\tau^- \rightarrow \ell^- \nu_{\tau} \bar{\nu}_{\ell}\gamma$ \\ {\small Belle preprint 2017-20, KEK preprint 2017-29}}
{~\\ \tiny ~}
\author{
\name{N.~Shimizu}{75}, \name{H.~Aihara}{75}, \name{D.~Epifanov}{3,59},
\name{I.~Adachi}{15,11},
\name{S.~Al~Said}{69,34}, \name{D.~M.~Asner}{61}, \name{V.~Aulchenko}{3,59}, \name{T.~Aushev}{49},
\name{R.~Ayad}{69}, \name{V.~Babu}{70}, \name{I.~Badhrees}{69,33}, \name{A.~M.~Bakich}{68}, \name{V.~Bansal}{61},
\name{E.~Barberio}{47}, \name{V.~Bhardwaj}{17}, \name{B.~Bhuyan}{19}, \name{J.~Biswal}{29}, \name{A.~Bobrov}{3,59},
\name{A.~Bozek}{56}, \name{M.~Bra\v{c}ko}{45,29}, \name{T.~E.~Browder}{14}, \name{D.~\v{C}ervenkov}{4}, \name{M.-C.~Chang}{9},
\name{P.~Chang}{55}, \name{V.~Chekelian}{46}, \name{A.~Chen}{53}, \name{B.~G.~Cheon}{13}, \name{K.~Chilikin}{40,48},
\name{K.~Cho}{35}, \name{S.-K.~Choi}{12}, \name{Y.~Choi}{67}, \name{D.~Cinabro}{80}, \name{T.~Czank}{73},
\name{N.~Dash}{18}, \name{S.~Di~Carlo}{80}, \name{Z.~Dole\v{z}al}{4}, \name{D.~Dutta}{70}, \name{S.~Eidelman}{3,59},
\name{J.~E.~Fast}{61}, \name{T.~Ferber}{7}, \name{B.~G.~Fulsom}{61}, \name{R.~Garg}{62},
\name{V.~Gaur}{79}, \name{N.~Gabyshev}{3,59}, \name{A.~Garmash}{3,59}, \name{M.~Gelb}{31}, \name{P.~Goldenzweig}{31},
\name{D.~Greenwald}{71}, \name{E.~Guido}{27}, \name{J.~Haba}{15,11}, \name{K.~Hayasaka}{58}, \name{H.~Hayashii}{52},
\name{M.~T.~Hedges}{14}, \name{S.~Hirose}{50}, \name{W.-S.~Hou}{55}, \name{T.~Iijima}{51,50}, \name{K.~Inami}{50},
\name{G.~Inguglia}{7}, \name{A.~Ishikawa}{73}, \name{R.~Itoh}{15,11}, \name{M.~Iwasaki}{60}, \name{I.~Jaegle}{8},
\name{H.~B.~Jeon}{38}, \name{S.~Jia}{2}, \name{Y.~Jin}{75}, \name{K.~K.~Joo}{5}, \name{T.~Julius}{47},
\name{K.~H.~Kang}{38}, \name{G.~Karyan}{7}, \name{T.~Kawasaki}{58}, \name{C.~Kiesling}{46}, \name{D.~Y.~Kim}{65},
\name{J.~B.~Kim}{36}, \name{S.~H.~Kim}{13}, \name{Y.~J.~Kim}{35}, \name{K.~Kinoshita}{6}, \name{P.~Kody\v{s}}{4},
\name{S.~Korpar}{45,29}, \name{D.~Kotchetkov}{14}, \name{P.~Kri\v{z}an}{41,29}, \name{R.~Kroeger}{25}, \name{P.~Krokovny}{3,59},
\name{R.~Kulasiri}{32}, \name{A.~Kuzmin}{3,59}, \name{Y.-J.~Kwon}{82}, \name{J.~S.~Lange}{10}, \name{I.~S.~Lee}{13},
\name{L.~K.~Li}{22}, \name{Y.~Li}{79}, \name{L.~Li~Gioi}{46}, \name{J.~Libby}{20}, \name{D.~Liventsev}{79,15},
\name{M.~Masuda}{74}, \name{M.~Merola}{26}, \name{K.~Miyabayashi}{52}, \name{H.~Miyata}{58}, \name{G.~B.~Mohanty}{70},
\name{H.~K.~Moon}{36}, \name{T.~Mori}{50}, \name{R.~Mussa}{27}, \name{E.~Nakano}{60}, \name{M.~Nakao}{15,11},
\name{T.~Nanut}{29}, \name{K.~J.~Nath}{19}, \name{Z.~Natkaniec}{56}, \name{M.~Nayak}{80,15}, \name{M.~Niiyama}{37},
\name{N.~K.~Nisar}{63}, \name{S.~Nishida}{15,11}, \name{S.~Ogawa}{72}, \name{S.~Okuno}{30}, \name{H.~Ono}{57,58},
\name{G.~Pakhlova}{40,49}, \name{B.~Pal}{6}, \name{C.~W.~Park}{67}, \name{H.~Park}{38}, \name{S.~Paul}{71},
\name{T.~K.~Pedlar}{43}, \name{R.~Pestotnik}{29}, \name{L.~E.~Piilonen}{79}, \name{V.~Popov}{49}, \name{M.~Ritter}{42},
\name{A.~Rostomyan}{7}, \name{Y.~Sakai}{15,11}, \name{M.~Salehi}{44,42}, \name{S.~Sandilya}{6}, \name{Y.~Sato}{50},
\name{V.~Savinov}{63}, \name{O.~Schneider}{39}, \name{G.~Schnell}{1,16}, \name{C.~Schwanda}{23}, \name{Y.~Seino}{58},
\name{K.~Senyo}{81}, \name{M.~E.~Sevior}{47}, \name{V.~Shebalin}{3,59}, \name{T.-A.~Shibata}{76}, \name{J.-G.~Shiu}{55}, \name{B.~Shwartz}{3,59}, \name{A.~Sokolov}{24}, \name{E.~Solovieva}{40,49}, \name{M.~Stari\v{c}}{29},
\name{J.~F.~Strube}{61}, \name{K.~Sumisawa}{15,11}, \name{T.~Sumiyoshi}{77}, \name{U.~Tamponi}{27,78}, \name{K.~Tanida}{28},
\name{F.~Tenchini}{47}, \name{K.~Trabelsi}{15,11}, \name{M.~Uchida}{76}, \name{T.~Uglov}{40,49}, \name{Y.~Unno}{13},
\name{S.~Uno}{15,11}, \name{Y.~Usov}{3,59}, \name{C.~Van~Hulse}{1}, \name{G.~Varner}{14}, \name{V.~Vorobyev}{3,59},
\name{A.~Vossen}{21}, \name{C.~H.~Wang}{54}, \name{M.-Z.~Wang}{55}, \name{P.~Wang}{22}, \name{M.~Watanabe}{58},
\name{E.~Widmann}{66}, \name{E.~Won}{36}, \name{Y.~Yamashita}{57}, \name{H.~Ye}{7}, \name{C.~Z.~Yuan}{22},
\name{Z.~P.~Zhang}{64}, \name{V.~Zhilich}{3,59}, \name{V.~Zhukova}{40,48}, \name{V.~Zhulanov}{3,59}, \name{A.~Zupanc}{41,29},
}

\address{
\affil{1}{University of the Basque Country UPV/EHU, 48080 Bilbao}
\affil{2}{Beihang University, Beijing 100191}
\affil{3}{Budker Institute of Nuclear Physics SB RAS, Novosibirsk 630090}
\affil{4}{Faculty of Mathematics and Physics, Charles University, 121 16 Prague}
\affil{5}{Chonnam National University, Kwangju 660-701}
\affil{6}{University of Cincinnati, Cincinnati, Ohio 45221}
\affil{7}{Deutsches Elektronen--Synchrotron, 22607 Hamburg}
\affil{8}{University of Florida, Gainesville, Florida 32611}
\affil{9}{Department of Physics, Fu Jen Catholic University, Taipei 24205}
\affil{10}{Justus-Liebig-Universit\"at Gie\ss{}en, 35392 Gie\ss{}en}
\affil{11}{SOKENDAI (The Graduate University for Advanced Studies), Hayama 240-0193}
\affil{12}{Gyeongsang National University, Chinju 660-701}
\affil{13}{Hanyang University, Seoul 133-791}
\affil{14}{University of Hawaii, Honolulu, Hawaii 96822}
\affil{15}{High Energy Accelerator Research Organization (KEK), Tsukuba 305-0801}
\affil{16}{IKERBASQUE, Basque Foundation for Science, 48013 Bilbao}
\affil{17}{Indian Institute of Science Education and Research Mohali, SAS Nagar, 140306}
\affil{18}{Indian Institute of Technology Bhubaneswar, Satya Nagar 751007}
\affil{19}{Indian Institute of Technology Guwahati, Assam 781039}
\affil{20}{Indian Institute of Technology Madras, Chennai 600036}
\affil{21}{Indiana University, Bloomington, Indiana 47408}
\affil{22}{Institute of High Energy Physics, Chinese Academy of Sciences, Beijing 100049}
\affil{23}{Institute of High Energy Physics, Vienna 1050}
\affil{24}{Institute for High Energy Physics, Protvino 142281}
\affil{25}{University of Mississippi, University, Mississippi 38677}
\affil{26}{INFN - Sezione di Napoli, 80126 Napoli}
\affil{27}{INFN - Sezione di Torino, 10125 Torino}
\affil{28}{Advanced Science Research Center, Japan Atomic Energy Agency, Naka 319-1195}
\affil{29}{J. Stefan Institute, 1000 Ljubljana}
\affil{30}{Kanagawa University, Yokohama 221-8686}
\affil{31}{Institut f\"ur Experimentelle Kernphysik, Karlsruher Institut f\"ur Technologie, 76131 Karlsruhe}
\affil{32}{Kennesaw State University, Kennesaw, Georgia 30144}
\affil{33}{King Abdulaziz City for Science and Technology, Riyadh 11442}
\affil{34}{Department of Physics, Faculty of Science, King Abdulaziz University, Jeddah 21589}
\affil{35}{Korea Institute of Science and Technology Information, Daejeon 305-806}
\affil{36}{Korea University, Seoul 136-713}
\affil{37}{Kyoto University, Kyoto 606-8502}
\affil{38}{Kyungpook National University, Daegu 702-701}
\affil{39}{\'Ecole Polytechnique F\'ed\'erale de Lausanne (EPFL), Lausanne 1015}
\affil{40}{P.N. Lebedev Physical Institute of the Russian Academy of Sciences, Moscow 119991}
\affil{41}{Faculty of Mathematics and Physics, University of Ljubljana, 1000 Ljubljana}
\affil{42}{Ludwig Maximilians University, 80539 Munich}
\affil{43}{Luther College, Decorah, Iowa 52101}
\affil{44}{University of Malaya, 50603 Kuala Lumpur}
\affil{45}{University of Maribor, 2000 Maribor}
\affil{46}{Max-Planck-Institut f\"ur Physik, 80805 M\"unchen}
\affil{47}{School of Physics, University of Melbourne, Victoria 3010}
\affil{48}{Moscow Physical Engineering Institute, Moscow 115409}
\affil{49}{Moscow Institute of Physics and Technology, Moscow Region 141700}
\affil{50}{Graduate School of Science, Nagoya University, Nagoya 464-8602}
\affil{51}{Kobayashi-Maskawa Institute, Nagoya University, Nagoya 464-8602}
\affil{52}{Nara Women's University, Nara 630-8506}
\affil{53}{National Central University, Chung-li 32054}
\affil{54}{National United University, Miao Li 36003}
\affil{55}{Department of Physics, National Taiwan University, Taipei 10617}
\affil{56}{H. Niewodniczanski Institute of Nuclear Physics, Krakow 31-342}
\affil{57}{Nippon Dental University, Niigata 951-8580}
\affil{58}{Niigata University, Niigata 950-2181}
\affil{59}{Novosibirsk State University, Novosibirsk 630090}
\affil{60}{Osaka City University, Osaka 558-8585}
\affil{61}{Pacific Northwest National Laboratory, Richland, Washington 99352}
\affil{62}{Panjab University, Chandigarh 160014}
\affil{63}{University of Pittsburgh, Pittsburgh, Pennsylvania 15260}
\affil{64}{University of Science and Technology of China, Hefei 230026}
\affil{65}{Soongsil University, Seoul 156-743}
\affil{66}{Stefan Meyer Institute for Subatomic Physics, Vienna 1090}
\affil{67}{Sungkyunkwan University, Suwon 440-746}
\affil{68}{School of Physics, University of Sydney, New South Wales 2006}
\affil{69}{Department of Physics, Faculty of Science, University of Tabuk, Tabuk 71451}
\affil{70}{Tata Institute of Fundamental Research, Mumbai 400005}
\affil{71}{Department of Physics, Technische Universit\"at M\"unchen, 85748 Garching}
\affil{72}{Toho University, Funabashi 274-8510}
\affil{73}{Department of Physics, Tohoku University, Sendai 980-8578}
\affil{74}{Earthquake Research Institute, University of Tokyo, Tokyo 113-0032}
\affil{75}{Department of Physics, University of Tokyo, Tokyo 113-0033}
\affil{76}{Tokyo Institute of Technology, Tokyo 152-8550}
\affil{77}{Tokyo Metropolitan University, Tokyo 192-0397}
\affil{78}{University of Torino, 10124 Torino}
\affil{79}{Virginia Polytechnic Institute and State University, Blacksburg, Virginia 24061}
\affil{80}{Wayne State University, Detroit, Michigan 48202}
\affil{81}{Yamagata University, Yamagata 990-8560}
\affil{82}{Yonsei University, Seoul 120-749}
}

\begin{abstract}
We present a measurement of the Michel parameters of the $\tau$ lepton,
 $\bar{\eta}$ and $\xi\kappa$, in the radiative leptonic
 decay $\tau^- \rightarrow \ell^- \nu_{\tau} \bar{\nu}_{\ell} \gamma$
 using 711~f$\mathrm{b}^{-1}$ of collision data collected 
with the Belle detector at the KEKB $e^+e^-$ collider.
 The Michel parameters are measured in an unbinned maximum likelihood fit 
to the kinematic distribution
 of $e^+e^-\rightarrow\tau^+\tau^-\rightarrow (\pi^+\pi^0 \bar{\nu}_\tau)(\ell^-\nu_{\tau}\bar{\nu}_{\ell}\gamma)$
 $(\ell=e$ or $\mu)$. The measured values of the Michel parameters
 are $\bar{\eta} = -1.3 \pm 1.5 \pm 0.8$ and $\xi\kappa = 0.5 \pm 0.4 \pm 0.2$,
 where the first error is statistical and the second
 is systematic. This is the first measurement of these parameters.
 These results are consistent
 with the Standard Model predictions within their uncertainties and
 constrain the coupling constants of the generalized weak interaction.
\end{abstract}

\subjectindex{C01, C07, C21} 
\maketitle

\section{Introduction \label{introductuion}}

In the Standard Model (SM), there are three flavors of charged leptons: 
$e, \mu$, and $\tau$. The SM has proven to be the fundamental theory in 
describing the physics of particles;
 nevertheless, precision tests may reveal the presence of physics 
beyond the Standard Model (BSM).
 In particular, a measurement of Michel parameters in leptonic and radiative leptonic
 $\tau$ decays is a powerful probe for the BSM contributions~\cite{cite_Michel,cite_michel2}.  

The most general Lorentz-invariant derivative-free matrix element 
of leptonic $\tau$ decay
 $\tau^-\rightarrow \ell^- \nu_{\tau} \bar{\nu}_{\ell} \gamma$~\footnote{Unless otherwise stated, use of charge-conjugate modes is implied
throughout the paper.} is represented as~\cite{MPform} \\ \\
\hspace{2mm}  \raisebox{1cm}{$\mathcal{M} = $} 
\begin{fmffile}{tautree}
\begin{fmfchar*}(90,60)
  \fmfleft{tm,antinu} \fmflabel{$\tau$}{tm} \fmflabel{$\nu_{\tau}$}{antinu}
  \fmf{fermion}{tm,Wi}
  \fmf{fermion}{Wi,antinu}
  \fmf{dashes,label=$W$}{Wi,Wf}
  \fmf{fermion,label=\rotatebox{60}{\vspace{-2mm}\hspace{-3mm}$~$}}{fb,Wf}
  \fmf{fermion}{Wf,f}
  \fmfright{f,fb} \fmflabel{$\nu_{\ell}$}{fb} \fmflabel{$\ell$}{f}
  \fmfdot{Wi,Wf}
\end{fmfchar*}
\end{fmffile}
\hspace{-0cm}\raisebox{1.0cm}{$=\displaystyle \frac{4G_F}{\sqrt{2}}\sum_{\substack{N=S,V,T \\ i,j=L,R}}g_{ij}^{N}\left[\overline{u}_{i}(\ell)\Gamma^{N}v_{n}(\nu_\ell)\right]\left[\overline{u}_{m}(\nu_{\tau})\Gamma_{N}u_{j}(\tau)\right],$ }(1)  \\ \\
where $G_F$ is the Fermi constant, $i$ and $j$ are the chirality
indices for the charged leptons, $n$ and $m$ are the chirality indices
of the neutrinos, $\ell$ is $e$ or $\mu$, $\Gamma^{S}=1$,
$\Gamma^{V}=\gamma^{\mu}$, and $\displaystyle
\Gamma^{T}={i}\left(\gamma^{\mu}\gamma^{\nu}-\gamma^{\nu}\gamma^{\mu}\right)/2\sqrt{2}$
are, respectively, the scalar, vector and tensor Lorentz structures
in terms of the Dirac matrices $\gamma^{\mu}$, $u_i$ and $v_i$ are the 
four-component spinors of  a particle and an antiparticle, respectively, 
and $g_{ij}^{N}$ are the corresponding dimensionless couplings. 
In the SM, $\tau^-$ decays into $\nu_{\tau}$ and a $W^-$-boson,
 the latter decays into $\ell^-$ and right-handed $\bar{\nu}_{\ell}$; 
\textit{i.e.,} the only non-zero coupling is $g_{LL}^{V}=1$.
 Experimentally, only the squared matrix element is observable and
 bilinear combinations of the $g_{ij}^{N}$ are accessible.
Of all such combinations, four Michel parameters, $\eta $, $\rho $, 
$\delta $, and $\xi $, can be measured in the leptonic decay of the $\tau$
 when the final-state neutrinos are not observed and the spin of the outgoing lepton is not measured~\cite{ordMP}:
\begin{eqnarray}
\displaystyle \rho&=&\frac{3}{4}-\frac{3}{4}\left(\left|g_{LR}^{V}\right|^{2}+\left|g_{RL}^{V}\right|^{2} +2\left|g_{LR}^{T}\right|^{2}+2\left|g_{RL}^{T}\right|^{2} +    \Re\left(g_{LR}^{S}g_{LR}^{T*}+g_{RL}^{S}g_{RL}^{T*} \right)\right)  \setcounter{equation}{2},\\
\displaystyle \eta&=&\frac{1}{2}\Re\left(6g_{RL}^{V}g_{LR}^{T*}+6g_{LR}^{V}g_{RL}^{T*}+g_{RR}^{S}g_{LL}^{V*}+g_{RL}^{S}g_{LR}^{V*}+g_{LR}^{S}g_{RL}^{V*}+g_{LL}^{S}g_{RR}^{V*}\right) ,\\
\displaystyle \xi&=&4\Re\left(g_{LR}^{S}g_{LR}^{T*}-g_{RL}^{S}g_{RL}^{T*}\right)+\left|g_{LL}^{V}\right|^{2}+3\left|g_{LR}^{V}\right|^{2}-3\left|g_{RL}^{V}\right|^{2}-\left|g_{RR}^{V}\right|^{2} \nonumber \\
&{}&+5\left|g_{LR}^{T}\right|^{2}-5\left|g_{RL}^{T}\right|^{2}+\frac{1}{4}\left(\left|g_{LL}^{S}\right|^{2}-\left|g_{LR}^{S}\right|^{2}+\left|g_{RL}^{S}\right|^{2}-\left|g_{RR}^{S}\right|^{2}\right) ,\\
\displaystyle \xi\delta&=&\frac{3}{16}\left(\left|g_{LL}^{S}\right|^{2}-\left|g_{LR}^{S}\right|^{2}+\left|g_{RL}^{S}\right|^{2}-\left|g_{RR}^{S}\right|^{2}\right) \nonumber \\
&{}&-\frac{3}{4}\left(\left|g_{LR}^{T}\right|^{2}-\left|g_{RL}^{T}\right|^{2}-\left|g_{LL}^{V}\right|^{2}+\left|g_{RR}^{V}\right|^{2}-\Re\left(g_{LR}^{S}g_{LR}^{T*}+g_{RL}^{S}g_{RL}^{T*}\right)\right).
\end{eqnarray}
\begin{figure}[]
\begin{center}
\begin{fmffile}{externalph1}
\begin{fmfchar*}(120,60)
  \fmfleft{tm3,antinu3} \fmflabel{$\tau$}{tm3} \fmflabel{$\nu_{\tau}$}{antinu3}
  \fmfright{g3,f3,fb3} \fmflabel{$\nu_{\ell}$}{fb3} \fmflabel{$\ell$}{f3} \fmflabel{$\gamma $}{g3}
  \fmf{fermion}{tm3,v23,Wi3,antinu3}
  \fmf{dashes,tension=1,label=$W$}{Wi3,Wf3}
  \fmf{photon,tension=0}{v23,g3}
  \fmf{fermion,label=\rotatebox{50}{\vspace{-2mm}\hspace{-3mm}$~$}}{fb3,Wf3}
  \fmf{fermion}{Wf3,f3}
  \fmfdot{Wi3,Wf3,v23}
\end{fmfchar*}
\end{fmffile}\vspace{10mm}
\begin{fmffile}{externalph2}
\begin{fmfchar*}(120,60)
  \fmfleft{tm4,antinu4} \fmflabel{$\tau$}{tm4} \fmflabel{$\nu_{\tau}$}{antinu4}
  \fmfright{g4,f4,fb4} \fmflabel{$\nu_{\ell}$}{fb4} \fmflabel{$\ell$}{f4} \fmflabel{$\gamma $}{g4}
  \fmf{fermion}{tm4,Wi4}
  \fmf{fermion,tension=3}{Wi4,antinu4}
  \fmf{dashes,tension=2.3,label=$W$}{Wi4,Wf4}
  \fmf{fermion,label=\rotatebox{13}{\vspace{-2mm}\hspace{-3mm}$~$}}{fb4,Wf4}
  \fmf{fermion}{Wf4,v24,f4}
  \fmf{photon,tension=0}{v24,g4}
  \fmfdot{Wi4,Wf4,v24}
\end{fmfchar*}
\end{fmffile} \vspace{10mm}
\begin{fmffile}{externalph3}
\begin{fmfchar*}(120,60)
  \fmfleft{tm4,antinu4} \fmflabel{$\tau$}{tm4} \fmflabel{$\nu_{\tau}$}{antinu4}
  \fmfright{g4,f4,fb4} \fmflabel{$\nu_{\ell}$}{fb4} \fmflabel{$\ell$}{f4} \fmflabel{$\gamma $}{g4}
  \fmf{fermion}{tm4,Wi4}
  \fmf{fermion,tension=3}{Wi4,antinu4}
  \fmf{dashes,tension=2.3,label=$W$}{Wi4,vtx}
  \fmf{dashes,tension=2.3,label=$W$,label.dist=-4.5mm}{vtx,Wf4}
  \fmf{fermion,label=\rotatebox{13}{\vspace{-2mm}\hspace{-3mm}$~$}}{fb4,Wf4}
  \fmf{fermion}{Wf4,f4}
  \fmf{photon,tension=0}{vtx,g4}
  \fmfdot{Wi4,Wf4,vtx}
\end{fmfchar*}
\end{fmffile}
\end{center}
\caption{Three Feynman diagrams of the tau radiative leptonic decay}\label{RD}
\end{figure}
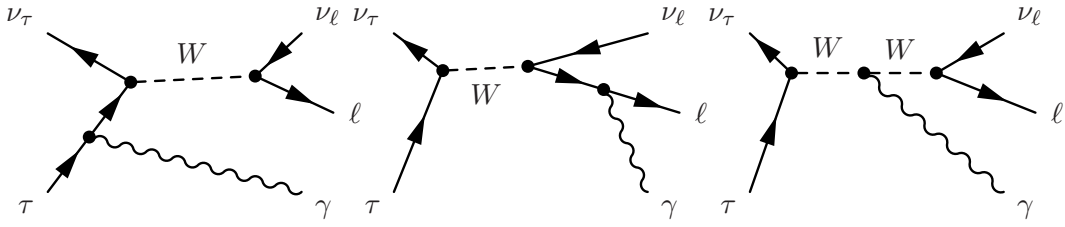

The Feynman diagrams describing the radiative leptonic decay of the
$\tau$ are presented in Fig.~\ref{RD}.
 The last amplitude is ignored because this contribution
 turns out to be suppressed by the very small
 factor $(m_\tau/m_W)^2$~\cite{w_supp}.
 As shown in Refs.~\cite{oldform,michel_form_arvzov},
 through the presence of a radiative photon in the final state,
 the polarization of the outgoing lepton is indirectly exposed;
 accordingly, three more Michel parameters,
 $\bar{\eta}$, $\eta^{\prime \prime}$, and $\xi\kappa$, become experimentally 
accessible:
\begin{eqnarray}
\bar{\eta}&=&\left|g_{RL}^{V}\right|^{2}+\left|g_{LR}^{V}\right|^{2}+\frac{1}{8}\left(\left|g_{RL}^{S}+2g_{RL}^{T}\right|^{2}+\left|g_{LR}^{S}+2g_{LR}^{T}\right|^{2}\right)+2\left(\left|g_{RL}^{T}\right|^{2}+\left|g_{LR}^{T}\right|^{2}\right) \label{etaform}, \\
\eta^{\prime\prime}&=&\Re\left\{ 24g_{RL}^{V}(g_{LR}^{S*}+6g_{LR}^{T*})+24g_{LR}^{V}(g_{RL}^{S*}+6g_{RL}^{T*})-8(g_{RR}^{V}g_{LL}^{S*}+g_{LL}^{V}g_{RR}^{S*})\right\}, \\
\xi\kappa&=&\left|g_{RL}^{V}\right|^{2}-\left|g_{LR}^{V}\right|^{2}+\frac{1}{8}\left(\left|g_{RL}^{S}+2g_{RL}^{T}\right|^{2}-\left|g_{LR}^{S}+2g_{LR}^{T}\right|^{2}\right)+2\left(\left|g_{RL}^{T}\right|^{2}-\left|g_{LR}^{T}\right|^{2}\right).
\end{eqnarray}
Both $\bar{\eta}$ and $\eta^{\prime \prime}$ appear in spin-independent terms 
in the differential decay width. Since all terms in Eq.~(\ref{etaform}) 
are strictly non-negative, the upper limit on $\bar{\eta}$ provides a 
constraint on each coupling constant.
 The effect of the nonzero value of $\eta^{\prime \prime}$
 is suppressed by a factor $m_\ell^2/m_\tau^2 \sim 10^{-7}$ for an electron
 mode and about $4\times 10^{-3}$ for a muon mode
 and so proves to be difficult to measure with the available statistics collected
 at Belle.
 In this study, we fix $\eta^{\prime \prime}$ at its SM value ($\eta^{\prime \prime} =0$).
 
To measure $\xi\kappa$, which appears in the spin-dependent part
 of the differential decay width, the knowledge of tau spin
 direction is required. Although the
 average polarization of a single $\tau$ is zero in experiments at $e^+ e^-$ colliders
 with unpolarized beams, the spin-spin correlation between the $\tau^+$ and $\tau^-$
 in the reaction $e^+ e^-\to\tau^+\tau^-$ can be exploited to measure $\xi\kappa$~\cite{Tsai}.
   
 According to Ref.~\cite{xiprelation},
 $\xi\kappa$ is related to another Michel-like parameter 
$\xi^\prime = -\xi -4\xi\kappa + 8\xi\delta/3$.
 Because the normalized probability that the $\tau^-$ decays
 into the right-handed charged daughter lepton $Q_R^\tau$
 is given by $Q_R^\tau=(1-\xi^\prime)/2$~\cite{xiprimeapll}, 
the measurement of $\xi\kappa$ provides
 a further constraint on the Lorentz structure of the weak current.
 The information on these parameters is summarized in Table~\ref{MPs}.

In muon decay, through the direct measurement of electron polarization in
 $\mu^+ \rightarrow e^+ \nu_{e} \bar{\nu}_{\mu}$, the relevant parameters
 $\xi^\prime$ and $\xi^{\prime \prime}=16\rho/3-4\bar{\eta}-3$ have been 
already measured. Those of the $\tau$ have not been measured yet.
 
 Using the statistically abundant data set of ordinary leptonic decays,
 previous measurements~\cite{etalep,rhoCLEO} have determined the Michel
 parameters $\eta$, $\rho$, $\delta$, and $\xi$ to an accuracy of a few percent
 and shows agreement with the SM prediction.
 Taking into account this measured agreement, the smaller data set of the 
radiative decay and its limited sensitivity, we focus in this analysis
 only on the extraction of $\bar{\eta }$ and $\xi\kappa$ by
 fixing $\eta$, $\rho$, $\delta$, $\xi$, and $\xi_\rho$ to the SM values.
 This represents the first measurement of the $\bar{\eta }$
 and $\xi\kappa$ parameters of the $\tau$ lepton.

\begin{table}[]
\caption{Michel parameters of the $\tau$ lepton $\dagger$ }\label{MPs}
\begin{center}
\begin{tabular}{ccccc} \hline \hline
Name &SM&Spin & Experimental& Comments and Ref. \\  
          &value& correlation &    result  \cite{PDG_paper}    &                  \\ \hline 
$\eta$& 0 & no & $0.013\pm0.020$ & (ALEPH) \cite{etalep} \\ 
$\rho $& $3/4$ & no & $0.745 \pm 0.008$ & (CLEO) \cite{rhoCLEO} \\
$\xi\delta $& $3/4$ & yes &$0.746\pm 0.021$ & (CLEO) \cite{rhoCLEO}\\
$\xi $& 1 & yes & $1.007\pm0.040$ & measured in leptonic decays (CLEO) \cite{rhoCLEO}\\
$\xi_h $& 1 & yes & $0.995\pm0.007$ & measured in hadronic decays (CLEO) \cite{rhoCLEO}\\
$\overline{\eta }$& 0 & no & not measured & ~~from radiative decay (RD)\\
$\xi\kappa$& 0 & yes & not measured& from RD\\  
$\eta^{\prime \prime}$& 0 & no & not measured & ~~~~~~~from RD, suppressed by $m_l^2/m_\tau^2$\\
$\xi^{\prime}$& 1 & yes & - &  $\xi^\prime = -\xi -4\xi\kappa + 8\xi\delta/3$ \\  
$\xi^{\prime \prime}$& 1 & no & - &  $\xi^{\prime \prime}= 16\rho/3 -4\bar{\eta} -3$ \\  \hline \hline
\end{tabular}
\end{center}
\begin{flushleft}\vspace{-0mm}$~^\dagger$~{Experimental results 
represent average values obtained by PDG~\cite{PDG_paper}.}\end{flushleft}
\end{table}

\section{Method}

\subsection{Unbinned maximum likelihood method}

The differential decay width for the radiative leptonic
 decay of $\tau^-$ with a definite spin direction $\bvec{S}^{*}_{-}$ is given by
\begin{equation}
\frac{\mydif\Gamma(\tau^-\rightarrow \ell^- \nu_{\tau} \bar{\nu}_{\ell} \gamma )}{\mydif E^*_\ell \mydif \Omega^*_\ell \mydif E^*_\gamma \mydif \Omega^*_\gamma } 
= \left( A_{0}^-+\bar{\eta}\,A_{1}^- \right)+ \left( \bvec{B}_{0}^- +\xi\kappa\,\bvec{B}_{1}^- \right)\cdot \bvec{S}^{*}_{-},\label{taum}
\end{equation}
where $A_{i}^{-}$ and $\bvec{B}_{i}^-$~($i=0,1$) are known functions of
the kinematics of the decay products\footnote{The detailed formulae
 of $A^-$, $\bvec{B}^-$ in Eq.~(\ref{taum}) and $A^+$, $\bvec{B}^+$ in Eq.~(\ref{taup})
 are given in the appendix.} with indices 
$i=0,1$ ($i$ is the function identifier),
 $\Omega_a$ stands for a set of $\{\mathrm{cos}\theta_{a}, \phi_{a}\}$ 
for a particle
 of the type $a$, and the asterisk means that the variable is defined in the
 $\tau^-$ rest frame.
 Equation~(\ref{taum}) shows that
 $\xi\kappa$ appears in the {\it spin-dependent} part of the decay
 width. This parameter can be measured by utilizing
 the well-known spin-spin correlation of the $\tau$ leptons in the
 $e^-e^+\rightarrow\tau^+\tau^-$ production:
\begin{eqnarray}
 \frac{\mathrm{d} \sigma \left( e^-e^+ \rightarrow \tau^- (\bvec{S}^{*}_{-}) \tau^+ ( \bvec{S}^{*}_{+} ) \right) }{\mathrm{d}\Omega_{\tau}}
= \frac{\alpha^2\beta_{\tau}}{64E_{\tau}^2}(D_{0}+{\textstyle \sum_{i,j}} D_{ij}(\bvec{S}^{*}_{-})_{i} (\bvec{S}^{*}_{+})_{j})
,\label{corr}
\end{eqnarray}
where $\alpha$ is the fine structure constant, $\beta_{\tau}$ and
$E_{\tau}$ are the velocity and energy of the $\tau^-$ in the center-of-mass 
system (c.m.s.), respectively,
$D_{0}$ is the spin-independent part of the cross section,
 and $D_{ij}$ $(i,j=0,1,2)$ is a tensor describing the spin-spin
correlation~(see Eq.~(4.11) in Ref.~\cite{Tsai}).
For the partner $\tau^+$, its spin information is
 extracted using the two-body decay $\tau^+\rightarrow \rho^+ \bar{\nu}_{\tau} \rightarrow\pi^+\pi^0\bar{\nu}_{\tau}$
 whose differential decay width is
\begin{equation}
\frac{\mathrm{d}\Gamma (\tau^+\rightarrow
  \pi^+\pi^0\bar{\nu}_\tau)}{\mydif \Omega_{\rho}^*\mydif m^2 \mydif \widetilde{\Omega}_{\pi}} = A^+ + \xi_\rho \bvec{B}^+\cdot \bvec{S}_{+}^{*}; \label{taup}
\end{equation}
$A^+$ and $\bvec{B}^+$ are known functions for the spin-independent and spin-dependent parts,
 respectively; the tilde indicates variables defined in the
 $\rho^+$ rest frame and $m$
 is the invariant mass of the $\pi\pi^0$ system, $m^2=(p_\pi+p_{\pi^0})^2$.
 As mentioned before, we use the SM value: $\xi_\rho=1$.
 Thus, the total differential cross section of
$e^+e^-\rightarrow\tau^-\tau^+\rightarrow
(\ell^-\nu_\tau \bar{\nu}_\ell\gamma)(\pi^+\pi^0\bar{\nu}_\tau)$
{(or, briefly, $(\ell^{-}\gamma, \pi^{+}\pi^0)$) can be written as:
\begin{equation}
\displaystyle \frac{\mathrm{d}\sigma(\ell^-\gamma,\pi^+\pi^0) }{ \mydif E^*_\ell \mydif \Omega^*_\ell \mydif E^*_\gamma \mydif \Omega^*_\gamma
  \mydif \Omega_{\rho}^* \mydif m^2 \mydif \widetilde{\Omega}_{\pi} \mydif \Omega_\tau} \propto \frac{\beta_{\tau}}{E_{\tau}^2}
\left[ D_{0} \left(A_{0}^-\! +A_{1}^-\!\cdot\!\bar{\eta} \right) A^+ + {\textstyle \sum_{i,j}} D_{ij} \!\left(\bvec{B}_{0}^-\!+\!
 \bvec{B}_{1}^-\!\cdot\!\xi\kappa \right)_i \! (\bvec{B}^+)_j \right].\label{totform}
\end{equation}
To extract the visible differential cross section, we transform the 
differential variables into ones defined
 in the c.m.s. using the Jacobian $J$: 
\begin{align}
&\hspace{3em}J = \left|\frac{\partial(E_{\ell}^{*}, \Omega_{\ell}^*)}{\partial(P_\ell, \Omega_\ell)} \right|
 \left|\frac{\partial(E_{\gamma}^{*}, \Omega_{\gamma}^*)}{\partial(P_\gamma, \Omega_\gamma)} \right|
 \left| \frac{\partial (\Omega_{\rho}^*, \Omega_{\tau})}{\partial ( P_{\rho}, \Omega_{\rho}, \Phi  )}\right| = \left(\frac{P_{\ell}^2}{E_\ell { P}_{\ell}^*}
 \right) \left( \frac{E_\gamma}{E_\gamma^*} \right)
 \left (\frac{m_{\tau} P_{\rho} }{ E_{\rho}P_{\rho}^* P_{\tau} } \right),
\end{align}
where the parameter $\Phi$ denotes the angle along the arc 
illustrated in Fig.~\ref{constraingedcone}.
\begin{figure}[]
\begin{center}
{\includegraphics[width=5.5cm]{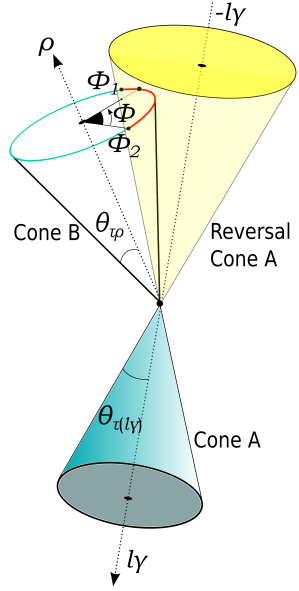}}
\caption[]{	\raggedright Kinematics of $\tau^+\tau^- \rightarrow (\rho^+(\rightarrow \pi^+\pi^0)\bar{\nu}_\tau)(\ell^-\nu_\tau\bar{\nu}_\ell\gamma)$ decay.
 Cones A and B are the surfaces that satisfy the c.m.s. conditions $(p_{\tau^-}-p_{\ell^-\gamma})^2=0$ and $(p_{\tau^+}-p_{\rho^+})^2=0$.
 The direction of $\tau^+$ is constrained to lie on an arc defined by the
 intersection of cone~B and the interior or exterior
 sector constrained by the reversal (\textit{i.e.}, mirror) cone~A.
 The arc (shown in red) is parametrized by the angle $\Phi\in[\Phi_1, \Phi_2]$. }\label{constraingedcone}
 \end{center}
\end{figure}

The visible differential cross section is, therefore, obtained by 
integration over $\Phi$:
\begin{align}
\displaystyle \frac{\mathrm{d}\sigma(l^-\gamma,\pi^+\pi^0) }{ \mydif P_\ell \mydif \Omega_\ell
 \mydif P_\gamma \mydif \Omega_\gamma \mydif P_\rho \mydif \Omega_\rho \mydif m^2 \mydif \widetilde{\Omega}_\pi   } 
&= \int_{\Phi_1}^{\Phi_2} \hspace{-0.5em} \mydif \Phi
\frac{\mathrm{d}\sigma(\ell^-\gamma,\pi^+\pi^0) }{\mydif \Phi \mydif P_\ell \mydif \Omega_\ell \mydif
 P_\gamma \mydif \Omega_\gamma \mydif P_\rho \mydif \Omega_\rho   \mydif m^2 \mydif \widetilde{\Omega}_{\pi} } \\
&= \int_{\Phi_1}^{\Phi_2} \hspace{-0.5em} \mydif \Phi
\frac{\mathrm{d}\sigma(\ell^-\gamma,\pi^+\pi^0) }{\mydif E^*_\ell \mydif \Omega^*_\ell
 \mydif E^*_\gamma \mydif \Omega^*_\gamma  \mydif \Omega_{\rho}^* \mydif m^2 \mydif \widetilde{\Omega}_{\pi} \mydif \Omega_\tau} J\label{totformwithJ} \\
&\equiv S(\bvec{x}),
\end{align}
where $S(\bvec{x})$ is proportional to the probability density function (PDF) 
of the signal and $\bvec{x}$ denotes the set of twelve measured variables:
 $\bvec{x} = \{ P_\ell, \Omega_\ell, P_\gamma, \Omega_\gamma, P_\rho, \Omega_\rho, m^2, \widetilde{\Omega}_\pi \}$.
 There are several corrections that must be incorporated in the
 procedure to take into account the real experimental situation.
 Physics corrections include electroweak higher-order corrections
 to the $e^+ e^-\to\tau^+\tau^-$ cross section
 \cite{Arbuzov:1997pj,Kuraev:1985hb,Berends:1982dy,Jadach:1985ac,Jadach:1984iy}.
 Apparatus corrections include the effect of the finite detection
 efficiency and resolution, the effect of the external bremsstrahlung
 for $(e^-\gamma,~\pi^+\pi^0)$ events, and the $e^\pm$ beam energy
 spread.


Accounting for the event-selection criteria and the contamination from 
identified backgrounds, the total visible (properly normalized) PDF
 for the observable $\bvec{x}$ in each event is given by
\begin{align}
P(\bvec{x}) = (1-\sum_{i}{\lambda_{i}}) \frac{S(\bvec{x})\varepsilon (\bvec{x})}{\int{\mathrm{d}\bvec{x} S(\bvec{x})\varepsilon (\bvec{x})}}
+\sum_{i}{\lambda_{i}\frac{B_{i}(\bvec{x})\varepsilon (\bvec{x})}{\int{\mathrm{d}\bvec{x} B_{i}(\bvec{x})\varepsilon (\bvec{x})}}},\label{generalpdf}
\end{align}
where $B_{i}(\bvec{x})$
 is the distribution of the $i^{\rm th}$ category of background,
 $\lambda_{i}$ is the fraction of this background, and 
$\varepsilon (\bvec{x})$ is the
 selection efficiency of the signal distribution.
 The categorization of $i$ is explained later (see the caption 
of Fig.~\ref{pi0dist}).
 In general, $B_i(\bvec{x})$ is evaluated as an integral of the $i^{\rm th}$ background
 PDF multiplied by the inefficiency that depends on the variables of missing particles.
 The PDFs of the dominant background processes are described
 analytically one by one, while the remaining background processes
 are described by one common PDF, tabulated from Monte Carlo (MC) simulation.

The denominator of the signal term in Eq.~(\ref{generalpdf}) 
represents normalization.
 Since $S(\bvec{x})$ is a linear combination of the Michel parameters
 $S(\bvec{x})=\mathcal{S}_0(\bvec{x}) + \mathcal{S}_1(\bvec{x}) \bar{\eta} + \mathcal{S}_2(\bvec{x}) \xi\kappa$,
 the normalization of signal PDF becomes 
\begin{align}
& ~~~ \int \hspace{-0.3em} \mydif \bvec{x} ~\Big( \mathcal{S}_0(\bvec{x}) + \mathcal{S}_1(\bvec{x}) \bar{\eta} + \mathcal{S}_2(\bvec{x}) \xi\kappa \Big) \varepsilon (\bvec{x}) \\
&= \mathcal{N}_0\int \hspace{-0.3em} \mydif \bvec{x} \left(\frac{\mathcal{S}_0(\bvec{x}) \varepsilon (\bvec{x})}{\mathcal{N}_0} \right)
 \cdot \frac{ \mathcal{S}_0(\bvec{x}) + \mathcal{S}_1(\bvec{x}) \bar{\eta} + \mathcal{S}_2(\bvec{x}) \xi\kappa }{\mathcal{S}_0(\bvec{x})} \\
 &=  \frac{\mathcal{N}_0 \bar{\varepsilon} }{N_{\rm sel}} \sum_{i:{\rm sel}} \frac{ \mathcal{S}_0(\bvec{x}^i) + \mathcal{S}_1(\bvec{x}^i) \bar{\eta}
 + \mathcal{S}_2(\bvec{x}^i) \xi\kappa }{\mathcal{S}_0(\bvec{x}^i)} \\
 &= \mathcal{N}_0 \bar{\varepsilon} \left[ 1+ \left< \frac{\mathcal{S}_1}{\mathcal{S}_0} \right> \bar{\eta} + \left< \frac{\mathcal{S}_2}{\mathcal{S}_0} \right> \xi\kappa \right],
\end{align}
where $\mathcal{N}_0$ is a normalization coefficient of the SM part defined by
 $\mathcal{N}_0=\int \hspace{-0.16em} \mydif \bvec{x}~\mathcal{S}_0(\bvec{x})$,
 $\bvec{x}^i$ represents a set of variables for the
$i^{\rm th}$ selected event of $N_{\rm sel}$ events,
 $\bar{\varepsilon}$ is an average selection efficiency,
 and the brackets $\left<~\right>$ indicate an average with respect 
to the selected SM distribution.
 We refer to $\mathcal{N}_0$ and 
$\left< {\mathcal{S}_i}/{\mathcal{S}_0} \right>$ ($i=1,2$) 
as absolute and relative normalizations, respectively.

From $P(\bvec{x})$, the negative logarithmic
 likelihood function (NLL) is constructed and the best estimators of the
 Michel parameters, $\bar{\eta}$ and $\xi\kappa$, are obtained by
 minimizing the NLL. The efficiency $\varepsilon(\bvec{x})$ is a common
 multiplier in Eq.~(\ref{generalpdf}) and does not depend on the Michel
 parameters. This is one
 of the essential features of the unbinned maximum likelihood method.
 We validated our fitter and procedures using a MC sample generated 
according to the SM distribution.
 The optimal values of the Michel parameters are consistent with their SM
 expectations within the statistical uncertainties.

\subsection{KEKB collider}

The KEKB collider (KEK laboratory, Tsukuba, Japan) is an 
energy-asymmetric $e^+e^-$ collider with beam energies of 3.5~GeV and 8.0~GeV 
for $e^+$ and $e^-$, respectively.
 Most of the data were taken at the c.m.s. energy of 10.58~GeV,
 corresponding to the mass of the $\Upsilon (4S)$, where a huge number of $\tau^+\tau^-$
 as well as $B\overline{B}$ pairs were produced.
 The KEKB collider was operated from 1999 to 2010 and accumulated 
1~${\rm ab^{-1}}$ of $e^+e^-$ collision data with the Belle detector.
 The achieved instantaneous luminosity of $2.11\times 10^{34}{\rm~cm^{-2}/s}$ 
is the world record.
 For this reason, the KEKB collider is often called a $B$-{\it factory}
 but it is worth considering it also as a $\tau$-{\it factory}, 
where $O(10^9)$ $\tau$ pair events have been produced.
 The world largest sample of $\tau$ leptons collected at Belle provides
 a unique opportunity to study radiative leptonic decay of $\tau$.
 In this analysis, we use 711~f$\mathrm{b}^{-1}$ of collision data
 collected at the $\Upsilon (4S)$ resonance energy~\cite{cite_KEKB1}.

\subsection{Belle detector}
The Belle detector is a large-solid-angle magnetic
spectrometer that consists of a silicon vertex detector,
a 50-layer central drift chamber (CDC), an array of
aerogel threshold Cherenkov counters (ACC),  
a barrel-like arrangement of time-of-flight
scintillation counters (TOF), and an electromagnetic calorimeter
comprised of CsI(Tl) crystals (ECL) located inside 
a superconducting solenoid coil that provides a 1.5~T
magnetic field.  An iron flux return located outside of
the coil is instrumented to detect $K_L^0$ mesons and to identify
muons (KLM). The detector
is described in detail elsewhere~\cite{citeBelle}.

\section{Event selection \label{labelevs}}

The event selection proceeds in two stages. At the preselection,
 $\tau^+\tau^-$ candidates are selected efficiently while suppressing
 the beam background and other physics processes like radiative Bhabha scattering,
 two-photon interaction, and radiative $\mu^+\mu^-$ pair production.
 The preselected events are then required to satisfy final selection
 criteria to enhance the purity of the signal events.
 \vspace{-4mm}
\subsection{Preselection}
\begin{itemize}
\item {There must be exactly two oppositely charged tracks in the event.
 The impact parameters of these tracks relative to the interaction point 
are required to be
 within $\pm 2.5$~cm along the beam axis and $\pm0.5$~cm in the transverse 
plane.
 The two-track transverse momentum must exceed $0.1$~GeV/$c$ and that 
of one track must exceed $0.5$~GeV/$c$. }
\item{Total energy deposition in the ECL in the laboratory frame must be lower 
than 9~GeV.}
\item{The opening angle $\psi$ of the two tracks must satisfy $20^\circ<\psi < 175^\circ$ in the laboratory frame.}
\item{The number of photons whose energy exceeds $80~$MeV in the c.m.s. must be fewer than five.}
\item{For the four-vector of missing momentum defined by 
$p_{\mathrm{miss}}=p_{\mathrm{beam}}-p_{\mathrm{obs}}$,
 the missing mass $M_{\rm miss}$ defined by 
 $M^2_{\mathrm{miss}} = p_{\mathrm{miss}}^2 c^2$ must lie in the range 
$1$ GeV$/c^2$ $\leq M_{\mathrm{miss}}\leq 7$ GeV$/c^2$, 
where $p_{\rm beam}$ and $p_{\rm obs}$
 are the four-momenta of the beam and all detected particles, respectively.}
\item{The polar angle of missing-momentum must satisfy 
$30^{\circ}\leq \theta_{\mathrm{miss}} \leq 150^\circ$ in the laboratory frame.}
\end{itemize}

\subsection{Final selection}
The candidates of the outgoing particles in 
$\tau^+\tau^- \rightarrow (\pi^+ \pi^0 \bar{\nu}_\tau)(\ell^-\nu_\tau\bar{\nu}_\ell\gamma )$,
 \textit{i.e.,} the lepton, photon, and charged and neutral pions, 
are assigned in each of the preselected events. 
\begin{itemize}
\item{The electron selection is based on the likelihood ratio cut, $P_e = {L_e}/(L_e + L_x) > 0.9$,
 where $L_e$ and $L_x$ are the likelihood values of the track for the electron 
and non-electron hypotheses, respectively.
 These values are determined using specific ionization
 ($\mydif E/\mydif x$) in the CDC, the ratio of ECL energy and CDC momentum 
$E/P$,
 the transverse shape of the cluster in the ECL, the matching
 of the track with the ECL cluster, and the light yield in the 
ACC~\cite{citeEID}.
 The muon selection uses the likelihood ratio 
$P_\mu = {L_\mu}/(L_\mu+L_\pi+L_K)> 0.9$, where the likelihood values are
 determined by the measured versus expected range for the $\mu$ hypothesis,
 and transverse scattering of the track in the KLM~\cite{citeMID}.
 The reductions of the signal efficiencies with
 lepton selections are approximately 10\% and 30\% 
for the electron and muon, respectively.
 The pion candidates are distinguished from kaons using $P_\pi = {L_\pi}/(L_\pi + L_K) > 0.4$,
 where the likelihood values are determined by the ACC response,
 the timing information from the TOF, and $\mydif E/\mydif x$ in the CDC.
 The reduction of the signal efficiency with pion selection is approximately 5\%.
}
 
\item{The $\pi^0$ candidate is formed from two photon candidates, 
where each photon satisfies
 $E_\gamma > 80$~MeV, with an invariant mass of $115$~MeV$/c^2$ $<M_{\gamma\gamma}<150$~MeV$/c^2$.
 Figure~\ref{pi0dist} shows the distribution of the invariant mass 
of the $\pi^0$ candidates. The reduction of the signal efficiency by the mass selection is approximately 50\%.
 In addition, when more than two $\pi^0$ candidates are found, the event is rejected.}
\item{The $\rho^+$ candidate is formed from a $\pi^+$ and a $\pi^0$ candidate, 
with an invariant mass of $0.5~\mathrm{GeV}/{c^2}<M_{\pi^+\pi^0}<1.5$~GeV$/{c^2}$.
 Figure~\ref{rhodist} shows the distribution of the invariant mass of the 
$\rho$ candidates.
 The reduction of the signal efficiency is approximately 3\%.}
\item{The c.m.s. energy of signal photon candidate must exceed $80$~MeV if 
within the ECL barrel ($31.4^\circ<\theta_\gamma < 131.5^\circ$) or
 $100$~MeV if within the ECL endcaps 
($12.0^\circ<\theta_\gamma < 31.4^\circ$ or $131.5^\circ<\theta_\gamma < 157.1^\circ$).
 As shown in Fig.~\ref{selcoslg}, this photon must lie in a cone determined 
by the lepton-candidate direction that is defined by 
cos$\theta_{e\gamma}>0.9848$ and
 cos$\theta_{\mu\gamma}>0.9700$ for the electron and muon mode, respectively,
 where $\theta_{\ell\gamma}$ ($\ell=e$ or $\mu$) is the angle between the lepton and the photon.
 The reductions of the signal efficiencies for the requirement
 on this photon direction are approximately 11\% and 27\% for the 
electron and muon mode, respectively.
 Furthermore, if the photon candidate and either of the photons 
from the $\pi^0$, which is a daughter of the $\rho^+$ candidate, 
form an invariant mass of the
 $\pi^0$ ($115$~MeV$/c^2$ $<M_{\gamma\gamma}<150$~MeV$/c^2$), 
the event is rejected. The additional selection reduces the signal 
efficiency by $1\%$.}
\item{The direction of the combined momentum of the lepton and photon in 
the c.m.s. must not belong to the hemisphere determined by the 
$\rho$ candidate: an event should satisfy 
$\theta_{(\ell^-\gamma)\rho^+}>90^\circ$, where $\theta_{(\ell^-\gamma) \rho^+}$ is the spatial
 angle between the $\ell^-\gamma$ system and the $\rho$ candidate. 
This selection reduces the signal efficiency by $0.4\%$.
}
\item{There must be no additional photons in the aforementioned cone 
around the lepton candidate; the sum of the energy in the laboratory frame 
of all additional photons that are not associated
 with the $\pi^0$ or the signal photon (denoted as 
$E^\mathrm{LAB}_{\mathrm{extra}\gamma}$) should not exceed 0.2~GeV and 0.3~GeV for
 the electron and muon mode, respectively. The reductions of the signal 
efficiencies for the requirement
 on the $E^\mathrm{LAB}_{\mathrm{extra}\gamma}$ are approximately 14\% and 6\% 
for the electron and muon mode, respectively.}
\end{itemize}

\begin{figure}[]
{\centering \subfloat[]{\includegraphics[width=8.0cm]{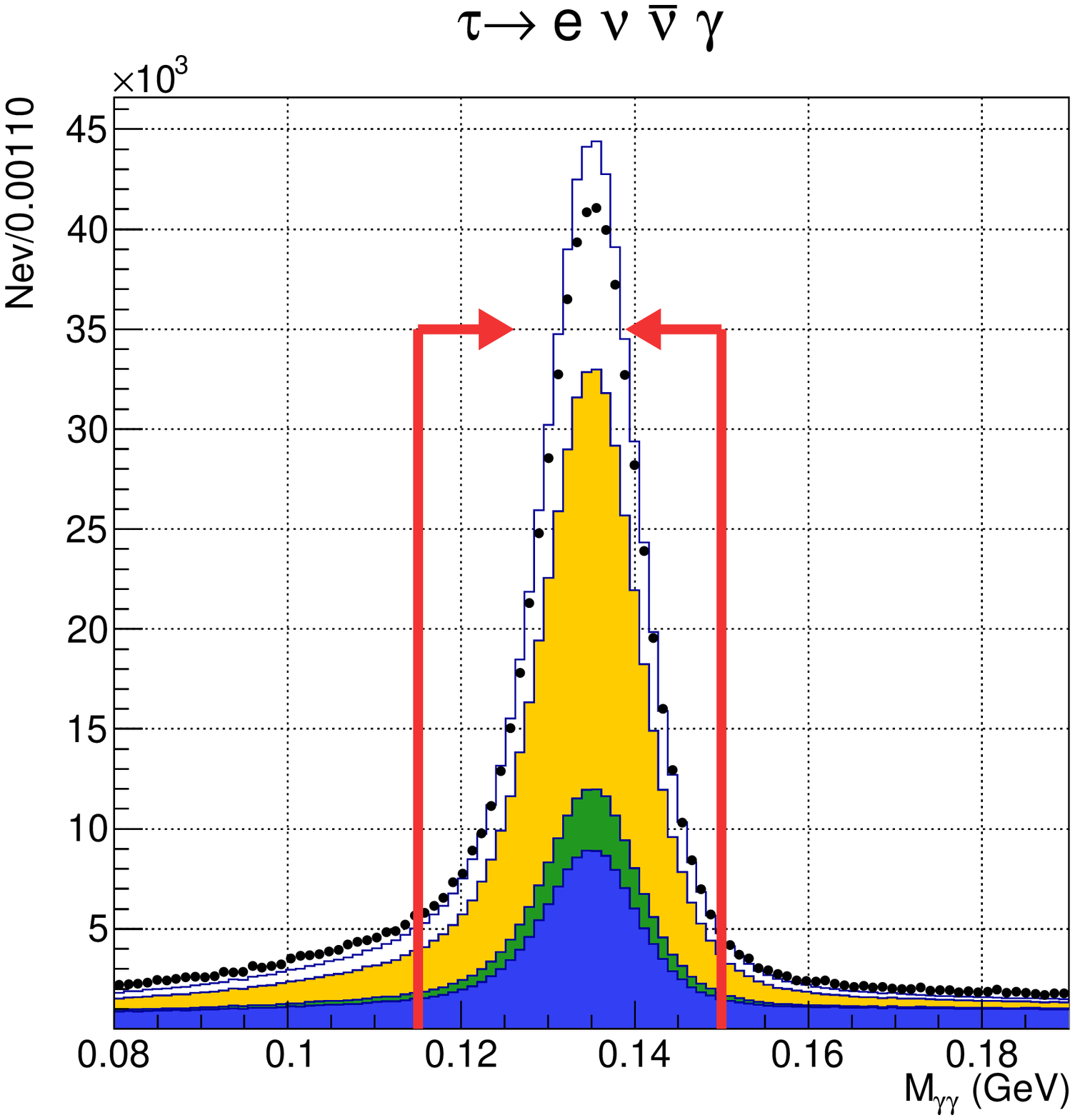}\label{pi0dist:pi0e}} }
{\centering \subfloat[]{\includegraphics[width=8.0cm]{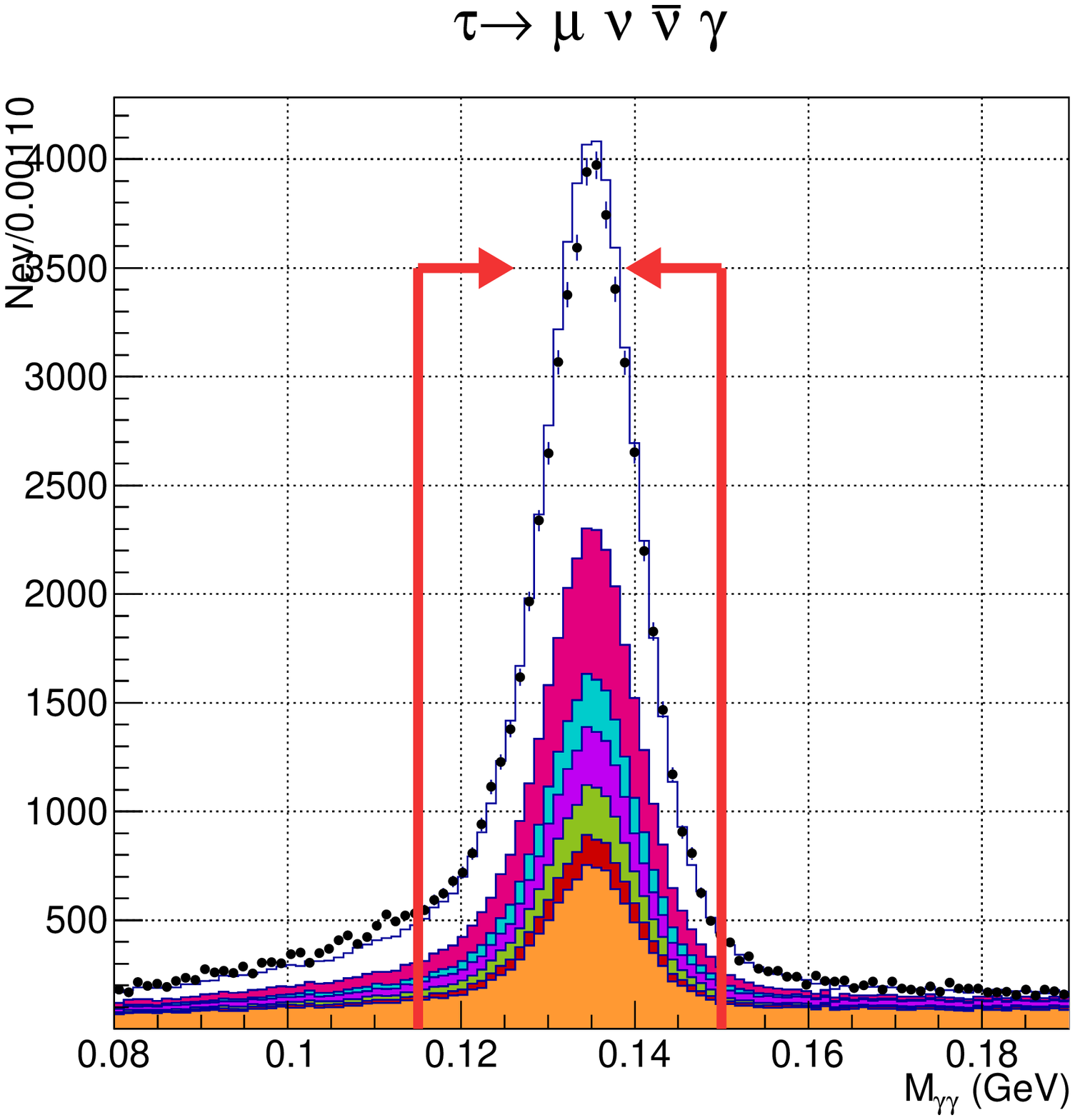}\label{pi0dist:pi0mu}}} 
\caption[]{\raggedright  Distribution of $M_{\gamma \gamma}$. Dots with uncertainties
 are experimental data and histograms are MC distributions.
 The MC histograms are scaled to the experimental one based on 
the yields just after the preselection.
 The red arrows indicate the selection window $115$~MeV$/c^2$ 
$<M_{\gamma\gamma}<150$~MeV$/c^2$.
  
~~(a) $\tau^- \rightarrow e^- \nu_\tau\bar{\nu}_e\gamma$ candidates:
 the open histogram corresponds to the signal, the yellow ($i=1$) and
 green ($i=2$) histograms represent ordinary leptonic decay plus extra bremsstrahlung due
 to the detector material and 
radiative leptonic decay plus bremsstrahlung,
 respectively, and the blue ($i=3$) histogram represents other processes such as radiative Bhabha,
 two-photon, and $e^+e^- \to q\bar{q}~(q=u,d,s,c)$ productions.
 
~~~(b) $\tau^- \rightarrow \mu^- \nu_\tau\bar{\nu}_\mu\gamma$ candidates: 
the open histogram corresponds to signal, the magenta ($i=1$) histogram represents 
ordinary leptonic decay plus beam background,
 the aqua ($i=2$) histogram represents ordinary leptonic decay plus ISR/FSR 
processes, the purple ($i=3$) histogram represents three-pion events where
 $\tau^+ \rightarrow\pi^+\pi^0\pi^0 \bar{\nu}_\tau$ is misreconstructed 
as a tagging $\tau^+\rightarrow \pi^+\pi^0 \bar{\nu}_\tau$ candidate, 
the green ($i=4$) histogram represents $\rho$-$\rho$ background
 where $\tau^- \rightarrow \pi^-\pi^0\nu_\tau$ is selected due to misidentification 
of pion as muon, the red ($i=5$) histogram represents 3$\pi$-$\rho$ events
 where $\tau^- \rightarrow \pi^-\pi^0\pi^0\nu_\tau$ is selected by 
misidentification similarly to the $\rho$-$\rho$ case, and the orange ($i=5$) 
histogram represents other processes (as in the electron mode).
 
~~~In Eq.~(\ref{generalpdf}) and
 the categories mentioned in this caption, $i \in \{1,\,2,\,3\}$ and 
$\{1,2,\ldots,6\}$ for the electron and muon modes, respectively.
 }\label{pi0dist}
\end{figure}

\begin{figure}[]
{\centering \subfloat[]{\includegraphics[width=8.0cm]{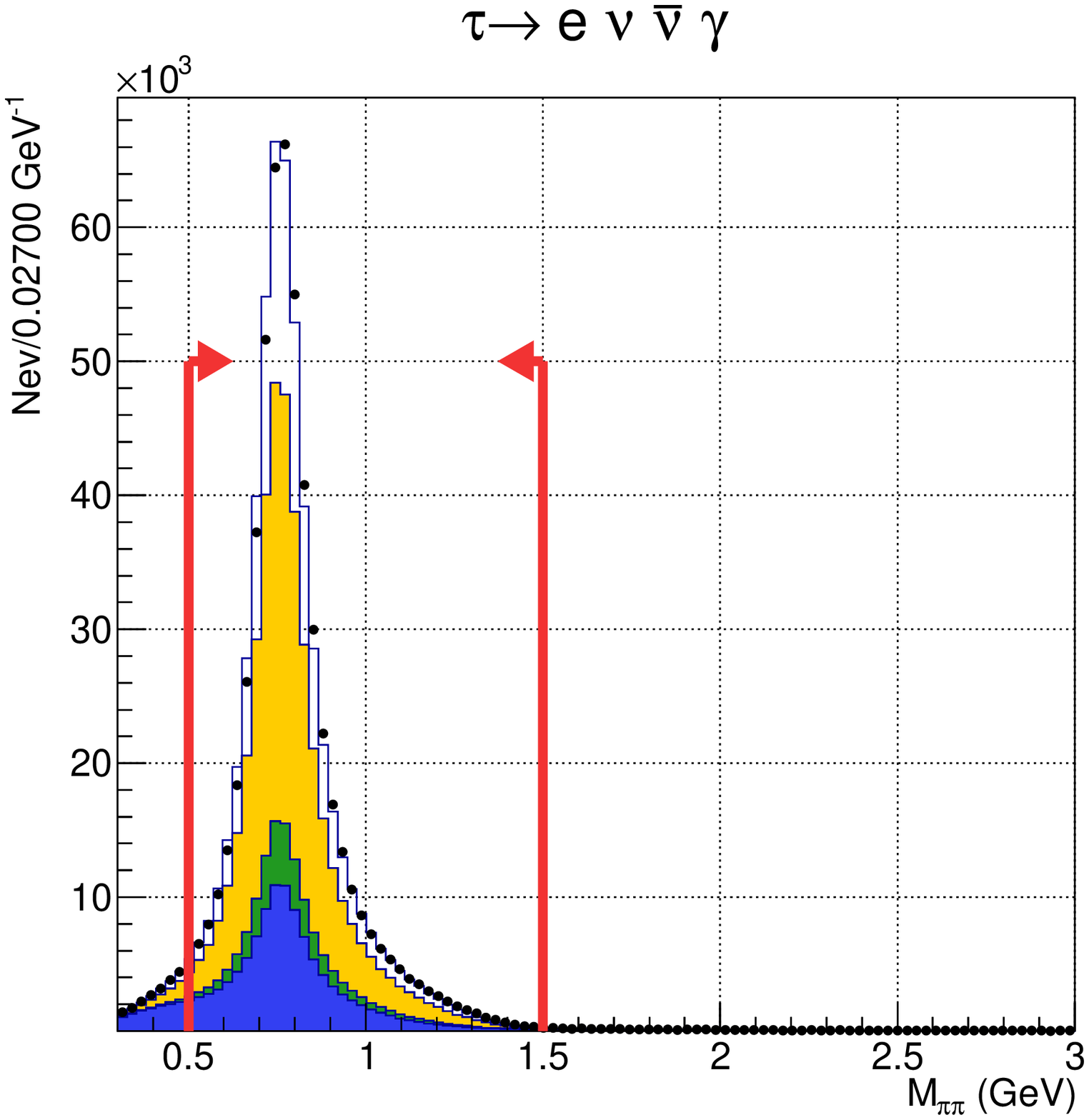}\label{pi0dist:rhoe}} }
{\centering \subfloat[]{\includegraphics[width=8.0cm]{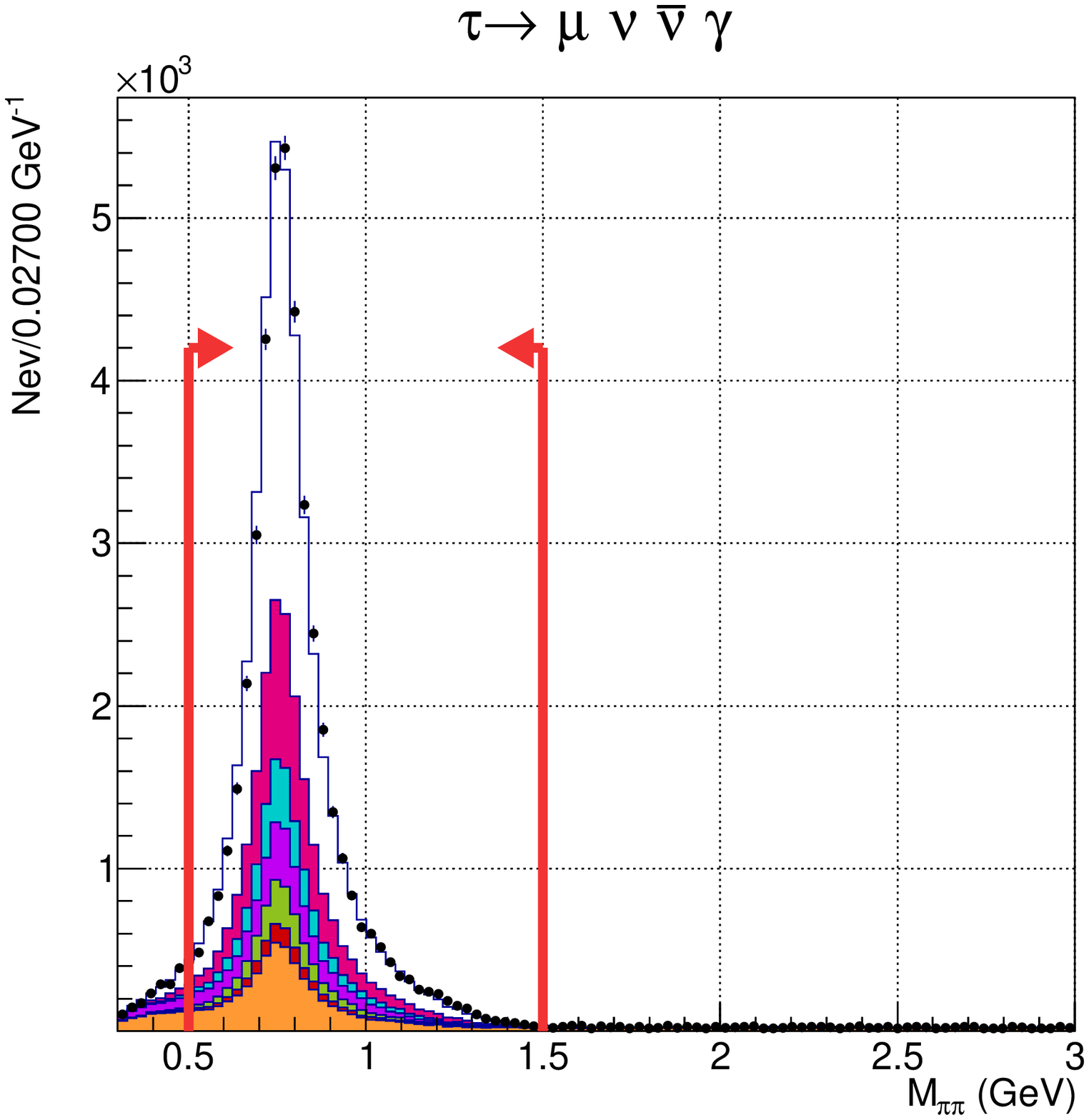}\label{pi0dist:rhomu}}}
\caption[]{\raggedright  Distribution of $M_{\pi \pi^0}$:
 (a) $\tau^- \rightarrow e^- \nu_\tau\bar{\nu}_e\gamma$ candidates and (b) $\tau^- \rightarrow \mu^- \nu_\tau\bar{\nu}_\mu\gamma$ candidates.
 Dots with uncertainties are experimental data and histograms are MC distributions.
 The color of each histogram is explained in Fig.~\ref{pi0dist}.
 The red arrows indicate the selection window 
$0.5$~GeV$/c^2$ $<M_{\pi\pi^0}<1.5$~MeV$/c^2$.}\label{rhodist}
\end{figure}

\begin{figure}[]
{\centering \subfloat[]{\includegraphics[width=8.0cm]{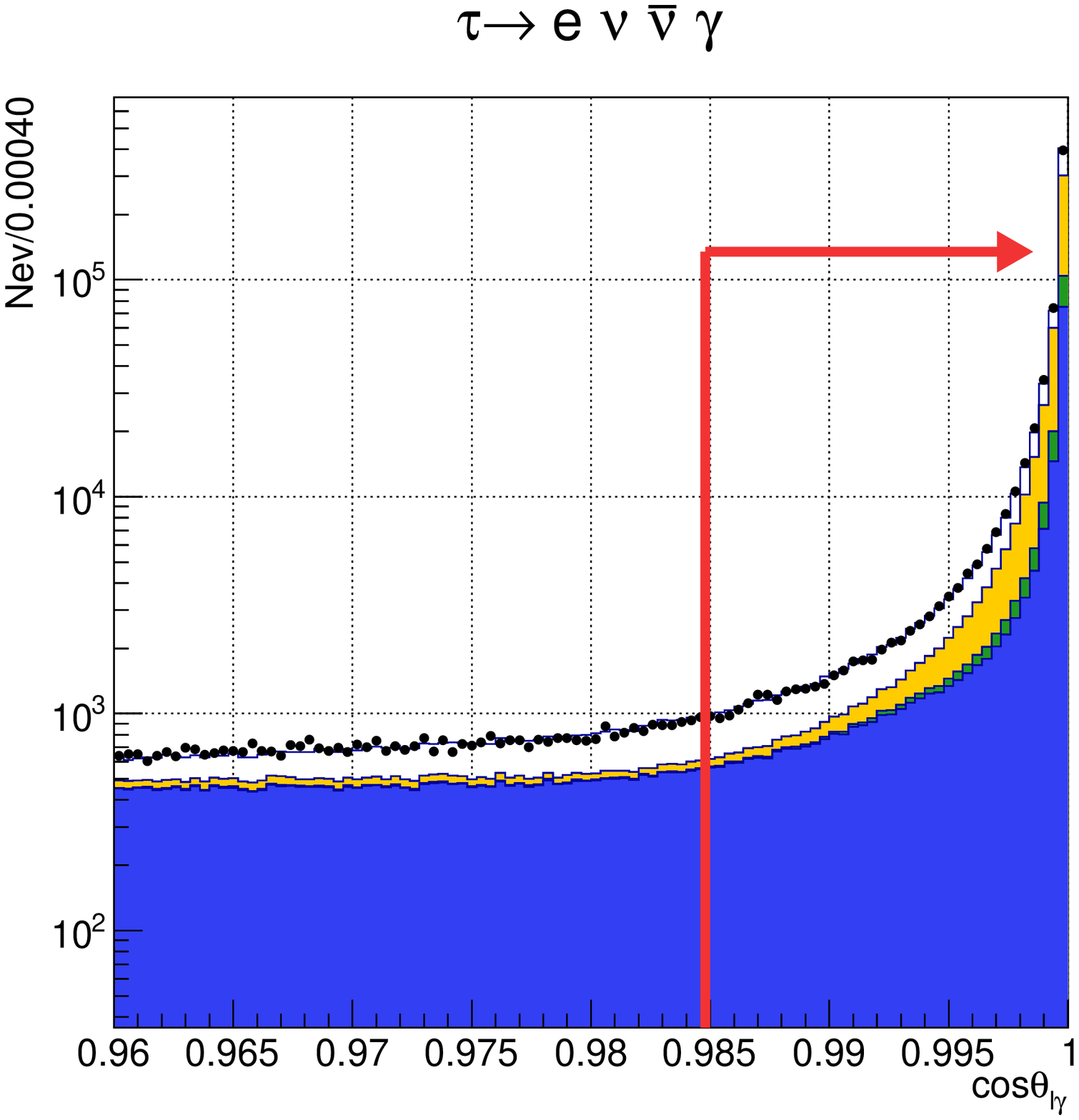}\label{selcoslg:e}} }
{\centering \subfloat[]{\includegraphics[width=8.0cm]{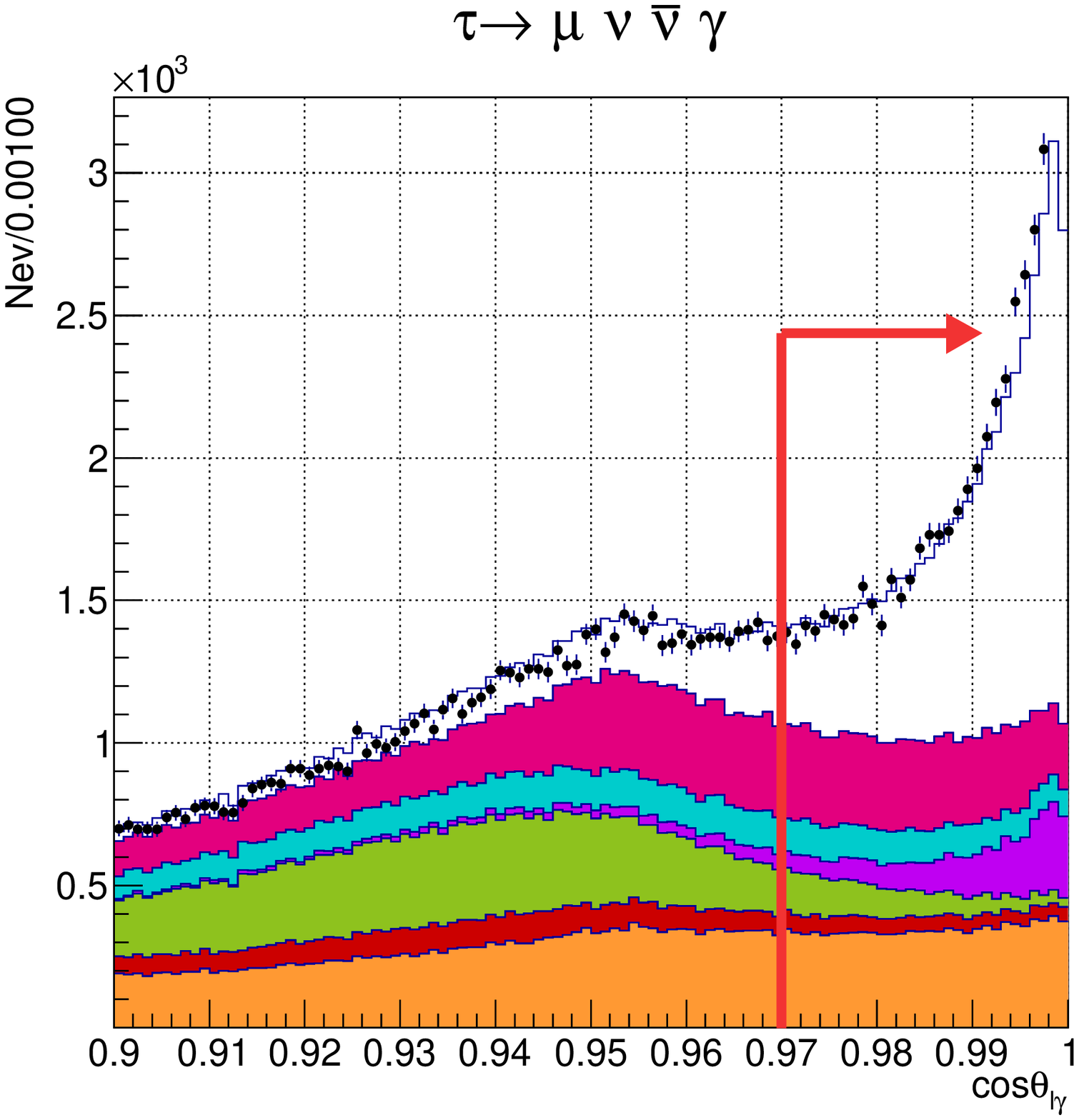}\label{selcoslg:mu}}} 
\caption[]{\raggedright  Distribution of $\cos\theta_{\ell\gamma}$:
 (a) $\tau^- \rightarrow e^- \nu_\tau\bar{\nu}_e\gamma$ candidates and (b) 
$\tau^- \rightarrow \mu^- \nu_\tau\bar{\nu}_\mu\gamma$ candidates.
 Dots with uncertainties are experimental data and histograms are MC distributions.
 The color of each histogram is explained in Fig.~\ref{pi0dist}.
 The red arrows indicate the selection condition cos$\theta_{e\gamma}>0.9848$ and
 cos$\theta_{\mu\gamma}>0.9700$ for the electron and muon mode, respectively.}\label{selcoslg}
\end{figure}

These selection criteria are optimized using MC simulation
 (five times as large as real data)
 where $e^+e^-\rightarrow \tau^+ \tau^-$ pair production and 
the successive decay of the $\tau$
 are simulated by the KKMC~\cite{citeKKMC} and 
TAUOLA~\cite{citeTAUOLA1,citeTAUOLA2}
 generators, respectively. The detector effects are simulated based 
on the GEANT3 package~\cite{citegeant3}.

\begin{figure}[]
{\centering \subfloat[$E_\gamma$]{\includegraphics[width=8.0cm]{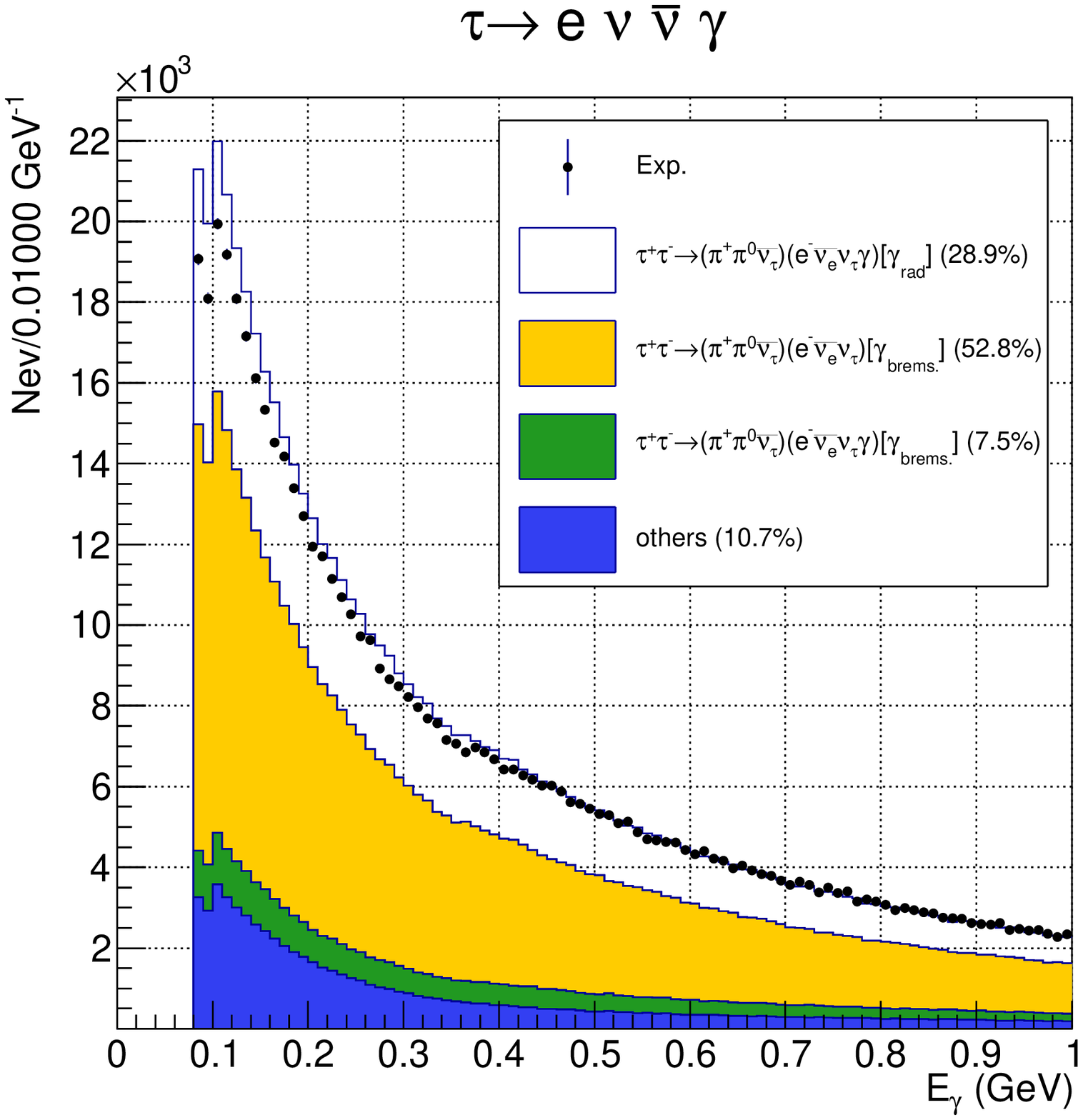}\label{selection:E_e}} }
{\centering \subfloat[$\theta_{e\gamma}$]{\includegraphics[width=8.0cm]{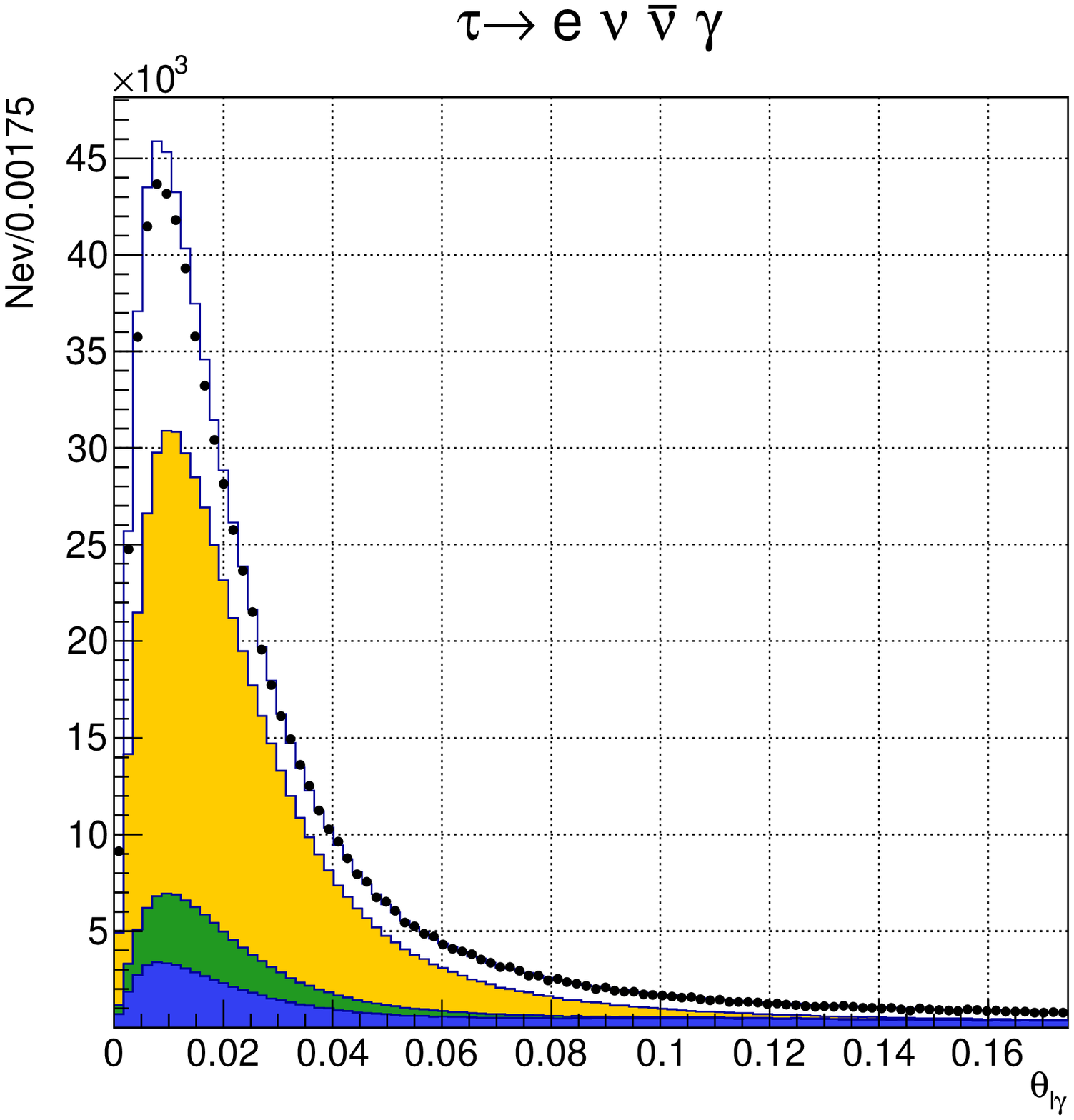}\label{selection:psi_e}}} 
\caption[]{\raggedright  Final distribution of (a) photon energy $E_\gamma$ and (b) $\theta_{e\gamma}$
 for the $\tau^+\tau^- \rightarrow (\pi^+ \pi^0 \bar{\nu}_\tau)(e^- \nu_\tau \bar{\nu}_e \gamma)$ decay candidates.
 Dots with uncertainties are experimental data and histograms 
are MC distributions.
 The color of each histogram is explained in Fig.~\ref{pi0dist}.
  }\label{comp_rem_e}
\end{figure}
\begin{figure}[]
{\centering \subfloat[$E_\gamma$]{\includegraphics[width=8.0cm]{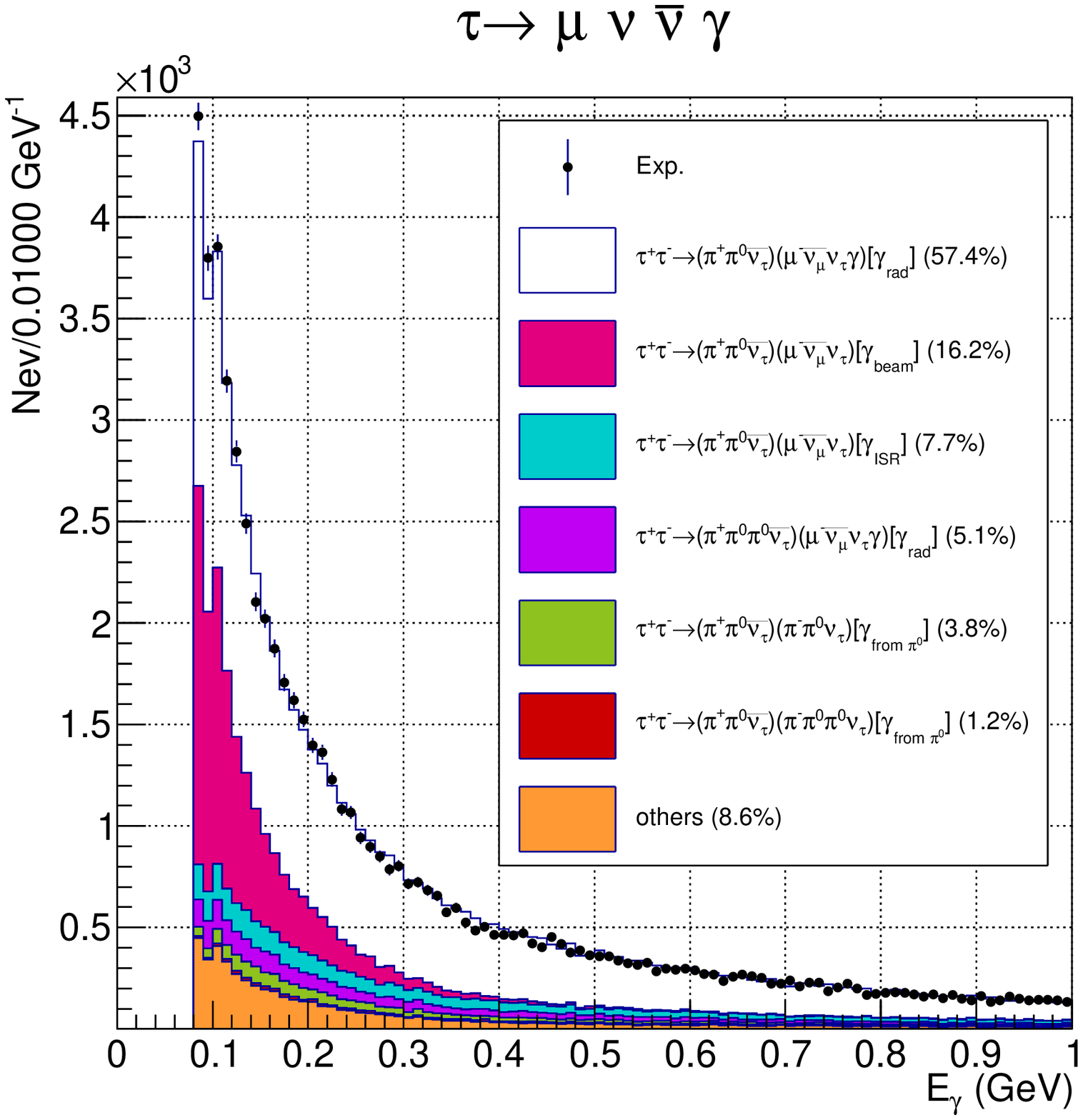}\label{selection:E_mu}} }
{\centering \subfloat[$\theta_{\mu\gamma}$]{\includegraphics[width=8.0cm]{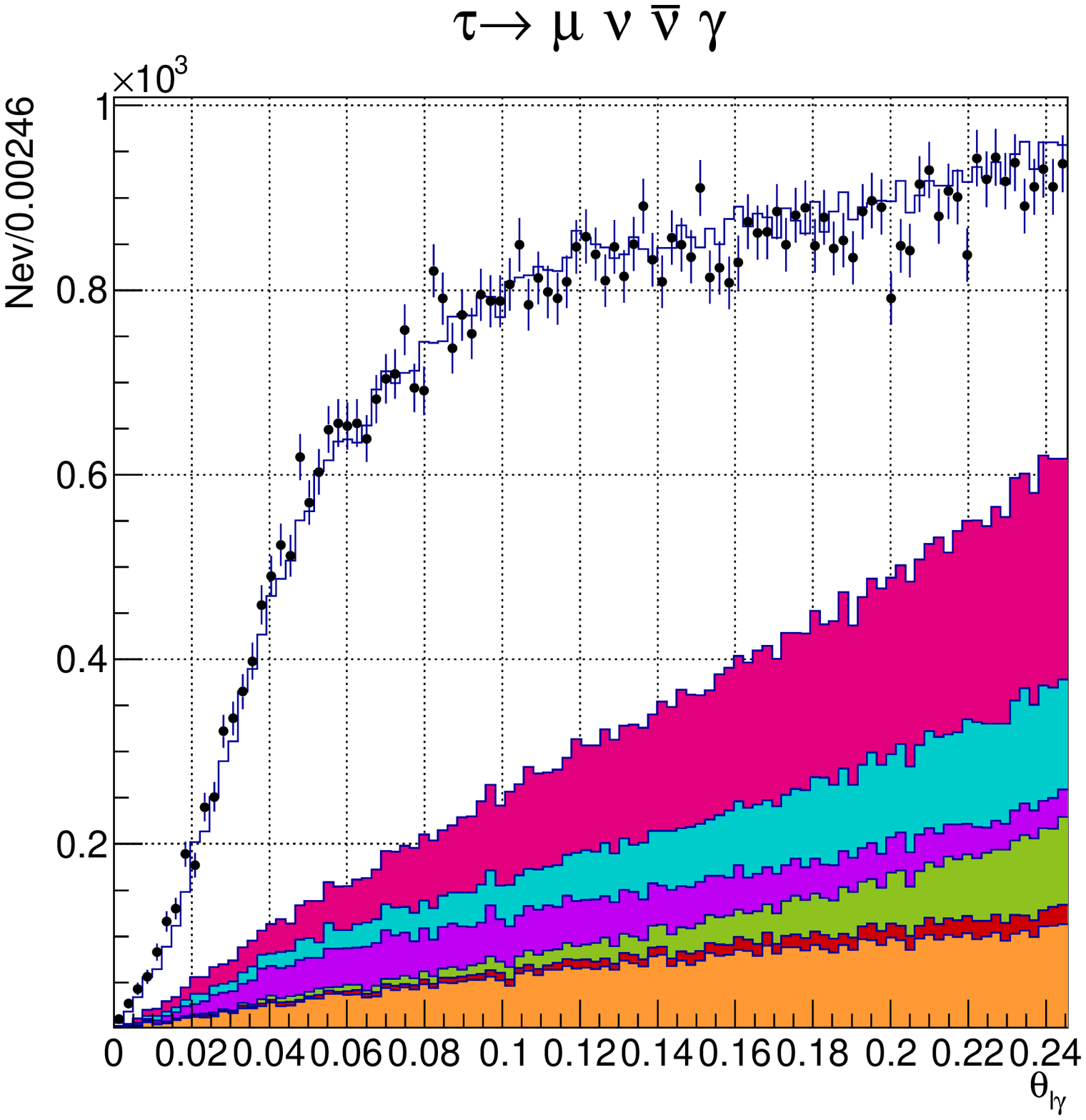}\label{selection:psi_mu}}}
\caption[]{\raggedright Final distribution of (a) photon energy $E_\gamma$
 and (b)~$\theta_{\mu\gamma}$ for the $\tau^+\tau^- \rightarrow (\pi^+\pi^0 \bar{\nu}_\tau)(\mu^- \nu_\tau \bar{\nu}_\mu \gamma)$ decay candidates.
 Dots with uncertainties are experimental data and histograms are MC distributions.
 The color of each histogram is explained in Fig.~\ref{pi0dist}.
 }\label{remgraphs_mu}
\label{comp_rem_mu}
\end{figure}

Distributions of the photon energy $E_\gamma$ and the angle between
 the lepton and photon, $\theta_{\ell\gamma}$, for the selected events are shown
 in Figs.~\ref{comp_rem_e} and \ref{comp_rem_mu} for
 $\tau^- \rightarrow e^- \nu_\tau \bar{\nu}_e \gamma$ and $\tau^- \rightarrow \mu^- \nu_\tau \bar{\nu}_\mu \gamma$
 candidates, respectively.
 
In the electron mode, the fraction of the signal decay in the selected 
sample is about $30\%$ due to the large external bremsstrahlung rate in 
the non-radiative leptonic
 $\tau$ decay events. In the muon mode, the
 fraction of the signal decay is about $60\%$; here,
 the main background arises from ordinary leptonic
 decay ($\tau^-\rightarrow \ell^- \nu_\tau \bar{\nu}_\ell$) events
 where either an additional photon is reconstructed from beam background
 in the ECL or a photon is emitted by the
 initial-state $e^+e^-$. The information
 is summarized in Table~\ref{evtsel_sum}.

As mentioned before, in the integration over 
$\Phi$ in Eq.~(\ref{totformwithJ}),
 the generated differential variables are varied according to the
 resolution function $\mathcal{R}$.
 Thus, the kinematic variables
 can extend outside the allowed phase
 space.
 For the unphysical values, we assign zero to
 the integrand because this implies negative neutrino masses.
 If such discarded trials in the integration exceed 20\% of the total 
number of iterations,
 we reject the event.
 This happens for events that lie near the kinematical boundary 
of the signal phase space.
 The corresponding reduction of the efficiency is approximately
 2\% and 3\% for the electron and muon mode, respectively.
 This additional decrease of the efficiency
 is not reflected in the values of Table~\ref{evtsel_sum}.
 
\begin{table}[]
\caption{Summary of event selection \label{evtsel_sum}}
\begin{center}
\scalebox{0.9}{
\begin{tabular}{l|cccc} \hline \hline
Item & $(e^- \nu_\tau\bar{\nu}_e\gamma)(\pi^+\pi^0\bar{\nu}_\tau)$ & $(e^+\nu_e\bar{\nu}_\tau\gamma )(\pi^-\pi^0{\nu}_\tau)$ & $(\mu^-\nu_\tau\bar{\nu}_\mu\gamma )(\pi^+\pi^0\bar{\nu}_\tau)$ & $(\mu^+\nu_\mu\bar{\nu}_\tau\gamma )(\pi^-\pi^0{\nu}_\tau)$ \\ \hline
$N_{\mathrm{sel}}$ & 391954  & 384880 & 35198 & 35973 \\ 
$\bar{\varepsilon}^{\dagger}$ (\%) & $4.28\pm0.24$ & $4.25\pm0.23$ & $3.58\pm0.19$  & $3.56\pm0.18$ \\
Purity (\%) & \multicolumn{2}{c}{$28.9\pm 0.8$} & \multicolumn{2}{c}{$57.4\pm 1.3$} \\ \hline \hline
\end{tabular}
}
\end{center}
\begin{flushleft}\vspace{-0mm}$~^\dagger$~{\small The efficiency is 
determined based on the photon
 energy threshold of $E_{\gamma}^{*}>10$~MeV in the 
$\tau$ rest frame.}\end{flushleft}
\end{table}

\section{Analysis of experimental data }

When we fit the Michel parameters for the real experimental data,
 the difference in selection efficiency between real data and MC simulation
 must be taken into account by the correction factor
 $R(\bvec{x})=\varepsilon^{\mathrm{EX}}(\bvec{x})/\varepsilon^{\mathrm{MC}}(\bvec{x})$
 that is close to unity; its extraction is described below.
 With this correction, Eq.~(\ref{generalpdf}) is modified to
\begin{equation}
P^{\mathrm{EX}}(\bvec{x}) = (1-\sum_{i}{\lambda_{i}})\cdot \frac{S(\bvec{x})\varepsilon^{\mathrm{MC}}
 (\bvec{x})R(\bvec{x})}{\int{\mathrm{d}\bvec{x} S(\bvec{x})\varepsilon^{\mathrm{MC}} (\bvec{x}) R(\bvec{x}) }}
+\sum_{i}{\lambda_{i}\frac{B_{i}(\bvec{x})\varepsilon^{\mathrm{MC}} (\bvec{x}) R(\bvec{x}) }
{\int{\mathrm{d}\bvec{x} B_{i}(\bvec{x})\varepsilon^{\mathrm{MC}} (\bvec{x})} R(\bvec{x}) }}\label{generalpdfR}.
\end{equation}
The presence of $R(\bvec{x})$ in the numerator does not affect the NLL 
minimization, but its presence in the denominator does.

We evaluate $R(\bvec{x})$ as the product of the measured corrections 
for the trigger,
 particle identification, track, $\pi^0$, and $\gamma$ reconstruction efficiencies:
\begin{align}
&R(\bvec{x})=R_\mathrm{trg}R_\ell(P_\ell, \cos\theta_\ell)R_{\gamma }(P_\gamma, \cos\theta_\gamma)R_\pi(P_\pi, \cos\theta_\pi)R_{\pi^0}(P_{\pi^0}, \cos\theta_{\pi^0}), \\
&R_\ell(P_\ell, \cos\theta_\ell) = R_\mathrm{trk}(P_\ell, \cos\theta_\ell) R_\mathrm{LID}(P_\ell, \cos\theta_\ell), \\
&R_\pi (P_\pi, \cos\theta_\pi) = R_\mathrm{trk}(P_\pi, \cos\theta_\pi) R_\mathrm{\pi ID}(P_\pi, \cos\theta_\pi ).
\end{align}

The lepton identification efficiency correction is estimated
 using two-photon processes $e^+e^-\rightarrow e^+e^-\ell^+\ell^-$ 
($\ell=e$ or $\mu$).
 Since the momentum of the lepton from
 the two-photon process ranges from the detector threshold
 to approximately $4$~GeV$/c$ in the laboratory frame,
 the efficiency correction factor
 can be evaluated for our signal process
 as a function of $P_\ell$ and $\cos\theta_\ell$.

The pion PID correction factor is obtained
 by the measurement of
 $D^{*+} \rightarrow D^0 \pi^+_{s} \rightarrow (K^- \pi^+)\pi^+_{s}$ decay
 (where the subscript $s$ indicates ``slow").
 The small momentum of the pion from $D^{*+}$ allows us to select this process.
 As a result, assuming the mass of $D^{0}$ meson, we can reconstruct $D^{*+}$ even if
 this $\pi^+$ is missed. 
 
The track reconstruction efficiency correction is extracted from 
$\tau^+\tau^- \rightarrow (\ell^+\nu_\ell\bar{\nu}_\tau)(\pi^-\pi^+\pi^- {\nu}_\tau)$ events.
 Here, we count the number of events $N_4$ ($N_3$) in which four (three) 
charged tracks are reconstructed.
 The three-track event is required to have a negative net charge 
($\pi^+$ is missing).
 Since the charged track reconstruction efficiency $\varepsilon$ 
is included as, respectively,
 $\varepsilon^4$ and $\varepsilon^3(1-\varepsilon)$ in $N_4$ and $N_3$,
 the value of $\varepsilon$ can be obtained by $\varepsilon = N_4/(N_4+N_3)$.
 The momentum and angular dependences of $\varepsilon$ are
 extracted by modifying $N_4\rightarrow \Delta N_4$, where
 $\Delta N_4$ is the number of observed events in a certain cell of
 the phase-space of reconstructed track.
 
The $\pi^0$ reconstruction efficiency correction is obtained by comparing
 the ratio of the number
 of selected events of $\tau^+\tau^- \rightarrow 
(\pi^+ \pi^0 \bar{\nu}_\tau)(\pi^- \pi^0 \nu_\tau)$
 and $\tau^+\tau^- \rightarrow (\pi^+ \pi^0 \bar{\nu}_\tau)(\pi^- \nu_\tau)$ between
 experiment and MC simulation. The momentum
 and angular dependence of the $\pi^0$ reconstruction efficiency is extracted
 by counting the number of events observed in a certain kinematic-variable
 cell of the $\pi^0$ phase space. 
 By randomly choosing either of the photon daughters from the $\pi^0$,
 the $\gamma$ reconstruction efficiency correction is
 extracted in the same manner.
 
The trigger efficiency correction has the largest impact among these factors.
 In particular, for the electron mode, because of the similar structure 
of our signal events and Bhabha events (back-to-back topology of 
two-track events),
 many signal events are rejected by the Bhabha veto in the trigger.
 The veto of the trigger results in a spectral distortion and a large 
systematic uncertainty.
 The correction factor is extracted using
 the charged and neutral subtriggers (denoted as $Z$ and $N$), which provide 
completely independent signals.
Since the trigger signal appears when at least one of the subtriggers
 fires (\textit{i.e.}, $Z$ OR $N$), its efficiency is given by
 $\varepsilon_{\rm trg} = 1 - (1-\varepsilon_Z)(1-\varepsilon_N)$, where
 $\varepsilon_Z = N_{Z\&N}/N_N$ and $\varepsilon_N = N_{Z\&N}/N_Z$ are
 the efficiencies of the charged and neutral subtriggers, respectively;
 $N_{Z\&N}$ is the number of events where both subtriggers fire
 (\textit{i.e.} $Z$ AND $N$), $N_Z$ ($N_N$) is a number of events triggered by $Z$ ($N$).
 Thus $R_{\rm trg}$ is obtained as the ratio of $\varepsilon_{\rm trg}$
 between the experiment and MC simulation.
 
Figure~\ref{corr_figs} shows the distribution of the momentum and the 
cosine of the polar angle of electron and muon events.
 In the figure, the effects of all corrections are seen mainly 
at cos$\theta_e<-0.6$ and cos$\theta_\mu<-0.6$.

\begin{figure}[]
{\centering \subfloat[$P_e^\mathrm{LAB}$]{\hspace{5mm}\includegraphics[width=6.2cm]{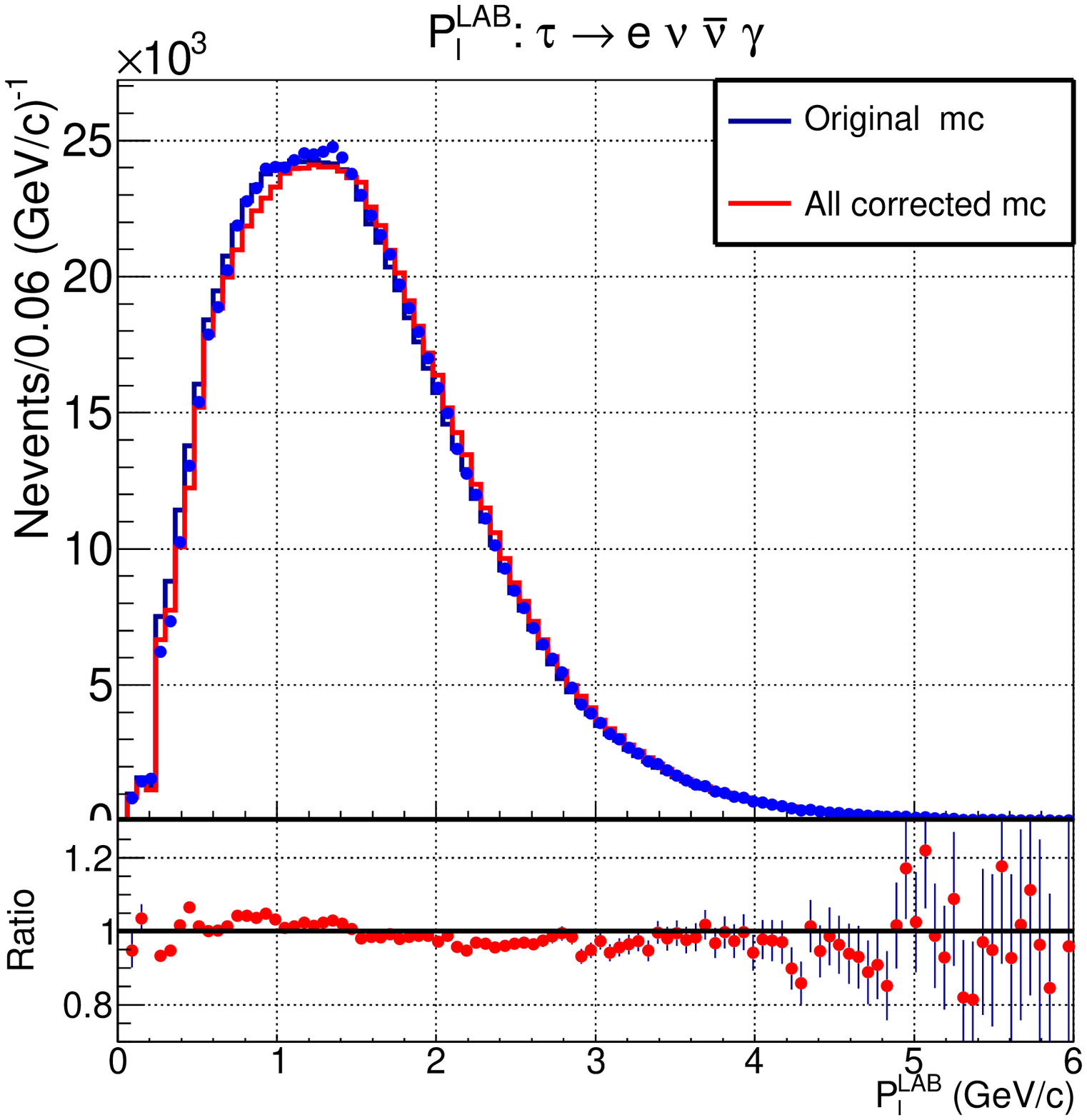}\label{selection:p_e}} }
\hspace{5mm}{\centering \subfloat[cos$\theta_{e}^\mathrm{LAB}$]{\includegraphics[width=6.2cm]{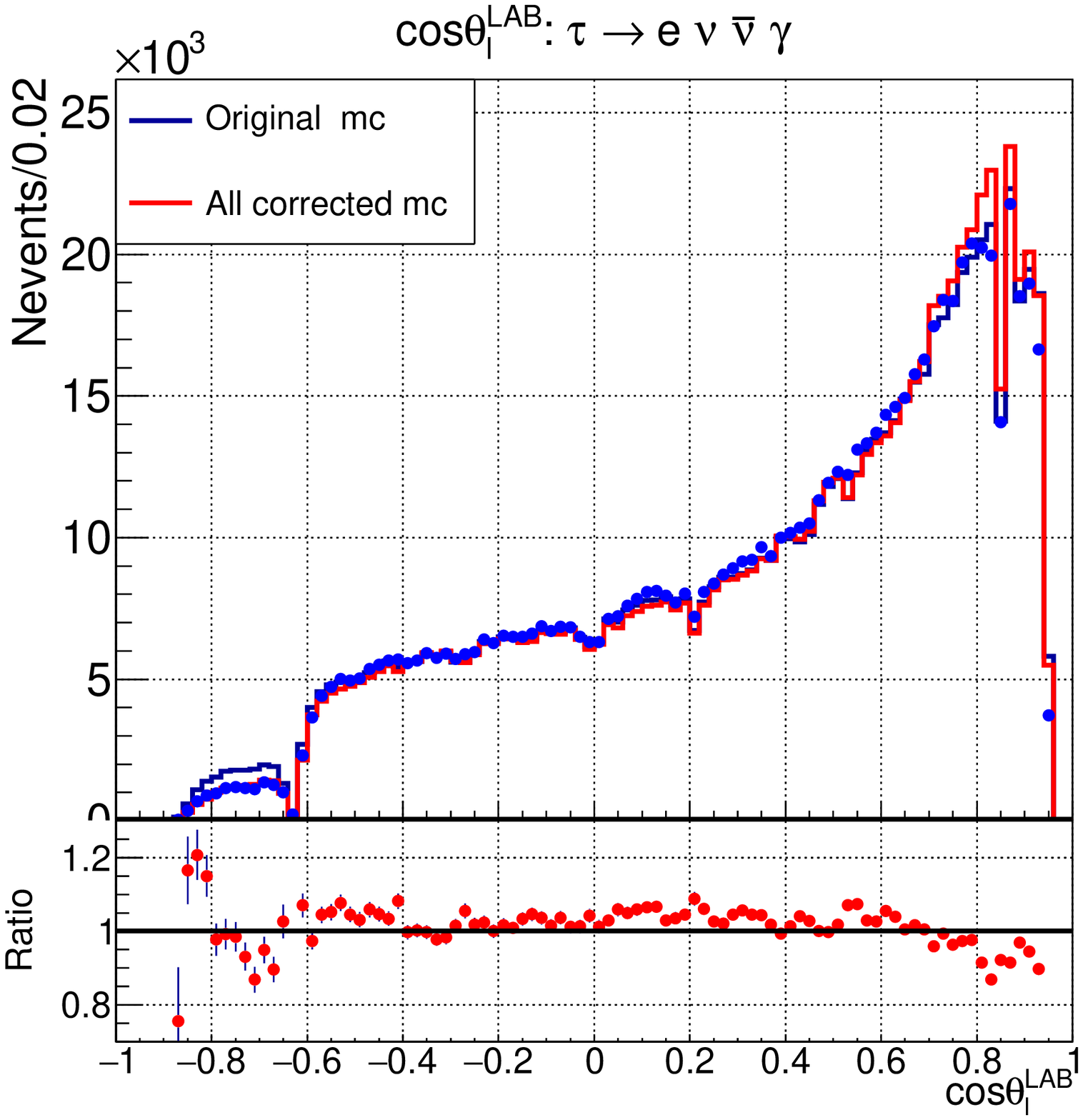}\label{selection:cth_e}} }\\
{\centering \subfloat[$E_\gamma^\mathrm{LAB}$ ($e$ mode)]{\hspace{5mm} \includegraphics[width=6.2cm]{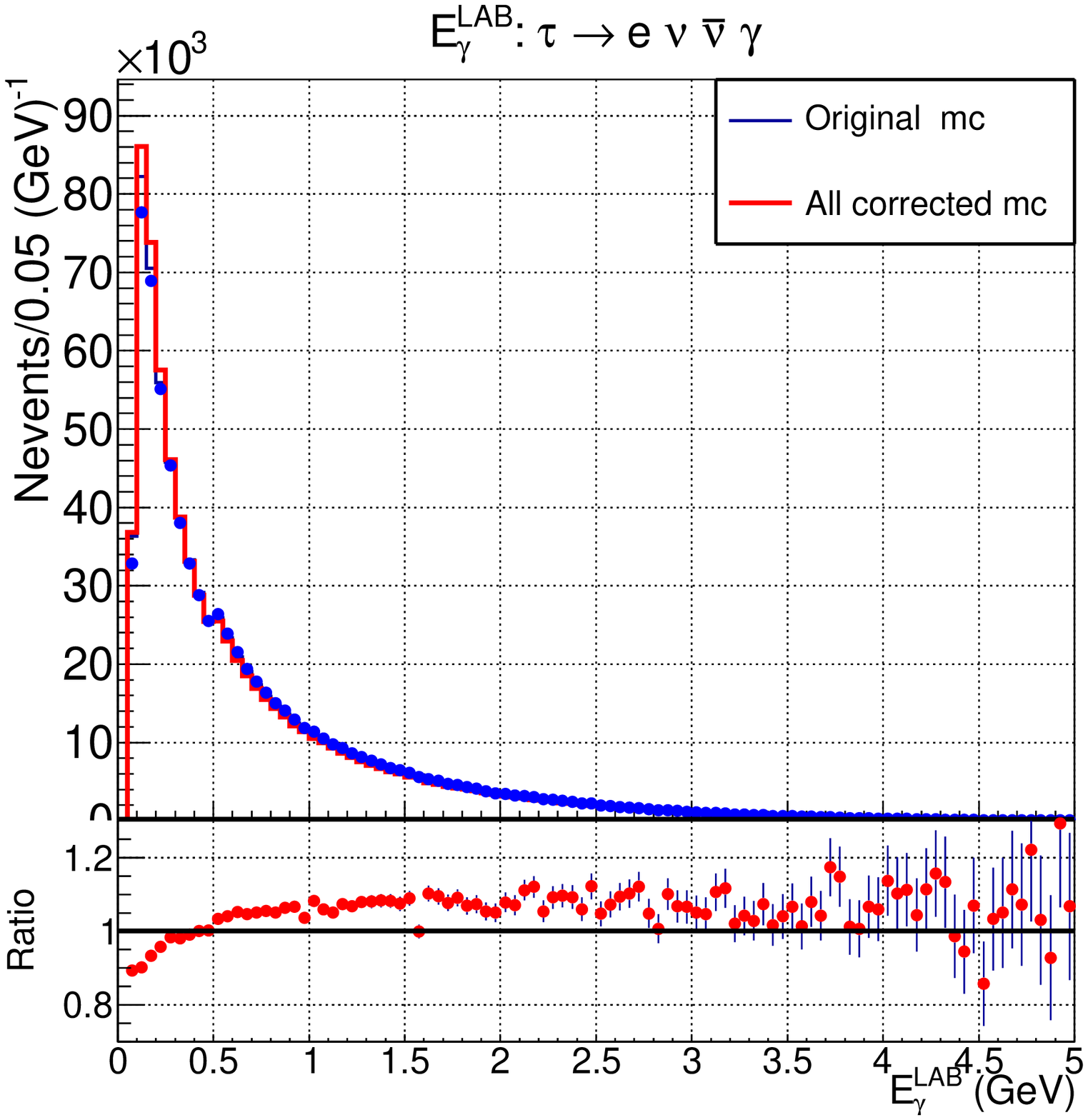}\label{selection:pg_e}} }
{\centering \subfloat[$P_\mu^\mathrm{LAB}$]{\hspace{5mm} \includegraphics[width=6.2cm]{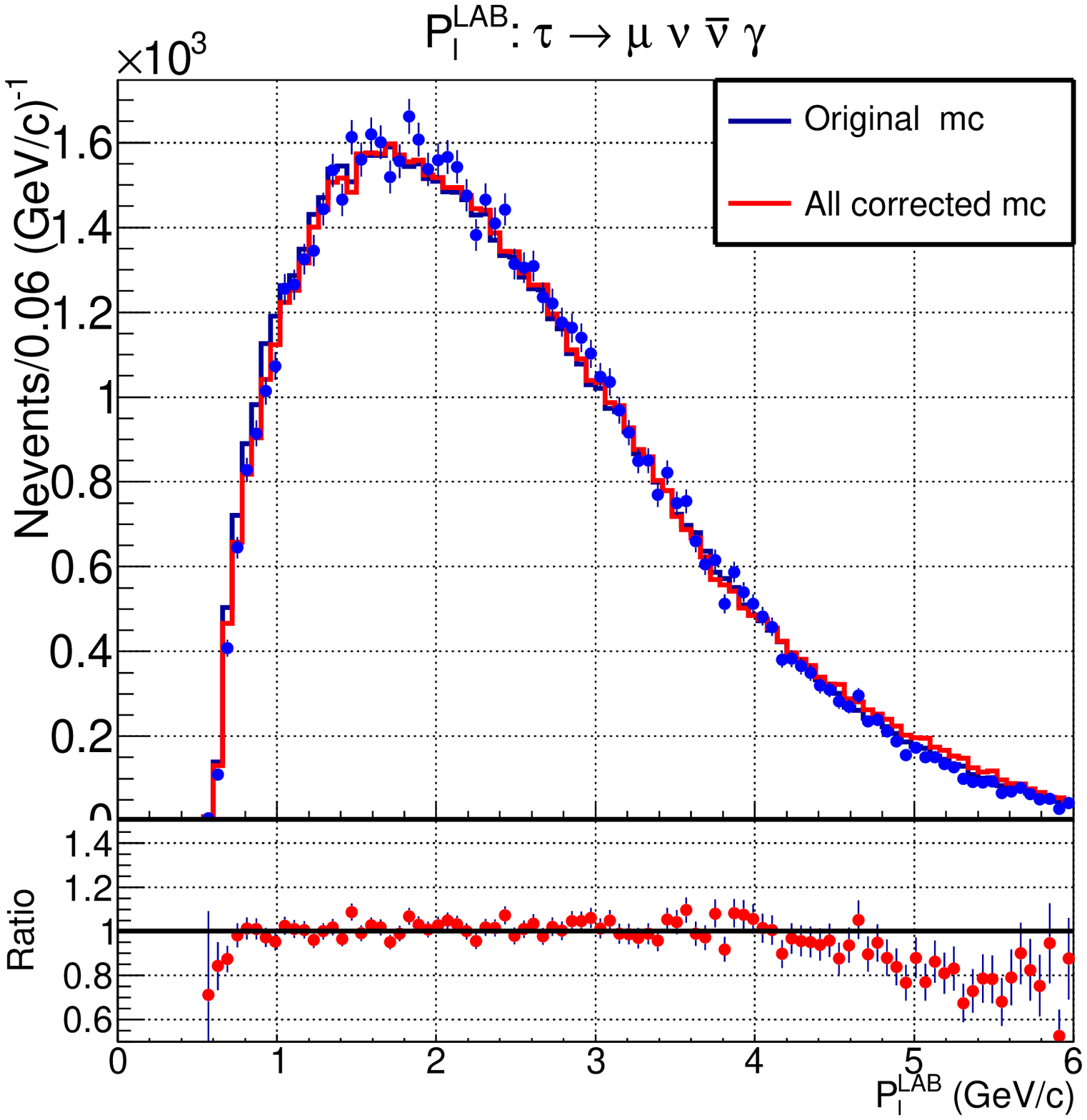}\label{selection:p_mu}} } \\
{\centering \subfloat[cos$\theta_{\mu}^\mathrm{LAB}$]{\includegraphics[width=6.2cm]{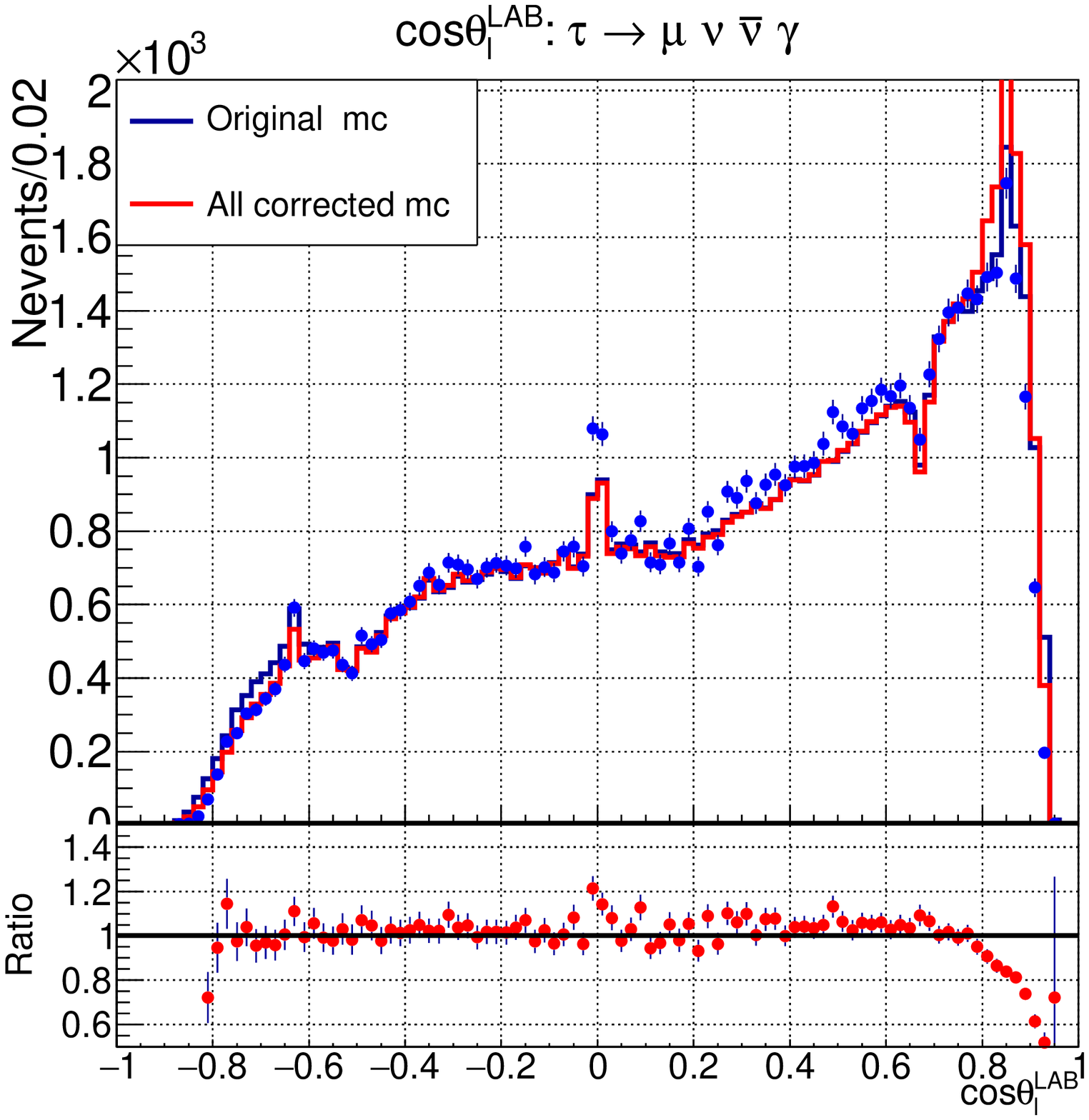}\label{selection:cth_mu}}}
\hspace{9mm}{\centering \subfloat[$E_\gamma^\mathrm{LAB}$ ($\mu$ mode)]{\includegraphics[width=6.2cm]{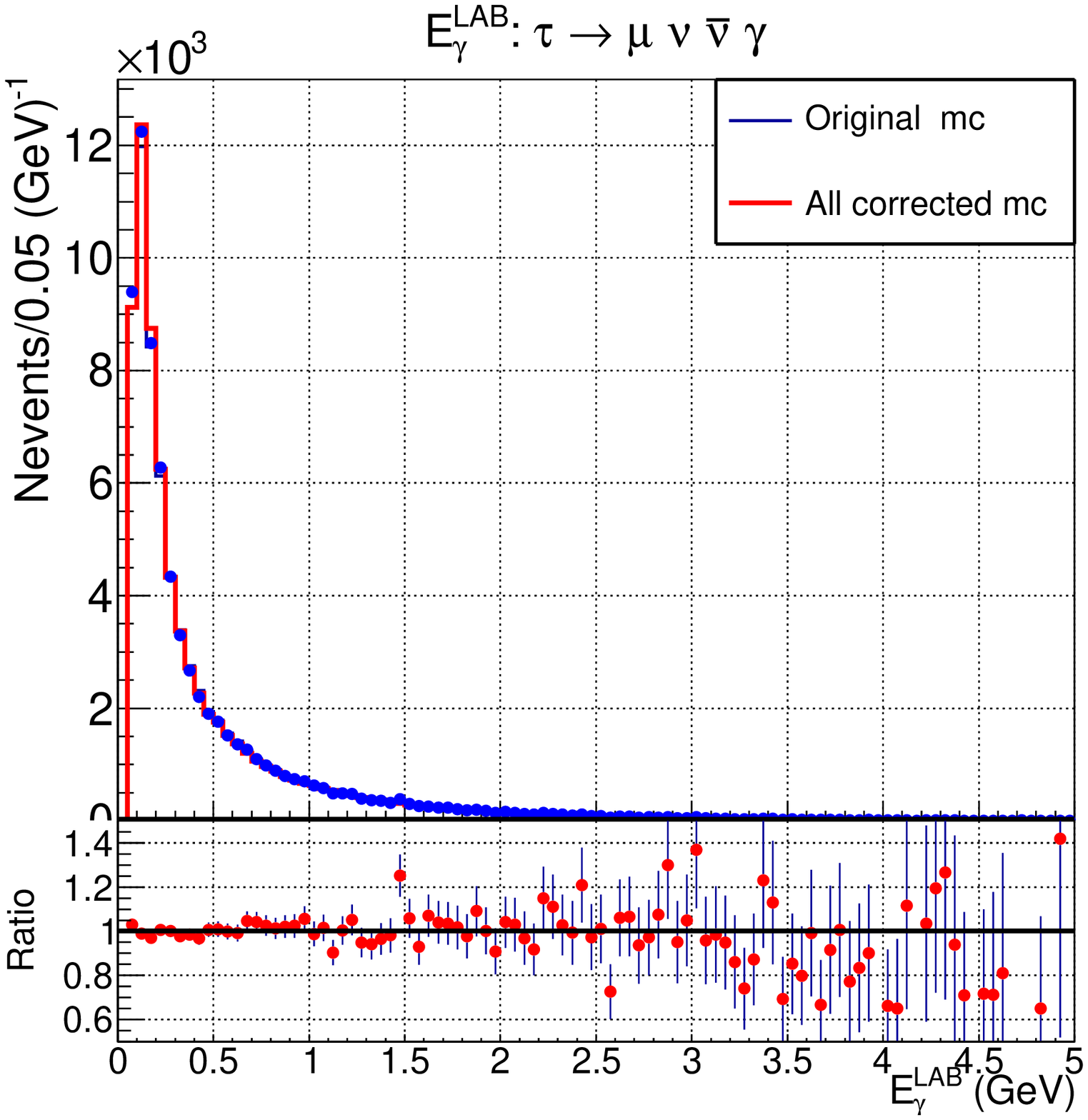}\label{selection:pg_mu}} }
\caption[]{\raggedright Distributions of momenta of leptons, cosine of angles and photon energy:
 (a)(b)(c) for the electron modes and (d)(e)(f) for the muon modes.
 Blue points with uncertainties represent the experimental data while
 the black and red lines represent the distributions of the original 
and corrected MC simulation, respectively. }\label{corr_figs}
\end{figure}

\section{Evaluation of systematic uncertainties}

\begin{table}[]
\caption{List of systematic contributions \label{sys_sum}}
\begin{center}
\begin{tabular}{l|cccc} \hline \hline
Item & $\sigma_{\bar{\eta }}^e$ & $\sigma_{\xi \kappa }^e$ & $\sigma_{\bar{\eta }}^\mu$ & $\sigma_{\xi \kappa }^\mu$  \\ \hline
Relative normalizations & $3.8$ & $0.69$  & 0.13    & 0.04   \\
Absolute normalizations & $1.0$ & $0.01$ &  $0.03$ & $0.001$\\
Formulation of PDFs & $2.5$ & $0.24$ & 0.67 & 0.22    \\
Input of branching ratio & $3.8$ & $0.05$ & $0.25$ & $0.01$ \\
Effect of cluster overlap in ECL & $2.2$ & $0.46$  & $0.02$ & $0.06$  \\
Detector resolution & 0.74 & 0.20 & 0.22 & 0.02 \\
Exp/MC corrections  & $1.9$ &  $0.14$ & $0.09$ &  $0.10$ \\
$E_{\gamma}$ cut & 0.91 & 0.22 & - & - \\ \hline 
Total & 6.8 & 0.93 & 0.77 & 0.25 \\ \hline \hline
\end{tabular}
\end{center}
\end{table}

In Table~\ref{sys_sum}, we summarize the contributions of the identified
 sources of systematic uncertainties.
 The dominant source for the electron mode is the calculation of the relative normalizations.
 Due to the peculiarity of the signal PDF when $m_\ell\rightarrow m_e$,
 the convergence of the relative normalization coefficients is quite
 slow and results in a notable effect. 
 For a given number of MC events $N$,
 the errors of the relative normalizations $\left< A_i/A_0 \right>$ ($i=1,2$) are
 evaluated by $\sigma^2 = {\rm Var}(A_i/A_0)/N$, where ${\rm Var(}X)$ represents the variance of a random variable $X$. 
 The resulting systematic effect on the Michel parameter is estimated by varying the normalizations.
 The effect of the absolute normalization is estimated in the same way.
 
The largest systematic uncertainty for the muon mode is due to the limited precision of the description
 of the background PDF that appears in Eq.~(\ref{generalpdfR}).
 As mentioned before, the remaining background sources are described
 by a common PDF, which is tabulated utilizing a large $\tau^+\tau^-$ generic MC sample.
 This effective description can generally discard information about correlations
 in the phase space and thereby give significant bias.
 The residuals of the fitted Michel parameters from the SM prediction obtained by the fit to the MC distribution
 are taken as the corresponding systematic uncertainties.

Other notable uncertainties come from the accuracy of the
 measured branching ratios.
 In particular, the uncertainties of the branching ratio of the radiative
 decay $\tau^-\rightarrow \ell^- \nu_\tau \bar{\nu}_\ell \gamma$ dominate the contribution.
 The systematic effects of the cluster merging in the ECL are evaluated as a function of the
 angle between the photon and lepton clusters at the front face of ECL ($\theta_{\ell\gamma}^{\mathrm{ECL}}$).
 The limit $\theta_{\ell\gamma}^{\mathrm{ECL}} \rightarrow 0$ represents the merger of the two clusters
 and the comparison of the distribution between experiment and MC gives us the corresponding bias.
 A systematic effect due to the detector resolution is evaluated by comparing Michel parameters
 obtained in the fit with and without account
 of the resolution function $R(\bvec{x}-\bvec{x}^\prime)$.
 
The error of the measured correction factor $R$ is estimated by varying the central
 values based on the uncertainty in each bin.
 Moreover, as can be observed in Fig.~\ref{corr_figs}d in the muon mode,
 there is a notable disagreement of efficiency in the forward domain (cos$\theta_{\mu}^\mathrm{LAB}>0.9$).
 This is due to the contamination of backgrounds in the extraction of the correction factor
 of $R_{\mu \mathrm{ID}}$.
 We excluded this region (reducing the statistics by 1.5\%) and
 checked the shift of the refitted Michel parameters.
 
In the electron mode, we observe the disagreement
 of the photon reconstruction efficiency
 in the low-energy region (Fig.~\ref{corr_figs}c).
 It could arise from a discrepancy in the simulation of extra bremsstrahlung.
 We excluded the events having a low energy photon $E_{\gamma}^\mathrm{LAB}<150$~MeV and
 compared the refitted values. Because this cut reduces the number of events
 by approximately 20\%, the statistical fluctuation is also
 reflected in the shifts.
 
 The effect of the beam-energy spread is estimated by varying the input of this
 value for the calculation of the PDF with respect to run-dependent uncertainties, and
 turns out to be negligible.
 
The effects from the next-to-leading-order (NLO) contribution were checked
 by adding the NLO formulae~\cite{radiativedecay_theory} to the signal PDF and refitting, and
 were found to be negligible.
 
\section{Results}

Because of the suppression of sensitivity due to the small mass of the electron,
the $\bar{\eta}$ parameter is extracted only from the $\tau^-\rightarrow \mu^- \nu_\tau \bar{\nu}_\mu \gamma$ mode.
 Using the 71171 selected $\tau^- \rightarrow \mu^- \nu_\tau \bar{\nu}_\mu \gamma$ candidates, 
 $\bar{\eta}$ and $\xi\kappa$ are simultaneously fitted to the kinematic distribution to be 
\begin{align}
\bar{\eta}^{\mu} &= -1.3 \pm  1.5 \pm  0.8, \label{etbresdayo} \\
(\xi\kappa)^{\mu} &= ~~0.8 \pm  0.5 \pm 0.3.
\end{align}
Figure~\ref{exp03_fit_mu} shows the contour of the likelihood
 function for $\tau^- \rightarrow \mu^- \nu_\tau \bar{\nu}_\mu \gamma$ events.

\begin{figure}[]
\centering {\includegraphics[width=8cm]{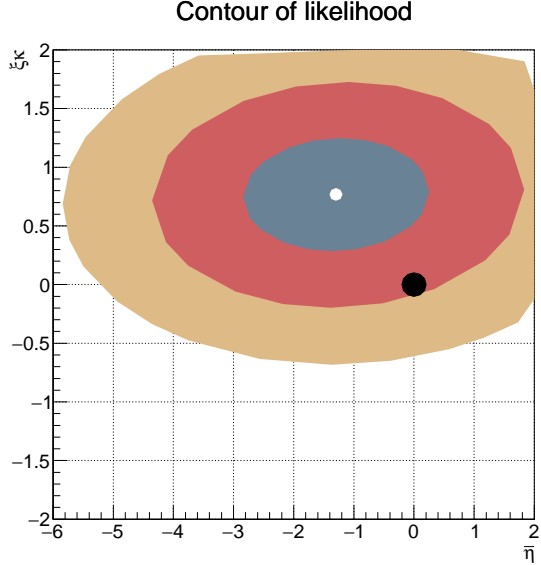}}
\caption[]{Contours of the likelihood function obtained using 71171
 events for $\tau^- \rightarrow \mu^- \nu_\tau \bar{\nu}_\mu \gamma$ candidates.
 The ovals are 1$\sigma$-, 2$\sigma$- and 3$\sigma$-contours of statistical deviation of the likelihood function from the best estimation.
 The black dot is the SM prediction. }\label{exp03_fit_mu}
\end{figure}

In the electron mode, $\xi\kappa$ is fitted by fixing the $\bar{\eta}$ value to the
 SM prediction of $\bar{\eta}=0$ and the optimal value is extracted using 776834 events to be
\begin{align}
(\xi\kappa)^e = -0.4 \pm 0.8 \pm 0.9.
\end{align}
In Equations \ref{etbresdayo}--\ref{xik_resdayo}, the first error is statistical and second systematic.
 The obtained values are consistent with the SM prediction.

Furthermore, the $\xi\kappa$ product is
 also obtained by fitting simultaneously to both electron and muon events as
\begin{align}
\xi\kappa &= 0.5 \pm 0.4 \pm 0.2\label{xik_resdayo},
\end{align}
Here, the systematic uncertainty
 is estimated from ${1}/\sigma^2_{\rm comb} = 1/\sigma^{2}_{e}+ 1/\sigma^{2}_{\mu}$
 by assuming they are uncorrelated.

We also obtain the dependence of the $E^{\mathrm{LAB}}_{\mathrm{extra \gamma}}$
 selection on the fitted Michel parameters as shown in Fig.~\ref{MC_etabarxikappa_Extra}.
 In the extraction of $\bar{\eta}$, we use $\tau^- \rightarrow \mu^- \nu_\tau \bar{\nu}_\mu \gamma$
 while, for $\xi\kappa$, we use the combined
 result for $\tau^- \rightarrow e^- \nu_\tau \bar{\nu}_e \gamma$
 and $\tau^- \rightarrow \mu^- \nu_\tau \bar{\nu}_\mu \gamma$ decays.
 We observe stability of the fitted Michel parameters within uncertainties.
 Figure~\ref{delL} shows the residual of the likelihood function
 $\Delta L = \rm{NLL}_{\rm min}-\rm{NLL}$ projected onto one axis.
 We observe a smooth and quadratic shape of the NLL around its minimum.

\begin{figure}[h]
\centering \subfloat[$\bar{\eta}$]{\includegraphics[width=8cm]{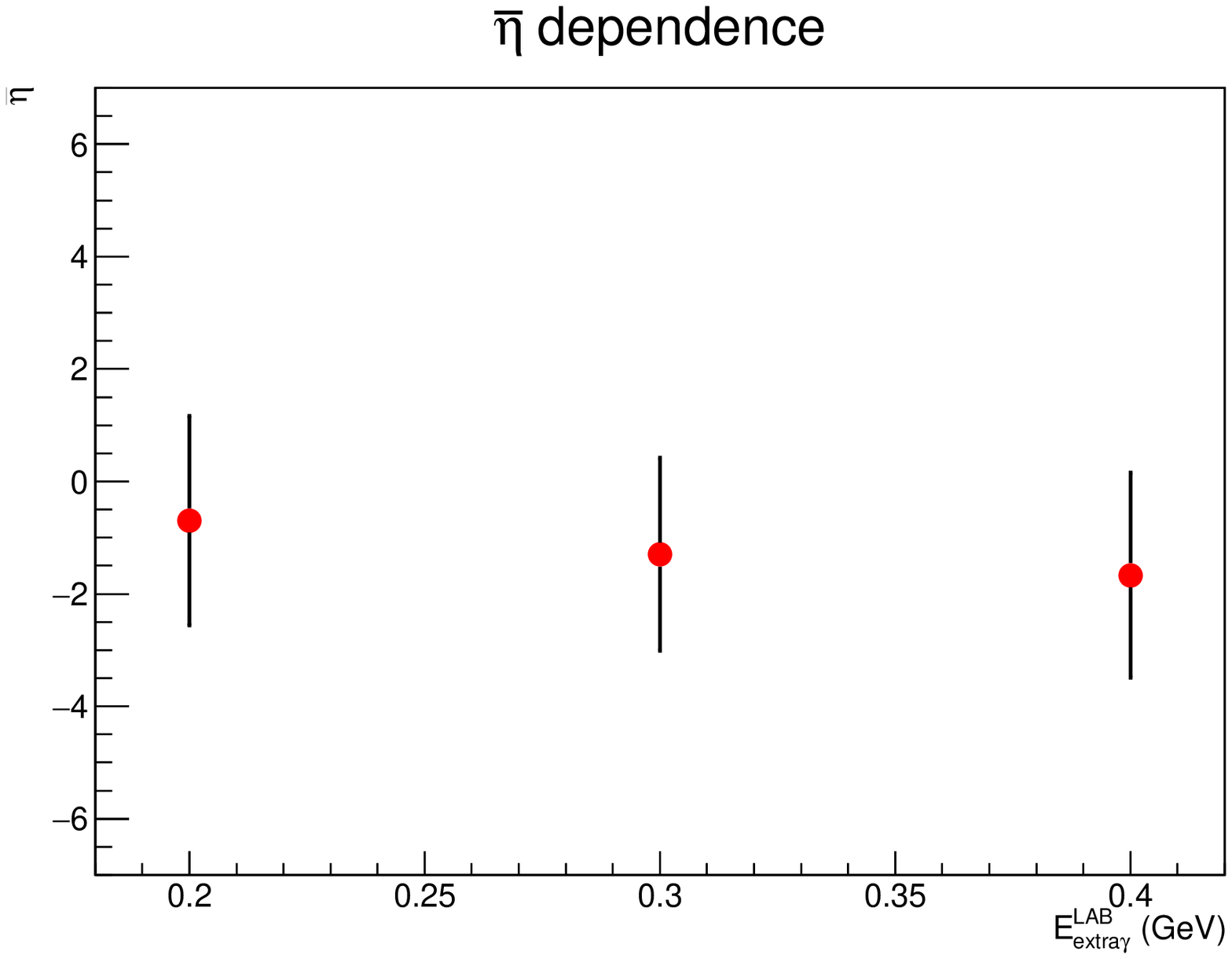}\label{EX_etabarxikappa_Extra:etabar}}
\centering \subfloat[$\xi\kappa$]{\includegraphics[width=8cm]{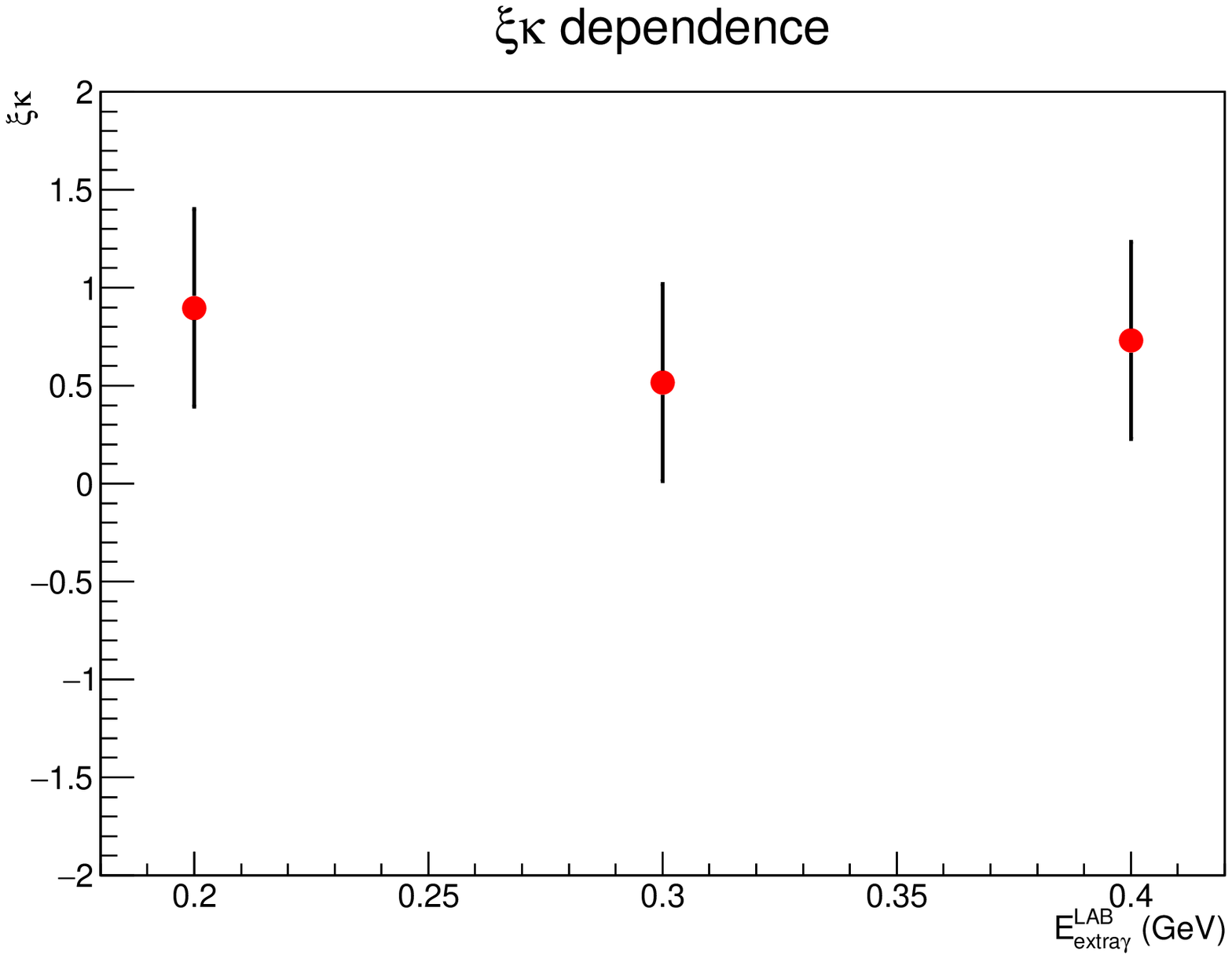}\label{EX_etabarxikappa_Extra:xikappa}}
\caption[]{Dependence of Michel parameters on
 the $E^{\mathrm{LAB}}_{\mathrm{extra \gamma}}$ selection (a) $\bar{\eta}$ (b) and $\xi\kappa$.
 The red markers with bars correspond to the optimal values of Michel parameters
 and their uncertainties, where both statistical and systematic uncertainties
 are considered.} \label{MC_etabarxikappa_Extra}
\end{figure} 

\begin{figure}[h]
\centering \subfloat[$\bar{\eta}^\mu$]{\includegraphics[width=5cm]{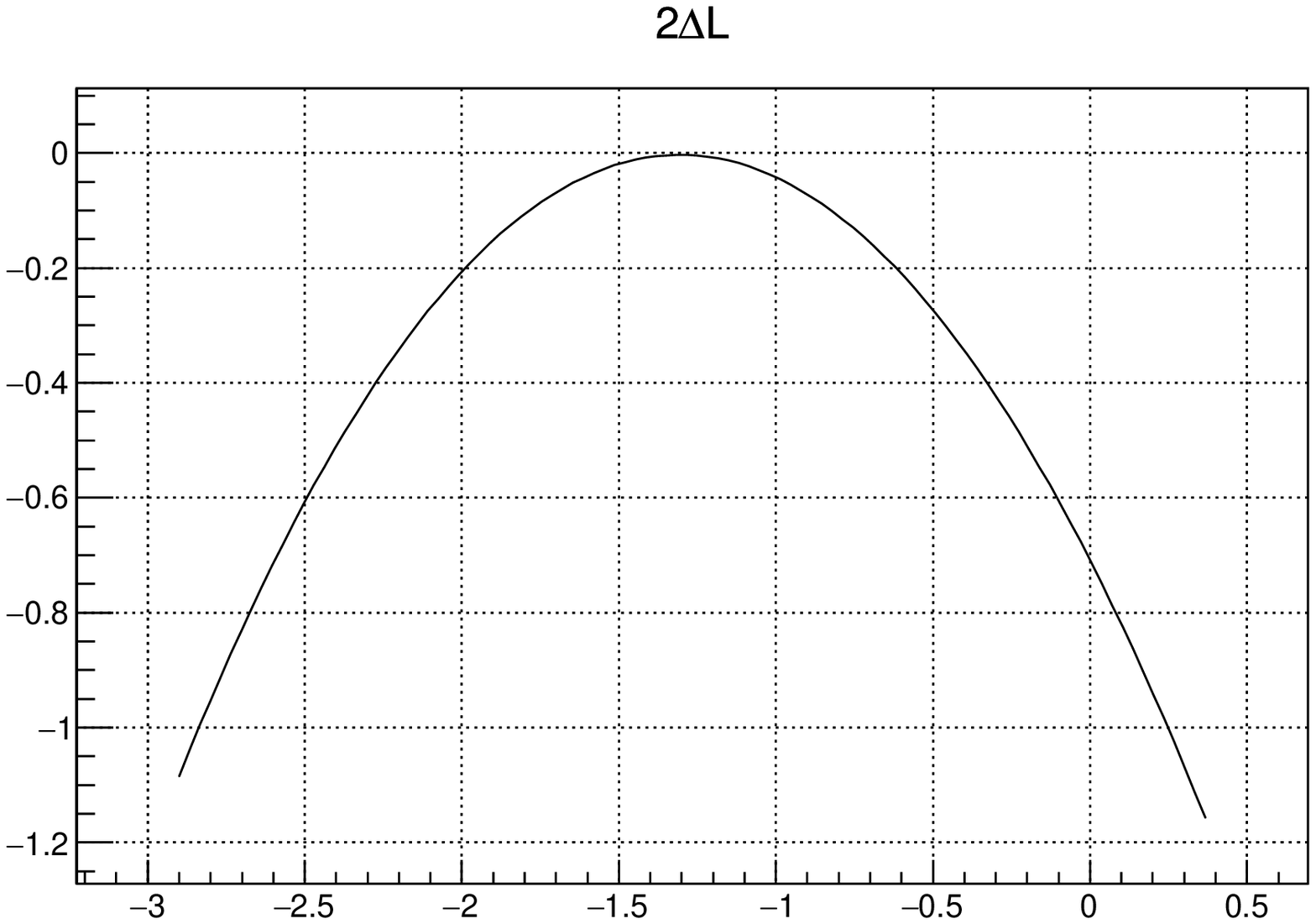}\label{delL:m_etabar}}
\centering \subfloat[$(\xi\kappa)^e$]{\includegraphics[width=5cm]{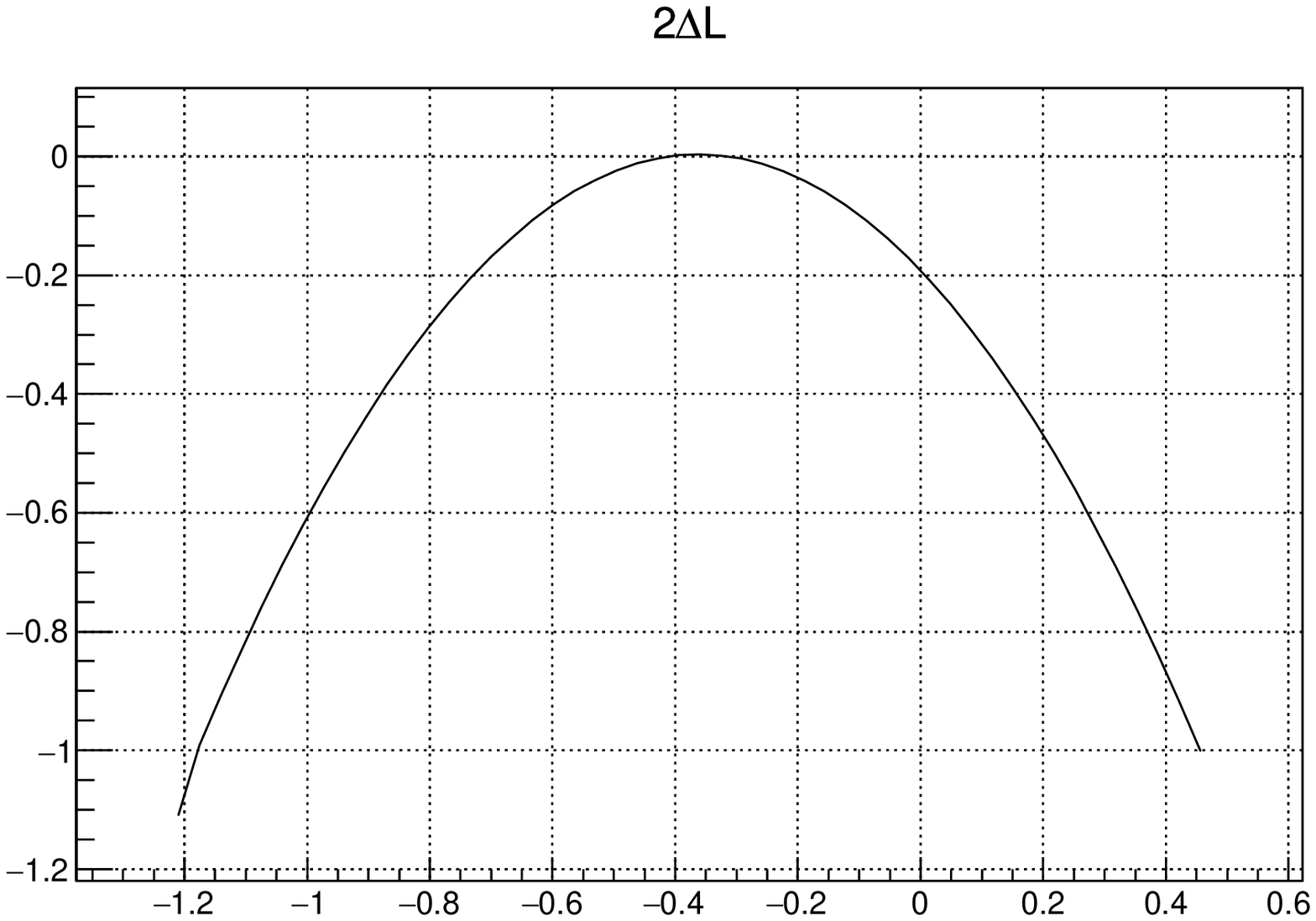}\label{delL:e_xikappa}}
\centering \subfloat[$(\xi\kappa)^\mu$]{\includegraphics[width=5cm]{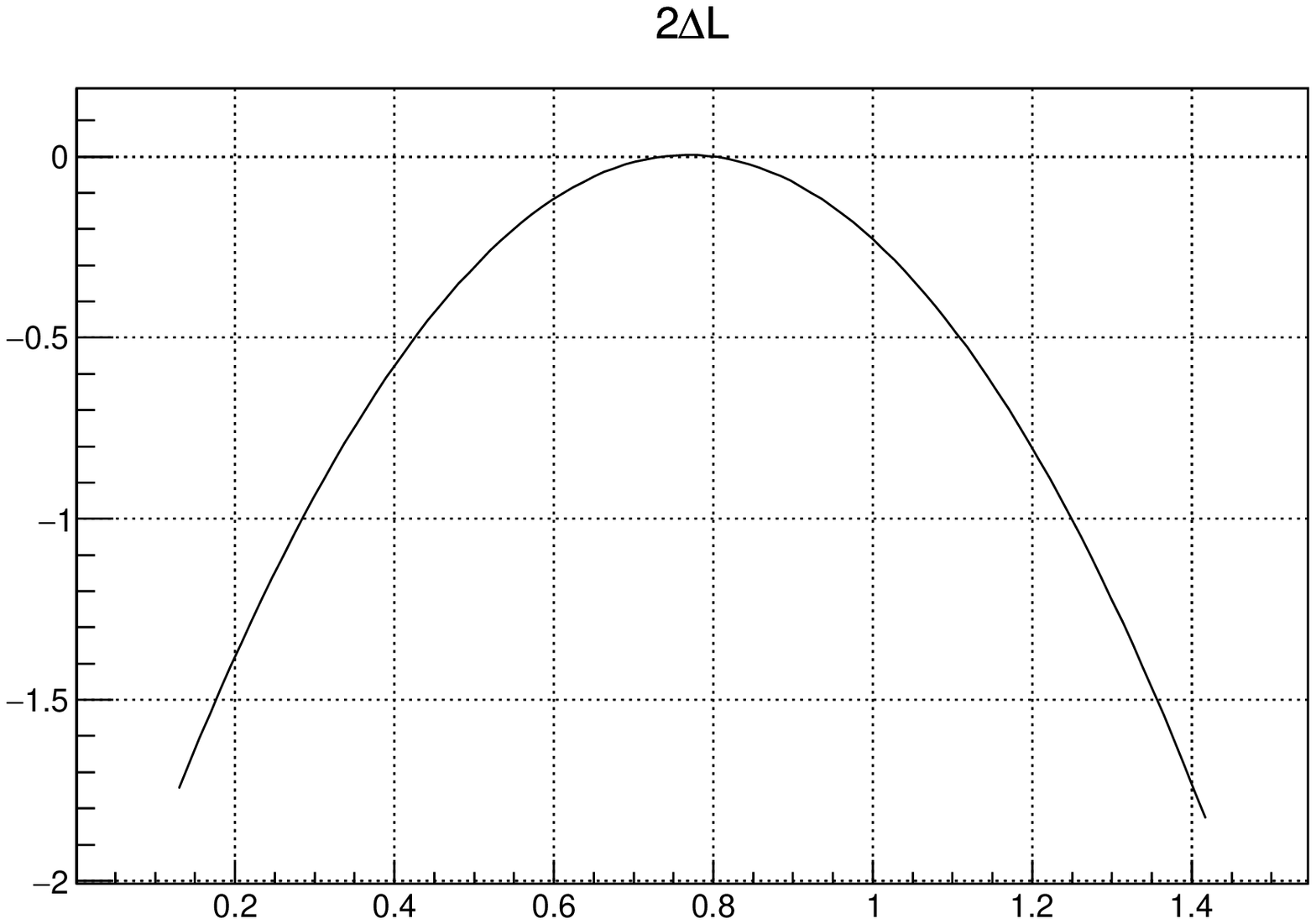}\label{delL:m_xikappa}}
\caption[]{Plot of $2\times\Delta L$ as a function of Michel parameters; (a) $-2\rm{NLL}(\bar{\eta}^{\mu})$ when $\xi\kappa $
 is set to the fitted value;
 (b) $-2\rm{NLL}(\xi\kappa^{e})$ when $\bar{\eta} = \bar{\eta}_{\rm SM} = 0$;
 (c) $-2\rm{NLL}(\xi\kappa^{\mu})$ when $\bar{\eta} $ is set to the fitted value.} \label{delL}
\end{figure}

\section{Measurement of the branching ratio $\mathcal{B}(\tau^- \rightarrow \ell^- \nu_\tau \bar{\nu}_\ell\gamma)$}
In addition to the Michel parameters, we have determined the 
branching ratios of the 
$\tau^- \rightarrow \ell^- \nu_\tau \bar{\nu}_\ell\gamma$~($\ell = e,\mu$) decays. 

Following the definition of Ref.~\cite{radiativedecay_theory}, we distinguish
 between two types of radiative decays in the NLO approximation: 
the exclusive radiative decay implies that only one hard photon is emitted in 
the event; in the inclusive radiative decay, at least one hard photon 
is emitted. Here, the hard photon energy threshold is 10~MeV in the 
$\tau^-$ rest frame.

In Ref.~\cite{radiativedecay_theory}, 
the precision measurement of the branching ratios of the radiative 
leptonic $\tau$ decays at BaBar is also discussed. 
While the measured branching ratios of both electron and muon modes agree 
with their leading-order (LO) theoretical predictions, 
the NLO exclusive branching ratio prediction for the 
$\tau^-\to e^-\nu_\tau\bar{\nu}_e\gamma$ decay differs from 
the BaBar result by 3.5 standard deviations. 
This is explained by the insufficient accuracy of the current 
MC simulation of the radiative and doubly-radiative leptonic 
$\tau$ decays. 
Neither an NLO correction to the radiative leptonic decay, nor the
 doubly-radiative leptonic mode itself, 
are incorporated in the current version of the TAUOLA MC generator. 
As a result, the detection efficiency is not precisely evaluated for the 
radiative decay. As well, the background from the doubly
 radiative decay 
is not subtracted at all. Finally, the second photon emission might
affect the efficiency of the photon veto and the shape of the neutral clusters 
in the calorimeter. Indeed, the ratio of the yield with two-photon 
emission to that with single-photon emission is approximately 
5\% and 1\% for the electron and muon modes, respectively. Thus, there is an
experimentally notable impact of the two-photon emission on the electron mode. 

In our measurement of the branching ratios, we do not take into account 
the up-to-date formalism of Ref.~\cite{radiativedecay_theory} since
 the main purpose of this study is a consistency check of our selection 
criteria and experimental efficiency corrections.

\subsection{Method}

The branching ratio is determined using
\begin{align}
\mathcal{B}(\tau^-\rightarrow\ell^-\nu_\tau\bar{\nu}_\ell\gamma) = \frac{N_{\mathrm{obs}}(1-f_{\mathrm{bg}})}{2\sigma_{\tau\tau}L\mathcal{B}(\tau^+\rightarrow\pi^+\pi^0\bar{\nu}_\tau)\bar{\varepsilon}^{\rm EX}},
\end{align}
where $\mathcal{B}(\tau^+\rightarrow\pi^+\pi^0 \bar{\nu}_\tau)=(25.52\pm0.09)\%$~\cite{PDG_paper}
 is the branching ratio of $\tau^+\rightarrow \pi^+\pi^0 \bar{\nu}_\tau$ decay,
 $N_{\mathrm{obs}}$ is the number of observed events, $f_{\mathrm{bg}}$ is the 
fraction of background events,
 $\sigma_{\tau\tau}=(0.919\pm 0.003)~\mathrm{nb}^{-1}$ is the cross section of the 
$e^+e^-\rightarrow \tau^+\tau^-$ process at $\Upsilon(4S)$~\cite{tautau_cross}, $L=(711\pm10)~{\rm fb}^{-1}$ 
is the integrated luminosity recorded at $\Upsilon(4S)$, and $\bar{\varepsilon}^{\rm EX}$ 
is the average detection efficiency of signal events.
The efficiency, $\bar{\varepsilon}^{\rm EX}$, is evaluated with help of the MC simulation.
 The correction factor, $R(\bvec{x})=\varepsilon^{\rm EX}(\bvec{x})/\varepsilon^{\rm MC}(\bvec{x})$, 
which is used to extract Michel parameters, is applied to compensate for the difference between experimental 
and MC efficiencies as follows: 
\begin{align}
\bar{\varepsilon}^{\rm EX} &= \int \hspace{-1mm} \mydif\bvec{x} ~ S(\bvec{x}) \varepsilon^{\rm EX} (\bvec{x})
= \int \hspace{-1mm} \mydif\bvec{x} ~S(\bvec{x}) \varepsilon^{\mathrm{MC}}(\bvec{x}) R(\bvec{x}) 
= \frac{\bar{\varepsilon}^{\mathrm{MC}}}{N_{\mathrm{sel}}} \sum_{i: \mathrm{sel~(MC)}} R(\bvec{x^{i}}) = \bar{\varepsilon}^{\mathrm{MC}} \bar{R},
\end{align}
where $S(\bvec{x})$ is the PDF of the signal events and $\bar{R}$ is
an average efficiency correction factor for the selected signal MC events.
Here, the average MC efficiency, $\bar{\varepsilon}^{\mathrm{MC}}$, is 
determined for the photon energy threshold of $10$~MeV in the $\tau$ rest frame. 
 
\subsection{Event selection}

We apply additional selection criteria to enhance the purity of the sample 
as well as to reduce systematic uncertainties.
 The extra-gamma-energy selection is released for the latter purpose but other
 selection criteria are common to those of Michel parameter measurement (see in Sec.~\ref{labelevs}).
For the electron mode, we apply the following selection criteria:
\begin{itemize}

\item{The uncertainty of the lepton identification efficiency in the forward 
and backward regions of the detector is large due to the notable background 
contamination of the control sample; thus,
 the electron polar angle in the laboratory frame must lie in the region defined by 
$\theta_{e}^{\rm LAB}<126^\circ $ as shown in Figs.~\ref{sel_br_opt_1}a and \ref{sel_br_opt_1}b.
}

\item{The electron identification is less precise at small momenta,
 so we apply the momentum threshold $P_{e}^{\rm LAB}>1.5~{\rm GeV}/c$ as shown in Fig.~\ref{sel_br_opt_1}c.
}

\item{After the final selections (explained in Sec.~\ref{labelevs}), 
the dominant background arises from the external bremsstrahlung on the material of the 
detector. It is effectively suppressed by applying the requirement on the invariant mass
of the electron-photon system, $M_{e\gamma}>0.1~{\rm GeV}/c^2$, as shown in Fig.~\ref{sel_br_opt_3}.}

\item{The extra gamma energy in the laboratory frame, $E^{\rm LAB}_{\mathrm{extra}\gamma}$, 
must be less than $0.2$~GeV.}

\end{itemize}

For the muon mode, we apply the following selection criteria:
\begin{itemize}
\item{The muon polar angle in the laboratory frame must satisfy
 $51^\circ < \theta_{\mu}^{\rm LAB} < 117^\circ$. }

\item{The spatial angle between $\mu$ and $\gamma$ in c.m.s. must satisfy $\cos\theta_{\mu\gamma}>0.99$. }

\item{The extra gamma energy in laboratory frame, $E^{\rm LAB}_{\mathrm{extra}\gamma}$, must be smaller 
than $0.3$~GeV.  }
\end{itemize}

\begin{figure}[]
\centering \subfloat[cos$\theta_e^{\rm LAB}$]{\hspace{-6mm} \includegraphics[width=5.5cm]{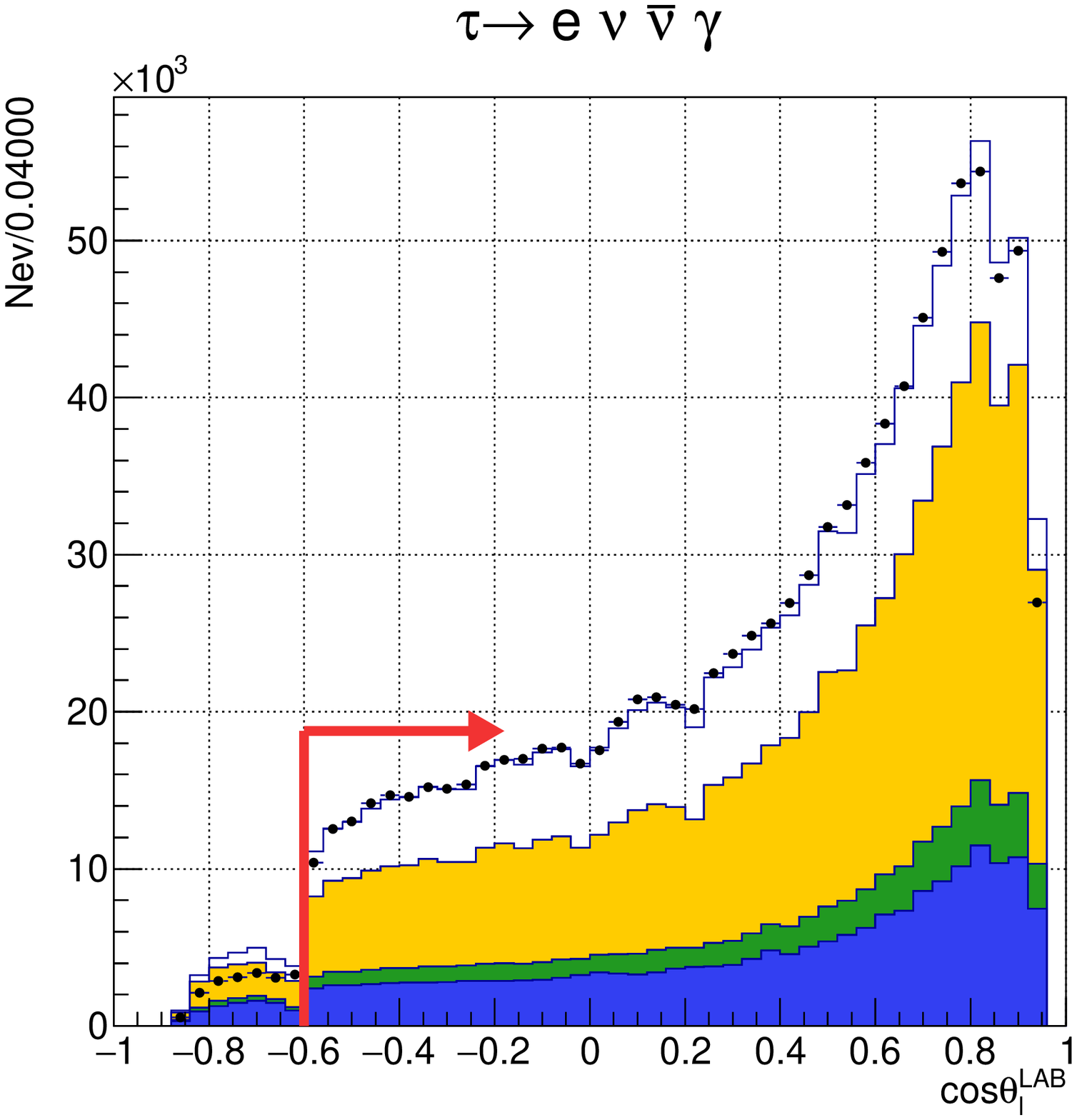}}
\centering \subfloat[cos$\theta_\mu^{\rm LAB}$]{\hspace{-3mm} \includegraphics[width=5.5cm]{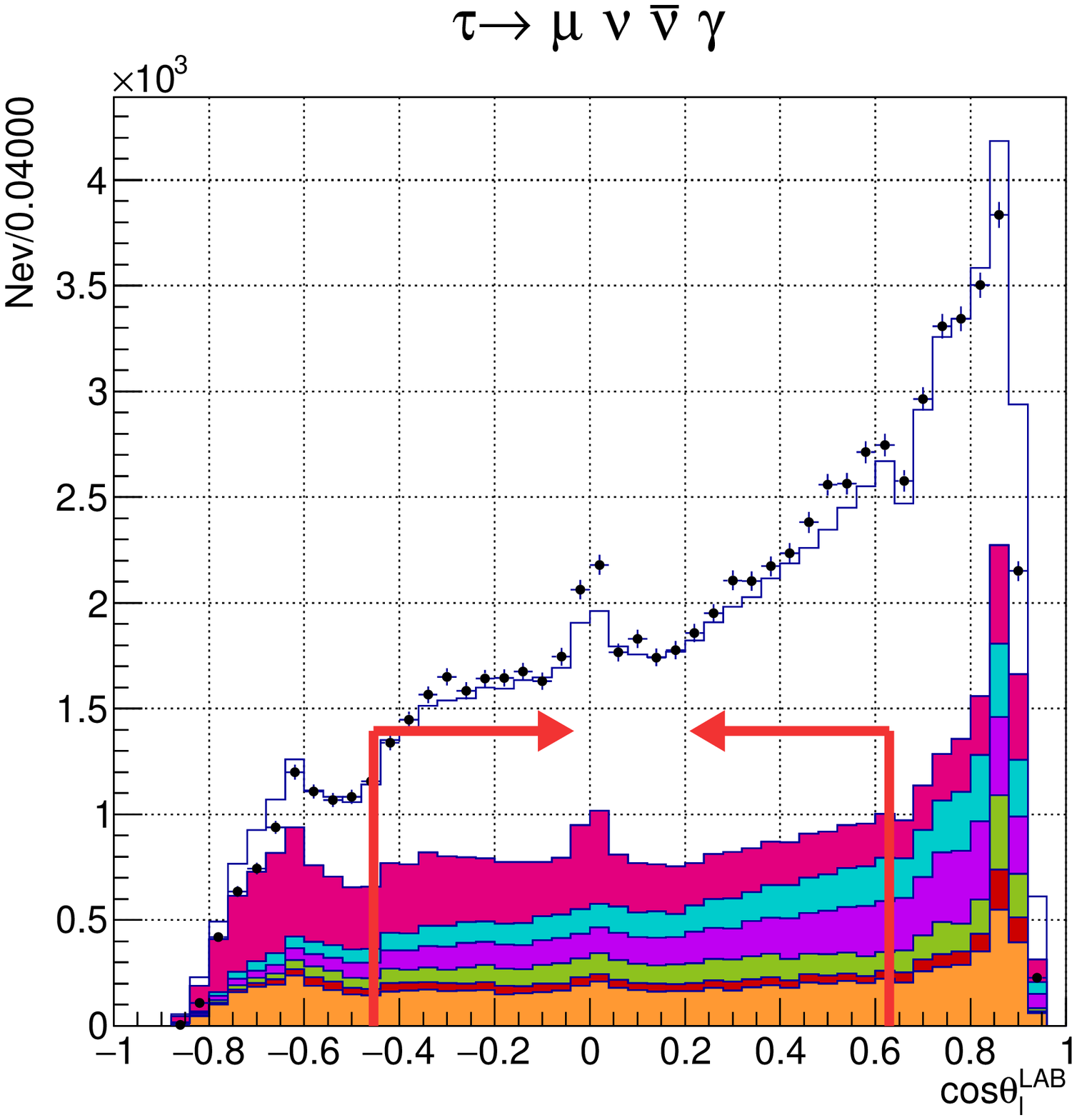}}
\centering \subfloat[$P_e^{\rm LAB}$]{\hspace{-3mm} \includegraphics[width=5.5cm]{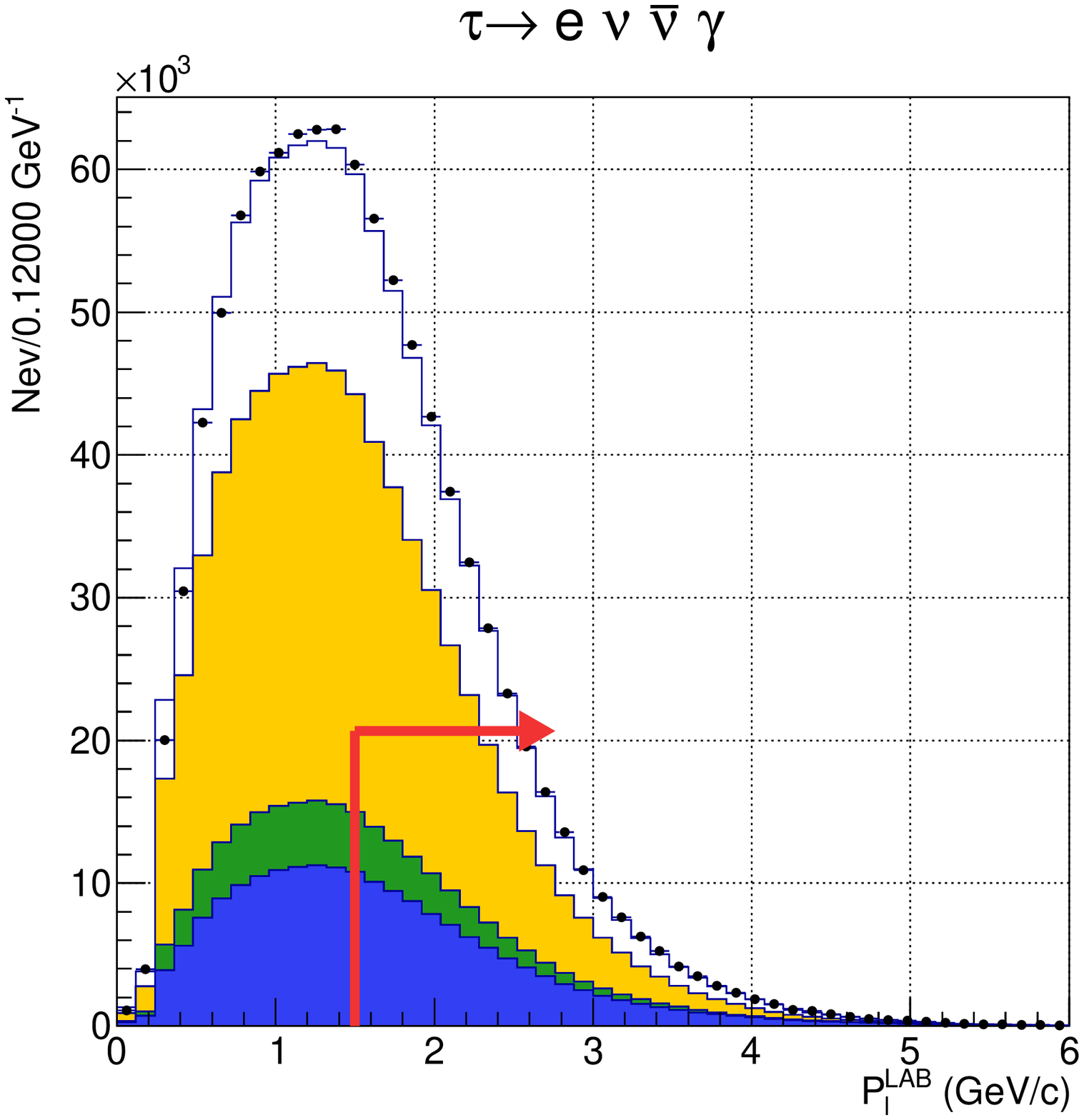}} 
\caption[]{Cosine of the polar angle for the electron (a) and muon (b) modes, and momentum of electron (c).
The color of each histogram is explained in the caption of Fig.~\ref{pi0dist} and the red arrows indicate the selection windows: 
(a) $\theta_{e}^{\rm LAB}<126^\circ$, (b) $51^\circ<\theta_{\mu}^{\rm LAB}<117^\circ$ 
and (c) $P_e^{\rm LAB}>1.5~$GeV$/c$. The relative drop of the efficiencies are approximately 2\%, 50\% and 36\% 
for (a), (b), and (c), respectively. The small peak around $\theta_{\rm LAB}\sim 90^\circ $ seen in (b) comes from the
 beam background.}\label{sel_br_opt_1}
\end{figure}

\begin{figure}[]
\centering \subfloat[]{\hspace{-3mm} \includegraphics[width=6.0cm]{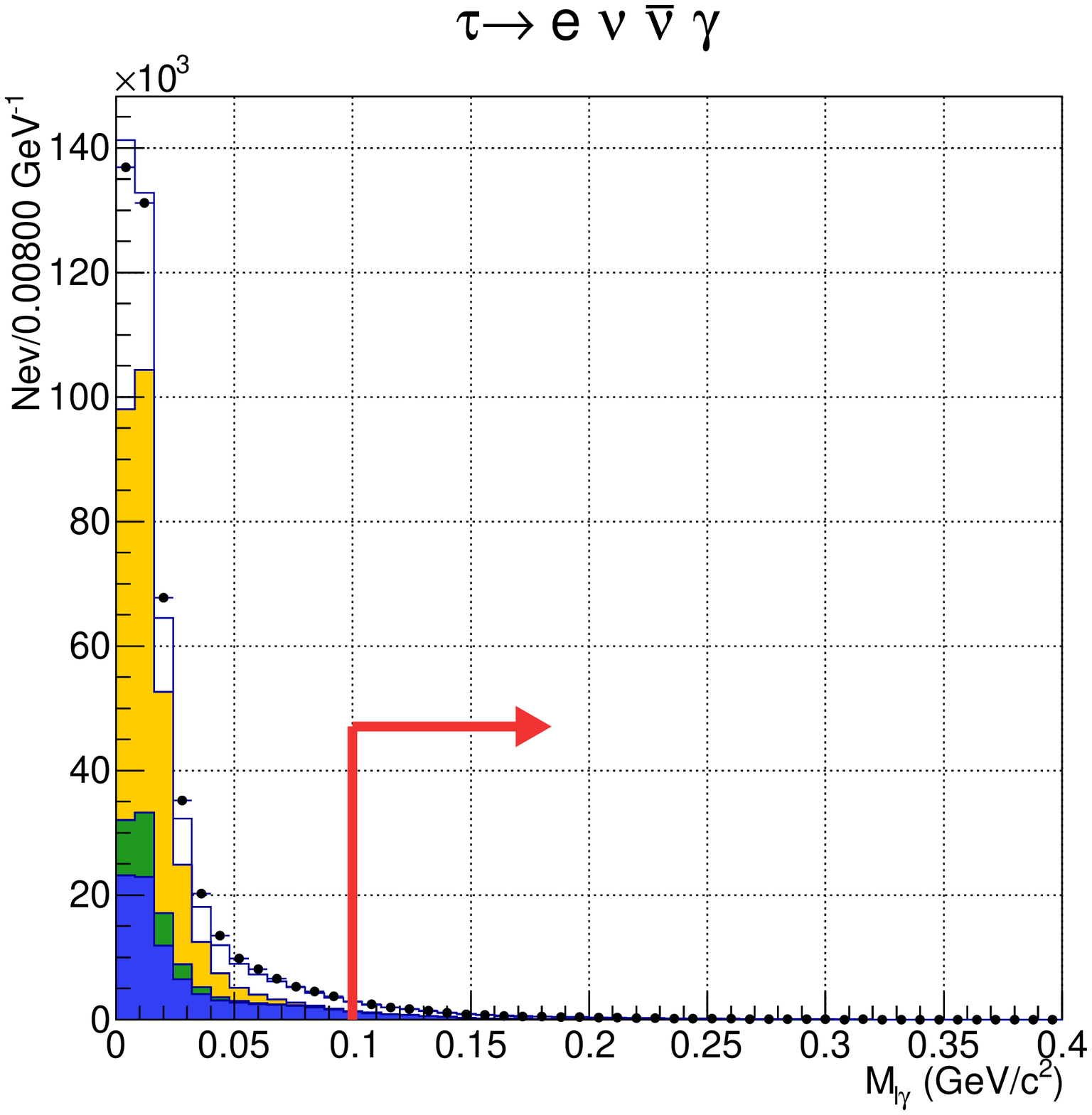}} 
\centering \subfloat[]{\hspace{-3mm} \includegraphics[width=6.0cm]{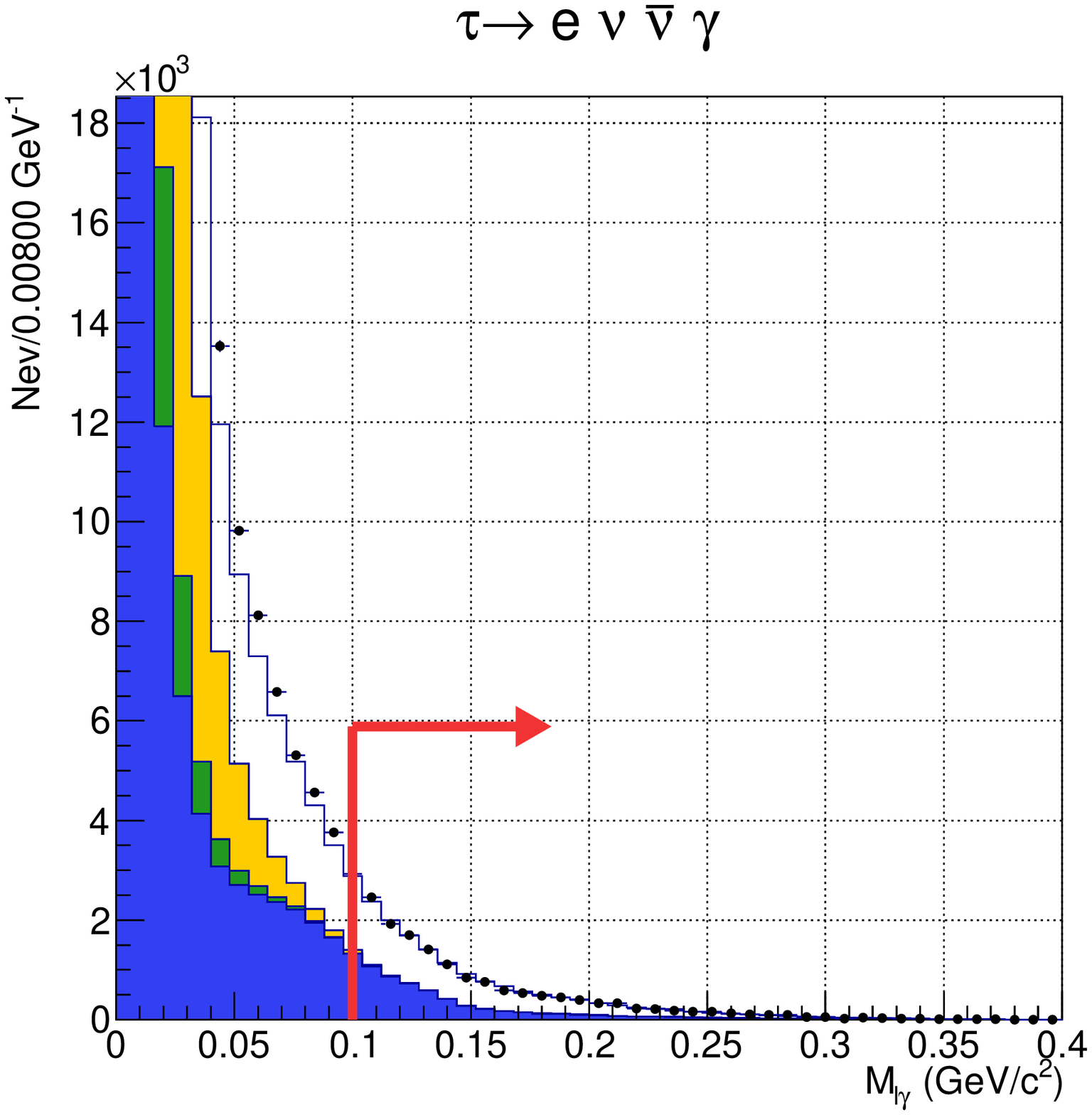}} 
\caption[]{Distribution of the invariant mass of $e-\gamma$ system, $M_{e\gamma}$.
The color of each histogram is explained in Fig.~\ref{pi0dist} and the red arrows indicate 
the selection windows. (a) overall view (b) enlarged view. The relative decrease of the 
efficiency is 93\%.}\label{sel_br_opt_3}
\end{figure}

\subsection{Evaluation of systematic uncertainties}

In Table~\ref{R_opt_tabl}, we summarize the sources of the systematic uncertainties of the branching ratios
 of the electron and muon modes. To estimate a systematic uncertainty from the efficiency correction, $\bar{R}$, 
we use the following method. The uncertainties of the $R_{\ell {\rm ID}}$ are
 determined by the finite statistics of $e^+e^- \to \ell^+ \ell^- e^+e^-$ sample,
 a comparison of $R_{\ell {\rm ID}}$ from $e^+e^- \to \ell^+ \ell^- e^+e^-$ and
 $R_{\ell {\rm ID}}$ from $J/\psi \rightarrow \ell^+ \ell^-$, and its time variation during the experiment.
 
$R_{\pi {\rm ID}}$ values are estimated from the finite statistics of a $D^{*+} \to D^0 (K^-\pi^+)\pi^+$ sample,
 the fit of the reconstructed mass distribution of $D^*$, and observation of time variation. 

The systematic uncertainties of the
 $R_{\pi^0 {\rm ID}}$, $R_{\gamma {\rm ID}}$,
 $R_{\rm trg}$, and $R_{\rm trk}$ values are estimated
 from a comparison between $\bar{R}$ and unity.

The uncertainty of $\mathcal{B}(\tau^+ \rightarrow \pi^+\pi^0  \bar{\nu}_\tau)$ 
is taken from the PDG average value~\cite{PDG_paper}
 and that of $\sigma(e^+e^-\rightarrow \tau^+\tau^-)$ is taken from
 Ref.~\cite{tautau_cross}.

The statistical uncertainty of MC events is ignored because
its fluctuation is small.
 
The evaluation of the systematic uncertainty of the purity $f_{\mathrm{bg}}$ is 
estimated based on sideband information.
 The sideband events are selected by the following criteria:
 $M_{e\gamma}<0.1$~GeV$/c^2$ and $0.90<$cos$\theta_{e\gamma}<0.94~(0.99)$ for the 
electron (muon) mode,
 where other selection criteria are common with those of the signal extraction.
 The difference of the background yield in the sideband region between 
MC simulation and the real experiment
 is $4.4\%$ (5.5\%) for the electron (muon) mode.
 Taking each fraction into account, we estimate that the resulting uncertainty is
 $1.3\%$ and $1.5\%$.
 
 
The effect of detector response is estimated
 by varying selection cut parameters.
 The effect due to variation of the photon energy threshold is
 based on the energy resolution at the threshold, and found to be 5\%~\cite{citeBelle}.
 The variation of other selection criteria is determined based on the 
propagation of the error matrix of the helix parameters.
 Of all the selection criteria, the requirement of 
$M_{e\gamma}>0.1~$GeV$/c^2$ has the largest impact.
 
As mentioned, to estimate the efficiency, we define the radiative events by
 the imposition of a photon energy threshold of $E_\gamma^* = 10$~MeV 
in the $\tau^-$ rest frame.
 However, in the real experiment, we
 cannot precisely determine this energy because the $\tau^- $
 momentum
 is not directly reconstructed. 
 Accordingly, there is a chance that an event that has a photon
 with an energy less than the threshold is also reconstructed
 as signal. This is only possible in a limited phase space, and
 such events are included in the selection with fractions of
 $1.1\%$ and $0.3\%$
 for electron and muon events, respectively.
 We take these fractions as sources of systematic effects due to
 the uncertainty of the experimental $E_\gamma^* $ threshold.
 
\begin{table}[]
\caption{
Systematic uncertainties (\%)
 on $\mathcal{B}(\tau^- \to \ell^- \nu_\tau \bar{\nu}_\ell \gamma )$
 for different configurations. \label{R_opt_tabl}
}
\begin{center}
\begin{tabular}{lcccc}  \hline \hline  
Item & $(e^-\gamma, \pi^+\pi^0)$ & $(e^+\gamma, \pi^-\pi^0)$ & $(\mu^-\gamma, \pi^+\pi^0)$ & $(\mu^+\gamma, \pi^-\pi^0)$  \\ \hline

$R_{\ell\mathrm{ID}}$ & $1.9$& $1.9$& $1.1$& $1.1$\\
$R_{\pi \mathrm{ID}}$ & $0.7$& $0.7$& $0.7$& $0.7$\\
$R_{\gamma \mathrm{ID}}$ & $1.0$ & $1.0$& $0.4$& $0.4$\\
$R_{\pi^0\mathrm{ID}}$ & $3.6$& $3.6$& $3.3$ & $3.3$\\
$R_{\mathrm{trg}}$ & $1.2$ & $1.2$ & $0.7$ & $0.7$\\
$R_{\rm trk.}$ & $0.7$& $0.7$& $0.7$ & $0.7$\\
Purity $(1-f_{\mathrm{bg}})$ & 1.3 & 1.3 & $1.5$ & $1.5$ \\ 
Detector response & 1.5 & 1.5 & 0.6 & 0.6 \\
Uncertainty of $E_\gamma^* $ threshold & 1.1 & 1.1 & 0.3 & 0.3 \\  
Luminosity & $1.4$ & $1.4$ & $1.4$ & $1.4$ \\
$\mathcal{B}(\tau^+\rightarrow \pi^+\pi^0\nu)$ & $0.4$& $0.4$ & $0.4$ & $0.4$\\
$\sigma(e^+e^-\rightarrow \tau^+\tau^-)$& $0.3$ & $0.3$ & $0.3$& $0.3$\\ \hline
Total & $5.3$ &  $5.3$ &  $4.3$&   $4.3$\\ \hline \hline
\end{tabular}
\end{center}
\end{table}

\subsection{Result}

In Table~\ref{sum_br2}, we show the result of measurements separately 
for the four following configurations:
 $(e^-\gamma, \pi^+\pi^0)$, $(e^+\gamma, \pi^-\pi^0)$, 
$(\mu^-\gamma, \pi^+\pi^0)$ and $(\mu^+\gamma, \pi^-\pi^0)$.
They are combined to give
\begin{align}
\mathcal{B}(\tau^- \rightarrow e^- \nu_\tau \bar{\nu}_e \gamma)_{E_{\gamma}^*>10~\mathrm{MeV}}   &= (1.79 \pm 0.02 \pm 0.10)\times 10^{-2}, \\
\mathcal{B}(\tau^- \rightarrow \mu^- \nu_\tau \bar{\nu}_\mu \gamma)_{E_{\gamma}^*>10~\mathrm{MeV}} &= (3.63 \pm 0.02 \pm 0.15) \times 10^{-3},
\end{align}
where the first error is statistical and the second is systematic.
 In Table~\ref{br_info}, we summarize the current experimental and theoretical 
 information on these decays. 
 While the LO theoretical calculations for these decays 
 were done long time ago, NLO corrections were 
 considered thoroughly only recently in Ref.~\cite{radiativedecay_theory}, 
 where the importance of taking into account the hard doubly-radiative decays 
 was emphasized.

We also obtain the dependence of Michel parameters on the
 $E_{\mathrm{extra}\gamma}^{\mathrm{LAB}}$ selection, as shown in Fig.~\ref{sel_br_opt_5}.
 These results are consistent with the LO theoretical prediction.
 
\begin{table}[]
\caption{Summary of results for the branching ratio measurement\label{sum_br2}}
\begin{center}
\scalebox{0.9}{ \begin{tabular}{lcccc}  \hline \hline  
Item & $(e^-\gamma, \pi^+\pi^0)$ & $(e^+\gamma, \pi^-\pi^0)$ & $(\mu^-\gamma, \pi^+\pi^0)$ & $(\mu^+\gamma, \pi^-\pi^0)$  \\ \hline
$N_{\mathrm{\mathrm{obs}}}$ & $6188\pm79$ & $6114\pm78$& $10458\pm102$ & $11170\pm106$ \\
$1-f_{\mathrm{bg}}\hspace{-1mm}$ ~($\%$)   & $70.2\pm 0.9$& $70.2\pm 0.9$ & $71.5\pm 1.0$ & $71.5\pm 1.0$ \\
$\bar{\varepsilon}^{\mathrm{MC}}$~($\%$) & $0.172 \pm 0.001 $& $ 0.169 \pm 0.001 $ & $ 1.26 \pm 0.01$ & $1.27 \pm 0.01$\\
$\bar{R}$ & $0.85 \pm 0.04$ & $0.85 \pm 0.04$ & $0.93 \pm 0.03$ & $0.93 \pm 0.03$\\ 
$\bar{\varepsilon}^{\mathrm{EX}}$~($\%$) & $0.146 \pm 0.007$& $0.144\pm 0.007$ & $1.28 \pm 0.05$ & $1.29 \pm 0.05$\\ \hline 
$\mathcal{B}$~$(\%)$&$1.79\pm0.02\pm0.10$& $1.80\pm0.02\pm0.10$ &$0.352\pm0.003\pm0.015$& $0.373\pm0.003\pm0.016$\\ \hline \hline
\end{tabular} }
\end{center}
\end{table}

\begin{table}[]
\caption{Information on the branching ratios of the radiative leptonic 
$\tau$ decays.\label{br_info}}
\begin{center}
\scalebox{0.95}{
\begin{tabular}{lcc}  \hline \hline 
 & $\tau^- \rightarrow e^- \nu_\tau \bar{\nu}_e\gamma$ & $\tau^- \rightarrow \mu^- \nu_\tau \bar{\nu}_\mu\gamma$ \\ \hline
This measurement & $(1.79 \pm 0.02 \pm 0.01)\times 10^{-2}$ & $(3.63 \pm 0.02 \pm 0.15)\times 10^{-3}$ \\ 
CLEO (experiment) \cite{taurad_cleo} & $(1.75 \pm 0.06 \pm 0.017)\times 10^{-2}$ & $(3.61 \pm 0.16 \pm 0.35)\times 10^{-3}$ \\
BaBar (experiment) \cite{babar_mes} & $(1.847 \pm 0.015 \pm 0.052)\times 10^{-2}$ & $(3.69 \pm 0.03 \pm 0.10)\times 10^{-3}.$\\ \hline
LO (theory) \cite{radiativedecay_theory}& $1.834 \times 10^{-2}$ & $3.663 \times 10^{-3}$ \\
NLO inclusive (theory) \cite{radiativedecay_theory} & $1.728 \times 10^{-2}$ & $3.605 \times 10^{-3}$ \\
NLO exclusive (theory) \cite{radiativedecay_theory} & $1.645 \times 10^{-2}$ & $3.572 \times 10^{-3}$ \\ \hline \hline
\end{tabular}
}
\end{center}
\end{table}

\begin{figure}[]
\centering \subfloat[]{\hspace{-3mm} \includegraphics[width=8cm]{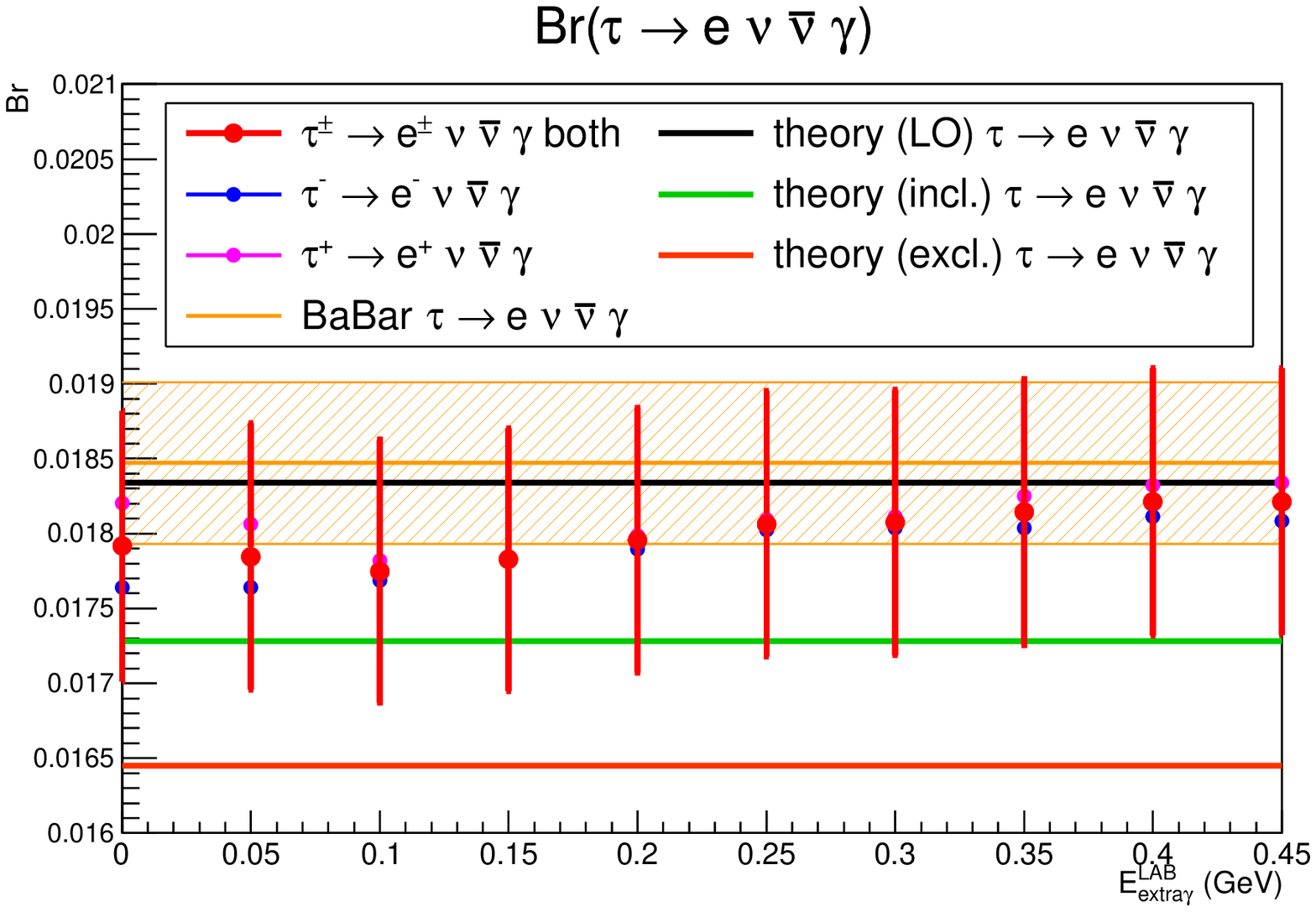}}
\centering \subfloat[]{\hspace{-3mm} \includegraphics[width=8cm]{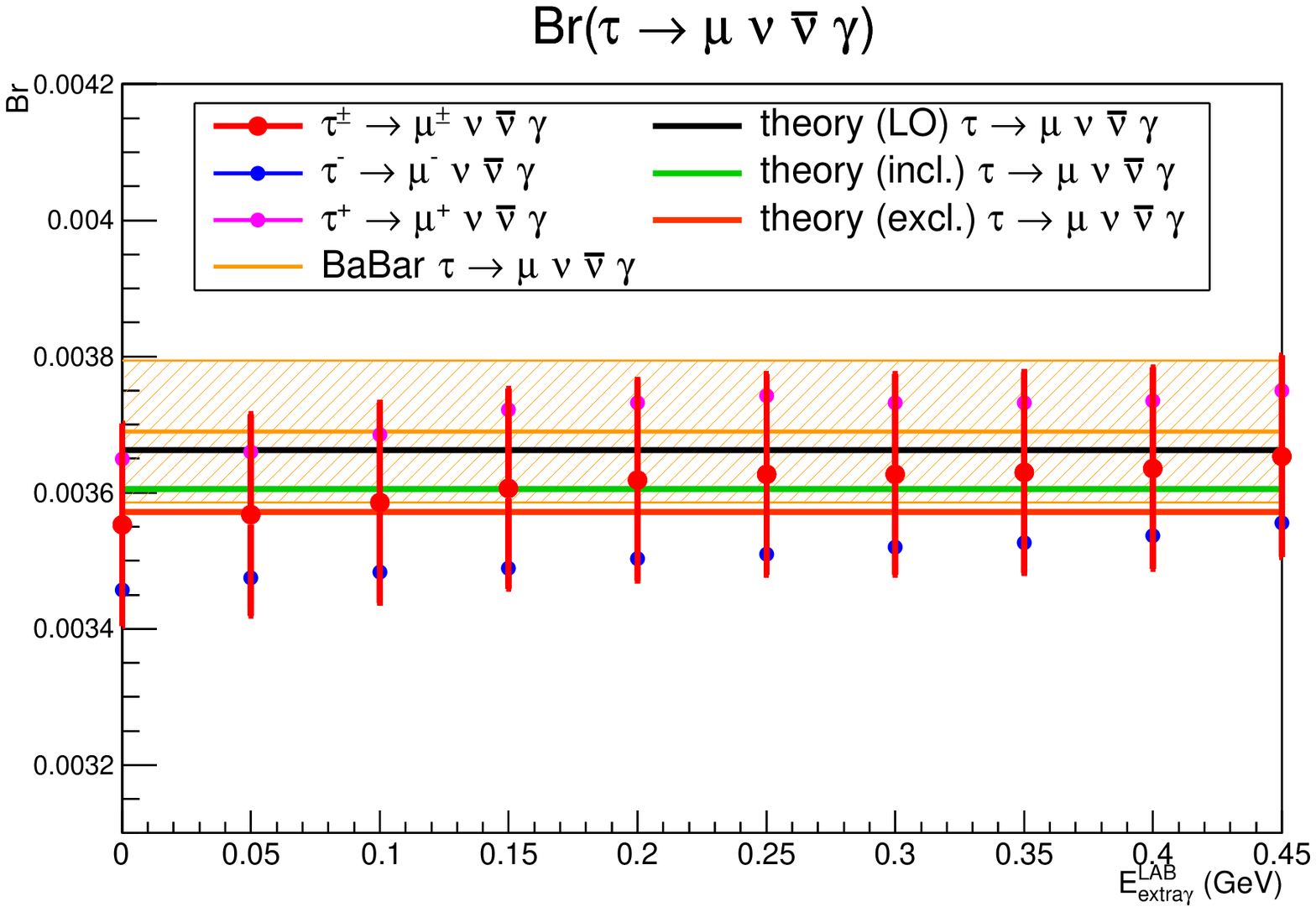}}
\caption[]{Branching ratio of $\tau^- \rightarrow \ell^- \nu_\tau \bar{\nu}_\ell \gamma$ 
decay as a function
 of $E_{\mathrm{extra}\gamma}^{\mathrm{LAB}}$ cut: (a) $\ell=e$ and (b) $\ell=\mu$.
 Red, blue, and magenta lines represent the measured branching ratio of
 $\tau^\pm \rightarrow \ell^\pm \nu \bar{\nu} \gamma$, $\tau^- \rightarrow \ell^- \nu_\tau \bar{\nu}_\ell \gamma$ and
 $\tau^+ \rightarrow \ell^+ \nu_\ell \bar{\nu}_\tau \gamma$, respectively.
 The bars represent uncertainties and are drawn only for the combined modes, where both statistical and systematic uncertainties are included.
 The orange band shows the BaBar measurement~\cite{babar_mes}.
 Black, green, and red lines are LO, NLO inclusive, and NLO exclusive
 theoretical predictions, respectively~\cite{radiativedecay_theory}.
}\label{sel_br_opt_5}
\end{figure}

As summarized in Table~\ref{R_opt_tabl},
 the dominant systematic contribution comes from
 the uncertainty of the efficiency correction for $\pi^0$.
 This uncertainty is canceled when
 we measure the ratio of branching fractions
 $\mathcal{Q} = \mathcal{B}(\tau^- \rightarrow e^- \nu_\tau \bar{\nu}_e \gamma)/\mathcal{B}(\tau^- \rightarrow \mu^- \nu_\tau \bar{\nu}_\mu \gamma)$.
 Moreover, other common systematic sources, namely $R_{\rm trk}$, $R_{\rm \pi ID}$, the integrated luminosity,
 the branching ratio of $\tau^+ \rightarrow \pi^+\pi^0 \bar{\nu}_\tau$ decay, and
 the cross section $\sigma(e^+e^- \rightarrow \tau^+\tau^-)$, also cancel.
 The obtained ratio is
\begin{align}
\mathcal{Q} = \frac{\mathcal{B}(\tau^- \rightarrow e^- \nu_\tau \bar{\nu}_e \gamma)_{E_{\gamma}^*>10~\mathrm{MeV}}}
{\mathcal{B}(\tau^- \rightarrow \mu^- \nu_\tau \bar{\nu}_\mu \gamma)_{E_{\gamma}^*>10~\mathrm{MeV}}} = 4.95 \pm 0.06 \pm 0.20,
\end{align}
where the first error is statistical and the second is systematic.
 In Table~\ref{Q_tbl}, we summarize the theoretical prediction and past experimental results
 for the ratio $\mathcal{Q}$.
  
\begin{table}[]
\caption{Comparison of the ratio $\mathcal{Q}$ (${E_{\gamma}^*>10~\mathrm{MeV}}$) \label{Q_tbl}} 
\begin{center}
\begin{tabular}{lc}  \hline \hline  
${\rm \bf Theory}$ \\ 
Leading order & 5.007 \\
Next-to-leading order incl. & 4.793 \\
Next-to-leading order excl. & 4.605 \\ \hline
${\rm \bf Experiment}$ \\
CLEO & 4.9 $\pm$ 0.6 \cite{taurad_cleo}\\
BaBar & 5.01 $\pm$ 0.20 $\dagger$ \cite{babar_mes}\\
This measurement & 4.95 $\pm$ 0.06 $\pm$ 0.20 \\ \hline \hline
\end{tabular}
\end{center}
\begin{flushleft}\hspace{3.1cm} $^\dagger${
\small Systematic uncertainty is calculated from the reference\\
\hspace{3.4cm} values, where cancellation is not taken into account.\\
\hspace{3.4cm} The statistical and systematic uncertainties are combined for\\
\hspace{3.4cm} the CLEO and BaBar measurements.}\end{flushleft}
\end{table}

\section{Conclusion}

We present the measurement of Michel parameters $\bar{\eta}$ and $\xi\kappa$ of
 the $\tau$ using 711~f$\mathrm{b}^{-1}$ of data collected with the Belle detector at the KEKB $e^+e^-$ collider.
 These parameters are extracted from the radiative leptonic decay $\tau^-\rightarrow \ell^- \nu_\tau \bar{\nu}_\ell \gamma$
 which is tagged by $\tau^+ \rightarrow \rho^+(\rightarrow \pi^+ \pi^0 ) \bar{\nu}_\tau$ decay of the partner $\tau^+$
 to exploit the spin-spin correlation in $e^+ e^- \to \tau^+ \tau^-$.
 Due to the small sensitivity to $\bar{\eta}$ in the electron mode,
 this parameter is extracted only from $\tau^-\rightarrow \mu^- \nu_\tau \bar{\nu}_\mu \gamma$
 to give $\bar{\eta} = -1.3 \pm 1.5 \pm 0.8$.
 The product $\xi\kappa$ is measured using both decays $\tau^-\rightarrow \ell^- \nu_\tau \bar{\nu}_\ell \gamma$ ($\ell=e$ and $\mu$)
 to be $\xi\kappa = 0.5 \pm 0.4 \pm 0.2$. The first error is statistical and the second
 is systematic. This is the first measurement of both parameters for the $\tau$ lepton.
 These values are consistent with the SM expectation within uncertainty. 
 
For a consistency check of the procedure to measure the Michel
 parameters, we measure the branching ratio of
 $\tau^-\rightarrow \ell^- \nu_\tau \bar{\nu}_\ell \gamma$ decay.
 The obtained values are consistent with the LO theoretical
 prediction and support the measurement by BaBar, which is known to
 deviate from the SM exclusive branching ratio by 3.5$\sigma$.
 Accounting for the agreement between our result and the BaBar measurement~\cite{babar_mes}, the implementation
 of the NLO formalism in the TAUOLA generator is required to carry out more precise measurements.


\section*{Acknowledgments}

We would like to express our deepest appreciation to
 A.Arbuzov, M.Fael, T.Kopylova, L.Mercolli and M.Passera
 for very useful discussions and theoretical support of this work.
We thank the KEKB group for the excellent operation of the
accelerator; the KEK cryogenics group for the efficient
operation of the solenoid; and the KEK computer group,
the National Institute of Informatics, and the 
PNNL/EMSL computing group for valuable computing
and SINET5 network support.  We acknowledge support from
the Ministry of Education, Culture, Sports, Science, and
Technology (MEXT) of Japan, the Japan Society for the 
Promotion of Science (JSPS), and the Tau-Lepton Physics 
Research Center of Nagoya University; 
the Australian Research Council;
Austrian Science Fund under Grant No.~P 26794-N20;
the National Natural Science Foundation of China under Contracts 
No.~10575109, No.~10775142, No.~10875115, No.~11175187, No.~11475187, 
No.~11521505 and No.~11575017;
the Chinese Academy of Science Center for Excellence in Particle Physics; 
the Ministry of Education, Youth and Sports of the Czech
Republic under Contract No.~LTT17020;
the Carl Zeiss Foundation, the Deutsche Forschungsgemeinschaft, the
Excellence Cluster Universe, and the VolkswagenStiftung;
the Department of Science and Technology of India; 
the Istituto Nazionale di Fisica Nucleare of Italy; 
the WCU program of the Ministry of Education, National Research Foundation (NRF)
of Korea Grants No.~2011-0029457, No.~2012-0008143,
No.~2014R1A2A2A01005286,
No.~2014R1A2A2A01002734, No.~2015R1A2A2A01003280,
No.~2015H1A2A1033649, No.~2016R1D1A1B01010135, No.~2016K1A3A7A09005603, No.~2016K1A3A7A09005604, No.~2016R1D1A1B02012900,
No.~2016K1A3A7A09005606, No.~NRF-2013K1A3A7A06056592;
the Brain Korea 21-Plus program, Radiation Science Research Institute, Foreign Large-size Research Facility Application Supporting project and the Global Science Experimental Data Hub Center of the Korea Institute of Science and Technology Information;
the Polish Ministry of Science and Higher Education and 
the National Science Center;
the Ministry of Education and Science of the Russian Federation and
the Russian Foundation for Basic Research;
the Slovenian Research Agency;
Ikerbasque, Basque Foundation for Science and
MINECO (Juan de la Cierva), Spain;
the Swiss National Science Foundation; 
the Ministry of Education and the Ministry of Science and Technology of Taiwan;
and the U.S.\ Department of Energy and the National Science Foundation.

\appendix

\section*{Appendix~A: Differential decay width of $\tau^- \rightarrow \ell^- \nu_\tau \bar{\nu}_\ell \gamma$ \label{app_dif_sig}}
The general differential cross section of $\tau^- \rightarrow \ell^- \nu_\tau \bar{\nu}_\ell \gamma$ decay is expressed
 as a sum of the two terms:

\begin{equation}
\frac{\mydif\Gamma( \tau^\mp \rightarrow l^\mp \nu_{\tau} \bar{\nu_{\ell}} \gamma )}{\mydif E^*_\ell
 \mydif \Omega^*_\ell \mydif E^*_\gamma \mydif \Omega^*_\gamma } 
=  A \mp  \bvec{B}\cdot \bvec{S}_{\tau^\mp},
\end{equation}
where $A$ and $\bvec{B}$ represent spin-independent and spin-dependent terms, $E^*_i$ ($i=\ell,\gamma$) is
 the energy in the $\tau$ rest frame and $\Omega_i$ ($i=\ell,\gamma$) is the
 solid
 angle defined by $\{\cos\theta_i, \phi_i\}$ ($i=\ell,\gamma$).
 These terms are functions of dimensionless kinematic parameters $x,y$ and $d$ defined as
\begin{align}
r &= \frac{m_\ell}{m_{\tau}}, \\
x &= \frac{2E_{\ell}^*}{m_{\tau}},~~(2r < x < 1+r^2) \\ 
y &= \frac{2E_{\gamma}^*}{m_{\tau}},~~(0 < y < 1-r) \\ 
d &= 1-\beta_{\ell}^* \cos \theta_{\ell \gamma}^*, \\ 
y &< \frac{2(1+r^2-x)}{2-x+\cos \theta_{\ell\gamma}^* \sqrt{x^2-4r^2}}, \label{Dynamicrelation} 
\end{align}
$A$ and $\bvec{B}$ are parametrized by the Michel parameters $\rho$, $\eta$, $\xi$, $\xi\delta$, $\bar{\eta}$, $\eta^{\prime \prime}$, and $\xi\kappa$. 

\begin{align}
A(x,y,d)&= \frac{4\alpha G_F^2 m_{\tau}^3 }{3(4\pi)^6 } \cdot \beta_\ell^* \sum_{i=0,1\ldots 5}{F_{i}r^i}, \\
\bvec{B}(x,y,d)&= \frac{4 \alpha G_F^2 m_{\tau}^3 }{3(4\pi)^6 } \cdot \beta_\ell^* \sum_{i=0,1\ldots 5}{(\beta_{l}^*G_{i}\bvec{n}_{l}^{*}+H_{i}\bvec{n}_{\gamma}^{*})r^i} \label{sigeq_1},
\end{align}
where $\bvec{n}_l^*$ and $\bvec{n}_\gamma^*$ are normalized directions of lepton and photon in the $\tau$ rest frame,
 respectively, and $\beta_\ell^*$ is a velocity of daughter lepton in this frame.
The $F_i$, $G_i$ and $H_i$ ($i=0,1\ldots 5$) are functions of $x$, $y$, $d$, and $r$ and their explicit formulae are given in the
 Appendix of Ref.~\cite{michel_form_arvzov}.

\section*{Appendix~B: Differential decay width of $\tau \rightarrow \rho \nu_\tau$}

We use the CLEO model to define the differential decay width of $\tau^\pm \rightarrow \rho^\pm \nu_\tau$ decay.
 This is expressed as a sum of the spin-independent and spin-dependent parts~\cite{rhoCLEO}:
\begin{align}
&\hspace{3cm}\frac{{\mathrm{d}\Gamma(\tau^\pm\rightarrow}
  \pi^\pm\pi^0\nu_\tau)}{\mydif \Omega_{\rho}^*\mydif m^2 \mydif \widetilde{\Omega}_{\pi}} = A^+ \mp \xi_\rho \bvec{B}^+ \cdot \bvec{S_{\tau^\pm}}, \\
 A^+ =& \frac{G_F^2 |V_{ud}|^2}{(4\pi)^5 } \cdot \Big[2(E_{\pi}^*-E_{\pi^0}^*)(p_{\nu}\cdot q)-E_{\nu}^*q^2\Big] \cdot \mathrm{BPS},\\
\bvec{B}^+ =& \frac{G_F^2 |V_{ud}|^2}{(4\pi)^5 } \cdot  \Big[
  \bvec{ P}_{\pi}^* \left\{ ( q\cdot q) +2(p_{\nu}\cdot q)\right\}
+\bvec{P}_{\pi^0}^* \left\{ ( q\cdot q) -2(p_{\nu}\cdot q)\right\} 
 \Big] \cdot \mathrm{BPS}, \label{sigeq_2}
\end{align}
where $V_{ud}$ is the corresponding element of the Cabibbo-Kobayashi-Maskawa matrix,
 $E_i^*$ and $\bvec{ P}^*_{i}$ ($i=\pi,\pi^0$) are energies and three-momenta measured in the $\tau$ rest frame,
 $\Omega_\rho^*$ is
 the solid angle of the $\rho$ meson in the $\tau$ rest frame, $\tilde{\Omega}_\pi$ is the solid angle of the pion in the $\rho$ rest frame,
 $q$ is a four-momentum defined by $q=p_{\pi}-p_{\pi^0}$, and $p_\nu$ is the four-momentum of the
 tau neutrino.
 The factor BPS stands for the square of a relativistic Breit-Wigner function and a Lorentz-invariant phase space, and is calculated as follows:
\begin{align}
\mathrm{BPS} =& \left|\mathrm{BW}(m^2) \right|^2\ \left( \frac{2P_{\rho}^{*}(m^2)}{m_\tau} \right) \left(\frac{2\tilde{P}_{\pi}(m^2)}{m_{\rho}} \right), ~~~
\mathrm{BW}(m^2) =\frac{\mathrm{BW}_\rho+\beta \mathrm{BW}_{\rho^{\prime}}}{1+\beta}, \\
\mathrm{BW}_{\rho}\left(m^2\right) =& \frac{m_{\rho 0}^{2}}{m_{\rho 0}^{2}-m^{2}-im_{\rho 0}\Gamma_{\rho }\left(m^{2}\right)}
,~~~ \Gamma_{\rho}\left(m^2\right) =\displaystyle \Gamma_{\rho 0}\frac{m_{\rho 0}}{\sqrt{m^{2}}} \left(\frac{\tilde{P_{\pi}}\left(m^{2}\right)}{\tilde{P_{\pi}}\left(m_{\rho 0}^{2}\right)} \right)^{3}, \\
\mathrm{BW}_{\rho^{\prime}}\left(m^2\right) =& \frac{m_{\rho^{\prime}0}^{2}}{m_{\rho^{\prime} 0}^{2}-m^{2}-im_{\rho^{\prime}0}\Gamma_{\rho^{\prime}}\left(m^{2}\right)}
,~~~\Gamma_{\rho^{\prime } }\left(m^2\right)=\displaystyle \Gamma_{\rho^{\prime } 0}\frac{m_{\rho^{\prime }0}}{\sqrt{m^{2}}}\left(\frac{\tilde{P_{\pi}}\left(m^{2}\right)}{\tilde{P_{\pi}}\left(m_{\rho^{\prime } 0}^{2}\right)}\right)^{3}, \\
\hspace{2cm}&\tilde{P}_{\rho}^*(m^2) = \frac{m_\tau^2- m^2}{2 m_\tau} \\
\hspace{2cm}&\tilde{P}_{\pi}(m^2) = \frac{\sqrt{\left[m^2-(m_\pi-m_{\pi^0})^2\right]\left[m^2-(m_\pi +m_{\pi^0})^2\right]}}{2m}.
\end{align}
The factor $\mathrm{BW}_a$ ($a = \rho$ or $\rho^\prime$) represents the Breit-Wigner function associated with the
 resonance mass shape, and the parameter $\beta$ specifies their relative coupling.
 The nominal masses of the two resonance states are 
 $m_{\rho0}$ and $m_{\rho^{\prime}0}$, and their nominal total decay widths are
 $\Gamma_{\rho0}$ and $\Gamma_{\rho^{\prime}0}$.

\end{document}